\documentclass[%
reprint,
nofootinbib,
amsmath,amssymb,
aps,
]{revtex4-2}

\usepackage{graphicx}
\usepackage{dcolumn}
\usepackage{bm}
\usepackage{hyperref}
\usepackage{xcolor}
\usepackage{multirow}
\usepackage[normalem]{ulem}

\newcommand{\Geffsq}[0]{G^2_\mathrm{eff}}
\newcommand{\Geff}[0]{G_\mathrm{eff}}
\newcommand{\logGeff}[0]{\log_{10}(G_\mathrm{eff}/\mathrm{MeV}^{-2})}
\newcommand{\kmsMpc}[0]{\mathrm{km\:s}^{-1}\mathrm{Mpc}^{-1}}

\newcommand{\aap}{{A\&A}}

\begin{document}

\preprint{APS/123-QED}

\title{Strong constraints on a simple self-interacting neutrino cosmology}

\author{David Camarena}
 \email{dcamarena93@unm.edu}
\author{Francis-Yan Cyr-Racine}%
\affiliation{%
 Department of Physics and Astronomy,\\ University of New Mexico, Albuquerque, New Mexico 87106, USA
}%
\date{\today}

\begin{abstract}

Some cosmic microwave background (CMB) data allow a cosmological scenario in which the free streaming of neutrinos is delayed until close to matter-radiation equality. Interestingly, recent analyses have revealed that large-scale structure (LSS) data also align with this scenario, discarding the possibility of an accidental feature in the CMB sky and calling for further investigation into the free-streaming nature of neutrinos. By assuming a simple representation of self-interacting neutrinos, we investigate whether this nonstandard scenario can accommodate a consistent cosmology for both the CMB power spectra and the large-scale distribution of galaxies simultaneously. Employing three different approaches --- a profile likelihood exploration, a nested sampling method, and a heuristic Metropolis-Hasting approximation --- we exhaustively explore the parameter space and demonstrate that galaxy data exacerbates the challenge
already posed by the Planck polarization data for
this nonstandard scenario. We find that the Bayes factor disfavor strong interactions among neutrinos over the $\Lambda$CDM and $\Lambda$CDM + $N_\mathrm{eff}$ + $\sum m_\nu$ models with odds of $7:10000$ and $23:1000$, respectively, providing large evidence against the simple self-interacting neutrino model. Our analysis emphasizes the need to consider a broader range of phenomenologies in the early Universe. We also highlight significant numerical and theoretical challenges ahead in uncovering the exact nature of the feature observed in the data or, ultimately, confirming the standard chronological evolution of the Universe.

\end{abstract}

\maketitle


\section{\label{sec:introd}Introduction}

Despite their weak interactions making them elusive, neutrinos significantly influence the intricate evolution of the Universe. Their gravitational interactions leave sizable imprints in various cosmological observables across different times and scales. These features contribute to unveiling the nature of neutrinos through cosmological observations, enabling tight constraints on pivotal parameters describing neutrino properties, such as the sum of neutrino masses,~$\sum m_\nu$, and the effective number of relativistic species,~$N_\mathrm{eff}$~\cite[see~Refs.~][for instance]{Planck:2018vyg,Lattanzi:2017ubx,Sakr:2022ans}. Yet, cosmological phenomena could help us achieve a more comprehensive and profound understanding of neutrino physics that extends beyond constraints on~$\sum m_\nu$ and~$N_\mathrm{eff}$.

Cosmology holds the potential to contribute to a better understanding of neutrino physics, as observables provide a path to test the free-streaming nature of neutrinos and, by extension, new physics in this sector. According to the Standard Model (SM), neutrinos decouple from the cosmic plasma and commence free streaming as the Universe cools to approximately $\sim 1.5$ MeV. In the free-streaming phase, neutrinos remain gravitationally coupled, tugging any particles in their paths and leaving distinctive imprints on the evolution of cosmological perturbations~\cite{Bashinsky:2003tk}. New physics in the neutrino sector, however, can significantly alter these imprints~\cite[see Ref.][for instance]{Baumann:2015rya}.

In this context, novel neutrino interactions have been employed to investigate how altering the neutrino free streaming impacts cosmological observables~\cite[see Refs.~][for instance]{Konoplich:1988mj,Berkov:1988sd,Belotsky:2001fb,Cyr-Racine:2013jua,Archidiacono:2013dua,Lancaster:2017ksf,Oldengott:2017fhy,Choi:2018gho,Song:2018zyl,Lorenz:2018fzb,Barenboim:2019tux,Forastieri:2019cuf,Smirnov:2019cae,Escudero:2019gvw,Ghosh:2019tab,Funcke:2019grs,Sakstein:2019fmf,Mazumdar:2019tbm,Blinov:2020hmc,deGouvea:2019qaz,Froustey:2020mcq,Babu:2019iml,Kreisch:2019yzn,Park:2019ibn,Deppisch:2020sqh,Kelly:2020pcy,EscuderoAbenza:2020cmq,He:2020zns,Ding:2020yen,Berbig:2020wve,Gogoi:2020qif,Barenboim:2020dmg,Das:2020xke,Mazumdar:2020ibx,Brinckmann:2020bcn,Kelly:2020aks,Esteban:2021ozz,Arias-Aragon:2020qip,Du:2021idh,CarrilloGonzalez:2020oac,Huang:2021dba,Sung:2021swd,Escudero:2021rfi,RoyChoudhury:2020dmd,Carpio:2021jhu,Orlofsky:2021mmy,Green:2021gdc,Esteban:2021tub,Venzor:2022hql,Taule:2022jrz,RoyChoudhury:2022rva,Loverde:2022wih,Kreisch:2022zxp,Das:2023npl,Venzor:2023aka,Sandner:2023ptm}. In the last decade, analyses of cosmological scenarios with self-interacting neutrinos have shown that some data not only agree with a paradigm allowing moderate interactions among neutrinos~(MI$\nu$) --- effectively recovering the SM limit --- but also with a radical scenario where strong interactions among neutrinos delay their free streaming until close to matter-radiation equality~(SI$\nu$)~\cite{Cyr-Racine:2013jua,Archidiacono:2013dua,Lancaster:2017ksf,Oldengott:2017fhy,Barenboim:2019tux,Das:2020xke,Mazumdar:2019tbm,Brinckmann:2020bcn,RoyChoudhury:2020dmd,Kreisch:2019yzn,Park:2019ibn,Kreisch:2022zxp,Das:2023npl}. Although these divergent pictures were initially found in the cosmic microwave background (CMB), a recent CMB-independent analysis demonstrated that large-scale structure (LSS) observations also support the SI$_\nu$ scenario~\cite{Camarena:2023cku}, discarding the presumption of an accidental feature in the CMB data (see also Ref.~\cite{He:2023oke}) and increasing the riddle surrounding the free-streaming nature of neutrinos. 

The intriguing aspect of the SI$_\nu$ mode is noteworthy not only because it could hint at new physics in the early Universe but also due to its markedly varying statistical significance in different cosmological data contexts. For instance, polarization data provided by Planck~\cite{Planck:2018vyg} largely disfavor this extreme scenario~\cite{Das:2020xke,Mazumdar:2020ibx,Brinckmann:2020bcn,RoyChoudhury:2020dmd}. In agreement with this, standard neutrino free streaming appears to be favored by the agreement found between the imprints predicted by the SM and the observed phase of both CMB peaks~\cite{Follin:2015hya} and baryon acoustic oscillations (BAO)~\cite{Baumann:2017lmt,Baumann:2019keh}. In contrast, data from the Atacama Cosmology Telescope (ACT)~\cite{ACT:2020gnv} notably favors a delayed neutrino free streaming cosmology~\cite{Kreisch:2022zxp,Das:2023npl}, and more recently, an analysis of diverse cosmological data, including Lyman-$\alpha$ power spectrum from SDSS DR14 BOSS and eBOSS quasars~\cite{Chabanier:2018rga}, largely favors the SI$_\nu$ scenario~\cite{He:2023oke}.

Complementing this picture, our previous CMB-independent analysis~\cite{Camarena:2023cku} not only reveals a modest preference for a delayed onset of neutrino free streaming but also indicates a comparable phenomenology to that observed in the CMB data. Both data exhibit similarities in the self-interaction strength and important correlations between the neutrino interaction and the amplitude, $A_{\rm s}$, and tilt, $n_{\rm s}$, of the primordial curvature power spectrum. These similarities could suggest the existence of a self-consistent picture capable of providing a simultaneously good fit to the CMB and galaxy power spectra.

In this work, we investigate whether self-interacting neutrinos can provide a consistent cosmological scenario for both the CMB power spectra and the large-scale distribution of galaxies simultaneously. By adopting the simplest cosmological representation for self-interacting neutrinos, an effective four-fermion interaction model, and performing an exhaustive exploration of the parameter space considering three different statistical approaches --- a profile likelihood exploration, a nested sampling method, and a heuristic Metropolis-Hasting approximation --- we determine the statistical significance of the SI$_\nu$ scenario in light of the galaxy and CMB power spectra data. Our analyses robustly demonstrate that the inclusion of galaxy power spectra data diminishes the statistical significance of a strongly self-interacting neutrino cosmology, exacerbating the challenge already posed by the Planck polarization data. As detailed below, this reduction is mainly driven by the fact that strongly self-interacting neutrinos require a suppression of the effective amplitude of the CMB power spectrum that is significantly penalized by the Planck polarization data. Additionally, we describe and discuss some of the numerical challenges encountered when exploring a radical non-standard cosmological scenario with state-of-the-art cosmological data.

This paper is organized as follows. In Sec.~\ref{sec:model},
we present the framework used to model neutrino interactions and briefly review its impacts in the galaxy and CMB power spectra. The data and methodology used in this paper are presented in Sec.~\ref{sec:data}, while our results and discussion are shown in Sec.~\ref{sec:result}. Finally, we conclude in Sec.~\ref{sec:conclu}. 

\section{\label{sec:model} A delayed free streaming scenario: self-interactions in the neutrino sector}

According to the SM, neutrinos decouple from the primordial bath and commence to free stream when the temperature of the Universe drops to about $1.5$ MeV. Due to their gravitational coupling, the supersonic free-streaming neutrinos tug on any particles in their paths, leaving distinctive imprints on the cosmological perturbations~\cite{Bashinsky:2003tk}. However, novel interactions in the neutrino sector could alter this free-streaming pattern, affecting expected signatures in the cosmological phenomena. While the exact phenomenology of these interactions dictates specific alterations in neutrino free streaming, on the whole, self-interacting neutrinos do not contribute to the anisotropic stress of the Universe. Investigating scenarios involving self-interacting neutrinos that delay neutrino free streaming is thus of great significance, considering that free-streaming neutrinos are the exclusive source of anisotropic stress at early times. In this work, we adopt a simple self-interacting neutrino model to explore whether a cosmological scenario with a delayed neutrino free streaming can simultaneously fit the observed CMB and galaxy power spectra. 

A delayed neutrino free streaming cosmological scenario can be achieved by the simplest representation of self-interacting neutrinos: an effective four-fermion interaction universally coupling to all neutrino species, characterized by a dimensionful Fermi-like constant, $\Geff$, and interaction rate:
\begin{equation}
    \Gamma \equiv a \Geffsq T^5_\nu\,, \label{eq:Gamma_nu}
\end{equation} 
where $a$ is the scale factor and $T_\nu$ the temperature of neutrinos. This representation, which can be thought of as a Yukawa-type interaction between neutrinos and a scalar mediator particle, $\varphi$, of mass $m_\varphi \gtrsim 1$~keV, (see e.g.~Ref.~\cite{Berryman:2022hds} for a review) serves as a proxy to assess the impact of a delayed neutrino free streaming in the evolution of cosmological perturbation. Yet, this scenario is unlikely to correspond to a realistic representation of non-standard neutrino interactions.

As discussed in previous papers \cite[see][for instance]{Kreisch:2019yzn,Camarena:2023cku}, the model adopted here faces stringent constraints beyond cosmology~\cite{Kolb:1987qy,Manohar:1987ec,Dicus:1988jh,Davoudiasl:2005fd,Sher:2011mx,Fayet:2006sa,Choi:1989hi,Blennow:2008er,Galais:2011jh,Kachelriess:2000qc,Farzan:2002wx,Zhou:2011rc,Jeong:2018yts,Chang:2022aas,Fiorillo:2023cas,Fiorillo:2023ytr,Ahlgren:2013wba,Huang:2017egl,Venzor:2020ova,Ng:2014pca,Ioka:2014kca,Cherry:2016jol,Bilenky:1992xn,Bardin:1970wq,Bilenky:1999dn,Brdar:2020nbj,Lyu:2020lps,Brdar:2020nbj,Lyu:2020lps,Lessa:2007up,Bakhti:2017jhm,Arcadi:2018xdd,Blinov:2019gcj}.When taken at face value, such constraints only allow interactions that barely delay the neutrino free streaming, leading to cosmological scenarios where observables remain unchanged. Moreover, Planck polarization data largely disfavor this simple representation, featuring instead a preference for phenomenologies that include a flavor-dependent coupling~\cite{Das:2020xke,Mazumdar:2020ibx,Brinckmann:2020bcn,RoyChoudhury:2020dmd}. We employ this simple model, however, to provide an exhaustive joint analysis of the CMB and galaxy data. By limiting the complexity of the self-interaction model, we focus on broadly exploring the relevant parameter space. As discussed in the following sections, the cosmologies explored here significantly deviate from the standard paradigm and pose diverse challenges that complicate the analysis of the state-of-the-art cosmological data.

The impacts that self-interactions in the neutrino sector produce in the evolution of the Universe are particularly notable in the evolution of cosmological perturbations. Given that the interactions among neutrinos suppress the exclusive source of anisotropic stress in the early Universe, perturbations crossing the horizon during the tight-coupling phase exhibit distinct behavior from those crossing after the onset of free streaming, inducing a rich scale-dependent phenomenology in the cosmological observables. These effects become observable in the linear and mildly nonlinear scales for sufficiently large values of the self-interaction strength $\logGeff \gtrsim -4.5$.

Observationally, the presence of self-interactions in the neutrino sector increases the amplitude of the CMB power spectrum while also inducing a phase shift towards smaller scales --- larger $\ell$. In the linear matter power spectrum, the delay on the onset of the neutrino free streaming yields conspicuous scale-dependent signatures. In particular, interactions among neutrinos yield a suppression (enhancement) of the linear power spectrum at scales that cross the horizon when neutrinos are tightly coupled (commence to free-streaming). For a more detailed description, see Ref.~\cite{Camarena:2023cku} and references therein.

We highlight that $\Geff$ is given in units of MeV$^{-2}$, and that the usual Fermi constant corresponds to the value $\Geff \sim \mathcal{O}(10^{-11})$ MeV$^{-2}$. In order to consider a broad phenomenology, besides using $\logGeff$ to control the onset of neutrino free streaming, our model also considers the effective number of relativistic species, $N_\mathrm{eff}$, and the sum of neutrino masses,~$\sum m_\nu$, as free parameters. For the sake of simplicity, all models considered in this work assume a single massive neutrino containing all the mass instead of several degenerate massive neutrinos. 

\section{\label{sec:data} Data and methodology}

We perform cosmological analysis using our modified version of \texttt{CLASS-PT}~\cite{Blas:2011rf,Chudaykin:2020aoj}~\footnote{Our modified version of \texttt{CLASS-PT} as well as a more detailed description of our numerical implementation are available at \url{https://github.com/davidcato/class-interacting-neutrinos-PT}.} along with the \texttt{montepython} code~\cite{Audren:2012wb,Brinckmann:2018cvx}. We exhaustively explore the parameter space using three different schemes: a profile likelihood approximation, a heuristic Metropolis-Hastings pooling method, and the nested sampling algorithm~\cite{Skilling:2006gxv}. Besides ensuring a proper sampling of the expected multi-modal posterior, the nested sampling algorithm provides a way to assess the statistical significance of cosmological models through the Bayes evidence. On the other hand, the profile likelihood approach allows us to identify possible volume effects and better understand the goodness of fit of the model. Complementary to this, the Metropolis-Hastings pooling method introduced here proposes a heuristic route to cross-check posterior inference provided by nested sampling. Given that the results provided by this complementary method do not change the overall conclusions drawn from the NS analysis, the definition of this method as well as its results are presented in Appendix~\ref{ap:pooling}. Unless otherwise stated, we assume the proper uniform priors presented in Table~\ref{tab:priors}. 

It is crucial to highlight that the CLASS-PT code uses the Eulerian perturbation theory along with Einstein-de Sitter (EdS) kernels to compute the galaxy power spectrum. While strictly speaking this approximation is no longer valid if massive neutrinos are considered, which induced a scale-dependent growth rate, their effects both in the real and redshift spaces can be mimicked by applying the standard formalism to the linear power spectrum of the cold dark matter-baryon fluid that includes the suppression produced by these particles at small scales~\cite{Chudaykin:2020aoj}. Additionally, as argued in Ref.~\cite{Camarena:2023cku}, the simply self-interacting neutrino model does not affect this paradigm as this model merely changes the initial shape of the transfer function but leaves unmodified the typical neutrino free-streaming scale imprint in the linear power spectrum.

\begin{table}[htb]
\caption{\label{tab:priors} Prior ranges used in the analyses.}
\begin{ruledtabular}
\begingroup
\renewcommand{\arraystretch}{1.25}
\begin{tabular}{lc}
Parameter  &  Prior \\
\hline
$\logGeff$ & $[-5.5,0.5]$ \\
$N_\mathrm{eff}$ & $[2.013,5.013]$ \\
$\sum m_\nu \left[\mathrm{eV}\right]$ & $[0.06,1.5]$ \\
$\omega_\mathrm{b}$ & $[0.018,0.03]$ \\
$\omega_\mathrm{cdm}$ & $[0.1,0.2]$ \\
$H_0 \left[\kmsMpc\right]$ & $[60,80]$ \\
$\ln(10^{10}A_{\rm s})$ & $[2,4]$ \\
$n_{\rm s}$ & $[0.85,1.1]$ \\
$\tau_\mathrm{reio}$ & $[0.01,0.25]$ \\
\end{tabular}
\endgroup
\end{ruledtabular}
\end{table}

\subsection{Profile likelihood \label{sub:profile}}

We profile the likelihood of $\logGeff$ in twenty-four equally linearly spaced values distributed in $\logGeff = [-5.5,0.5]$. In contrast to our earlier analysis~\cite{Camarena:2023cku}, we observe that relying solely on a minimization algorithm for profiling the likelihood is highly inefficient when considering both CMB and galaxy power spectra observation simultaneously. Conversely, modified Metropolis-Hastings methods (as seen in Ref.~\cite{Schoneberg:2021qvd}) may become computationally burdensome when applied to the twenty-four assumed value of $\logGeff$. To overcome this issue, we employ a hybrid approach.

Starting with a prior $\logGeff = [-2.5,0.5]$, we obtain an initial best-fit through a Metropolis-Hastings exploration, demanding $R - 1 \lesssim 0.25$ for all parameters, where $R$ represents the Gelman-Rubin estimator~\cite{10.1214/ss/1177011136}. Employing this result as the initial input for the Derivative-Free Optimizer for Least-Squares (DFOLS) package~\cite{10.1145/3338517}, we then minimize the likelihood across the values in the range $\logGeff = [-2.5, 0.5]$, finding the corresponding best-fits. We repeat the same process for the remaining parameter space, i.e., considering the prior $\logGeff = [-5.5,-2.5]$. This analysis produces a collection of discrete points outlining the profile likelihood along the parameter space of interest. In Sec.~\ref{sec:result}, we supplement this result with a smooth representation obtained through a cubic spline interpolation.

\subsection{Nested sampling exploration \label{sub:nested}}

To determine the statistical significance of the cosmological scenarios analyzed here, we sample the parameter space using the nested sampling (NS) algorithm~\cite{Skilling:2006gxv}. In particular, we employ \texttt{PyMultiNest}~\cite{Buchner:pymul}, a Python implementation of \texttt{MultiNest}~\cite{Feroz:2007kg,Feroz:2008xx}. Our main analyses optimize for estimation of the Bayesian evidence by using 2000 live points, a sampling efficiency of 0.3, and a minimum precision of $0.05$ in log-evidence. 

Although NS is suitable for our study, it is crucial to bear in mind that this algorithm has potential limitations that can affect the analysis. These limitations, extensively discussed in the literature~\cite{Skilling:2006gxv,Higson:2018aaa,Higson:2018cqj,Fowlie:2020gfd,Buchner:2021,Ashton:2022grj}, depend on factors such as the NS configuration and the nature of the likelihood under scrutiny. Identifying or addressing such deficiencies often requires multiple runs, resulting in a significant increase in computational time and resources used in the analysis. Here, we evaluate the potential impact of such limitations in our analyses using a heuristic approach based on a Metropolis-Hasting method. This method, thoroughly described in the following section, primarily aims to assess the accuracy of the inference of the posterior provided by NS sampling. However, it can also help to identify potential biases in the estimation of the Bayesian evidence. When performing multiple runs is computationally feasible, however, we assess possible NS deficiencies conducting additional analyses with varied NS configurations. The results of such analyses are presented in Appendix~\ref{ap:NS}.

\subsection{Cosmological data \label{sub:data}}

We perform cosmological analyses considering different combinations of galaxy power spectrum, CMB, and Baryon Acoustic Oscillation (BAO) data. Specifically, we employ:
\begin{enumerate}
    \item $\mathrm{TT}$: low-$\ell$ and high-$\ell$ CMB temperature power spectrum from Planck 2018~\cite{Planck:2018vyg}. 
    \item $\mathrm{EE, TE}$: low-$\ell$ and high-$\ell$ CMB E-mode polarization and their temperature cross correlation from Planck 2018~\cite{Planck:2018vyg}. 
    \item FS: the multipoles of the galaxy power spectrum $P_\ell(k,z)$ $(\ell = 0,2,4)$~\cite{Philcox:2021kcw,Chudaykin:2020ghx} and the real space power spectrum estimator $Q_0$~\cite{Ivanov:2021fbu} obtained from the window-free galaxy power spectrum analyses~\cite{Philcox:2020vbm,Philcox:2021kcw} of the BOSS-DR12 galaxy sample~\cite{SDSS:2011jap,BOSS:2012dmf,BOSS:2016wmc}, at redshift $0.38$ and $0.61$ for the north and south Galactic cap. We adopt $k_\mathrm{min} = 0.01\: h/\mathrm{Mpc}$ and $k_\mathrm{max} = 0.2\: h/\mathrm{Mpc}$ for $P_\ell$ and $k_\mathrm{min} = 0.2\: h/\mathrm{Mpc}$ and $k_\mathrm{max} = 0.4\: h/\mathrm{Mpc}$ for $Q_0$. In both cases, we use a bin width of $\Delta k = 0.005\: h/\mathrm{Mpc}$. We also use the reconstructed power spectrum to constraint the Alcock-Paczynski parameters~\cite{Philcox:2020vvt}. Similarly to our previous analysis~\cite{Camarena:2023cku}, we adopt the likelihood that analytically marginalizes over the nuisance parameters that enter linearly in the power spectrum and uses MultiDark-Patchy 2048~\cite{Kitaura:2015uqa,Rodriguez-Torres:2015vqa} simulations to compute the covariance matrix~\cite{Philcox:2021kcw}.
    \item BAO: Baryon Acoustic Oscillations distances measurements from 6dFGS at $z=0.106$~\cite{Beutler:2011hx},  SDSS MGS at $z=0.15$~\cite{Ross:2014qpa}, and BOSS DR12 at $z=0.38,\:0.51$ and $0.61$~\cite{BOSS:2016wmc}.
\end{enumerate}

To reduce the number of free parameters in analyses employing the NS and MH pooling methods, we use the nuisance-marginalized versions of the Planck 2018 likelihoods for high-$\ell$, i.e., \verb|plik_lite_v22_TT| and \verb|plik_lite_v22_TTTEEE| for the $\mathrm{TT}$ and $\mathrm{TT,TE,EE}$ data, respectively. However, in order to unveil possible volume effects generated by unknown correlations between CMB foreground and nuisance parameters with the effect of self-interacting neutrinos, we also consider profile-likelihood analyses employing the non-marginalized version of the aforementioned likelihoods, i.e., \verb|plik_rd12_HM_v22_TT.clik| and \verb|plik_rd12_HM_v22b_TTTEEE.clik| for the $\mathrm{TT}$ and $\mathrm{TT,TE,EE}$ data, respectively. 

\begin{figure*}[htb]
\includegraphics[width=0.975\textwidth]{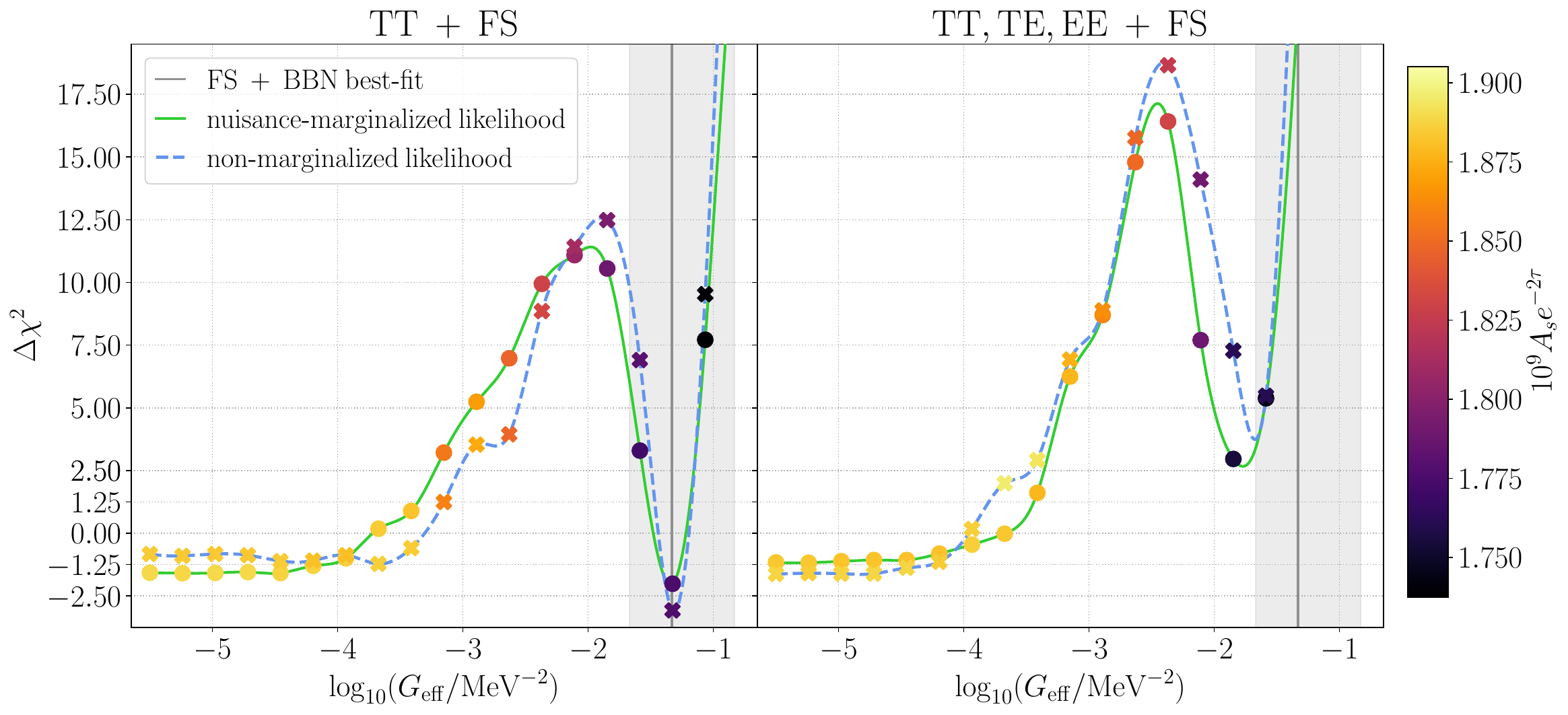}
\caption{\label{fig:profile_likelihood_all_alt} Profile likelihood of the $\logGeff$ parameter obtained from the analyses of $\mathrm{TT, EE, TE}\:+\:\mathrm{FS}$ (right) and $\mathrm{TT}\:+\:\mathrm{FS}$ (left) data when considering the nuisance-marginalized (circles and solid green lines) and non-marginalized (cross marks and dashed light blue lines) versions of Planck likelihoods. The color scheme of the circles and cross marks indicates the best-fit value of $10^9 A_{\rm s} e^{-2\tau}$ at the respective $\logGeff$. The horizontal solid gray lines and bands denote the $\logGeff$ best-fit and $68\%$ CL interval obtained from the CMB-independent analysis~\cite{Camarena:2023cku}, respectively.}
\end{figure*}

\begin{figure*}[htb]
\includegraphics[width=0.95\textwidth]{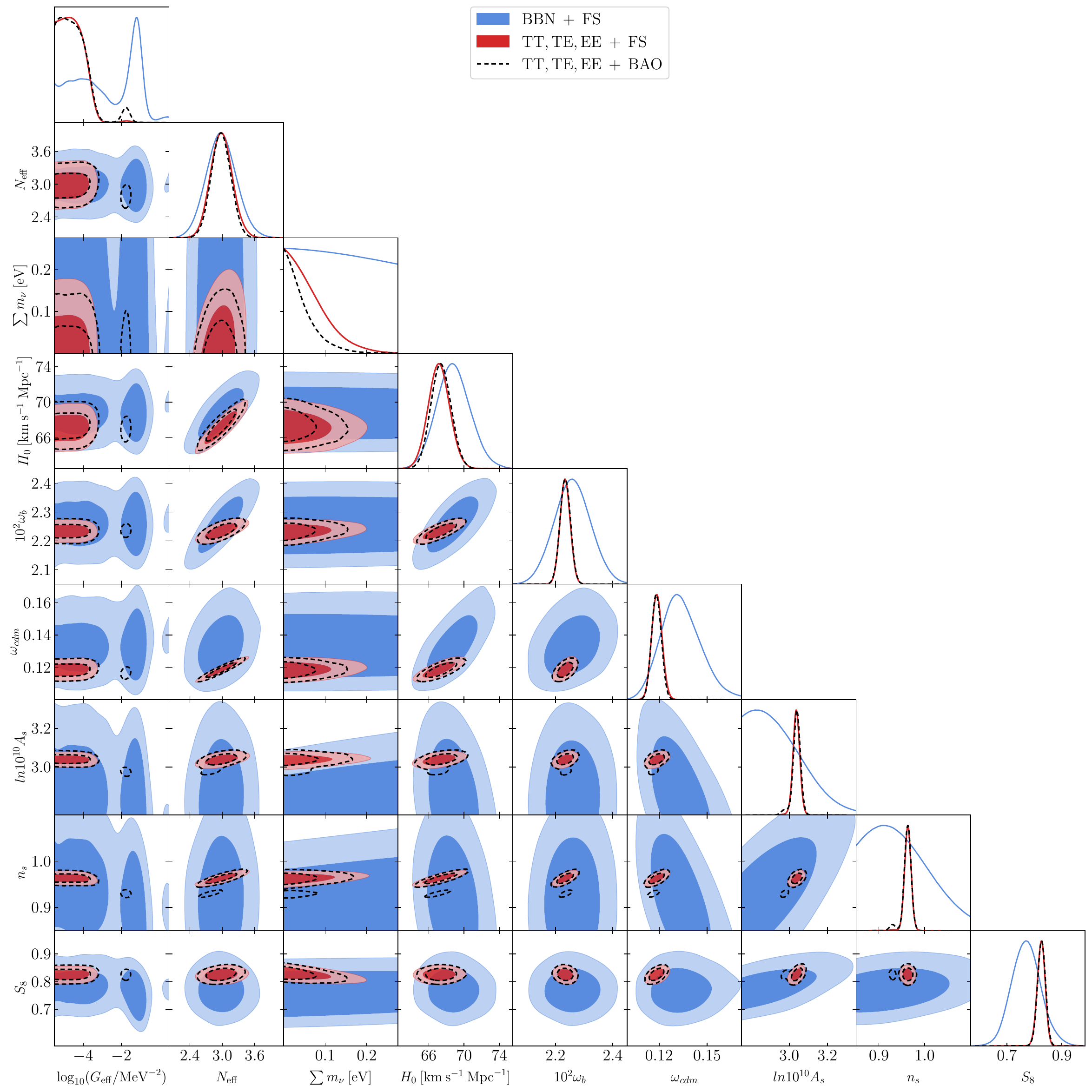}
\caption{\label{fig:triplot_no_modes} Marginalized constraints, at $68\%$ and $95\%$ confidence levels (CL), on the cosmological parameter of the self-interacting neutrino scenario when considering different combination of data.}
\end{figure*}

\section{\label{sec:result} Results and discussion}

Whenever appropriate, we compare the results from the analysis of the self-interacting neutrino cosmologies with the results from the analysis of the $\Lambda$CDM model, for which we assume $\sum m_\nu = 0.06\:\mathrm{eV}$. To distinguish the effects produced by delaying the neutrino free streaming from the effects induced by $N_\mathrm{eff}$ and $\sum m_\nu$, we additionally analyze an extension of the standard cosmological model, where the effective number of relativistic species, $N_\mathrm{eff}$, and the sum of neutrino masses,~$\sum m_\nu$, are considered as free parameters. We dubbed this extension as $\Lambda$CDM + $N_\mathrm{eff}$ + $\sum m_\nu$. We use \texttt{GetDist}~\cite{Lewis:2019xzd} to obtain most of the figures and tables displayed in this section. 

\subsection{\label{sub:volu} Volume effects}

We profile the $\logGeff$ likelihood considering both the $\mathrm{TT}\:+\:\mathrm{FS}$ and $\mathrm{TT, EE, TE}\:+\:\mathrm{FS}$ data combination. To assess possible volume effects resulting from the marginalization of the CMB foreground and nuisance parameters, we perform the analysis employing both the nuisance-marginalized and non-marginalized versions of the CMB likelihoods, see Sec.~\ref{sec:data}. Results of our analyses are shown in Fig.~\ref{fig:profile_likelihood_all_alt}. We quantify the goodness of fit of the self-interacting neutrino scenario relative to the $\Lambda$CDM model using $\Delta \chi^2 = \chi^2_\mathrm{model} - \chi^2_{\Lambda\mathrm{CDM}}$.

First, we observe that analyses employing either the nuisance-marginalized (solid green lines) or non-marginalized (dashed blue lines) versions of the Planck likelihoods yield comparable results. This similarity suggests that astrophysical foregrounds and instrumental modeling parameters are unlikely to exhibit significant degeneracy with the effects produced by neutrino interactions. Consequently, assumptions made to provide a more compact CMB-only Planck likelihood~\cite{Planck:2019nip} are expected to hold for a delayed neutrino free streaming scenario.

Second, in agreement with earlier studies~\cite[see][for instance]{Kreisch:2019yzn}, we note that the SI$_\nu$ mode offers a competitive fit to the combination of the CMB temperature and galaxy power spectra as compared to $\Lambda$CDM. As shown by the left panel of Fig.~\ref{fig:profile_likelihood_all_alt}, the best-fit value obtained when analyzing $\mathrm{TT}\:+\:\mathrm{FS}$ data not only yields $\Delta \chi^2 \approx -2.5$ but also agrees very well with the results of our previous CMB-independent analysis; in both cases the best-fit value features $\logGeff \approx - 1.3$. However, as displayed in the right panel, the SI$_\nu$ mode is disfavored compared to $\Lambda$CDM with $\Delta \chi^2 \approx 2.5$ once the Planck polarization data is included.

The fact that a strongly self-interacting neutrino scenario can provide a good fit to galaxy and CMB temperature power spectra data but fails to explain the Planck polarization observations can be understood by examining how cosmological data constrain the amplitude of the CMB power spectrum, denoted by $10^9 A_{\rm s} e^{-2\tau}$. In the absence of polarization data, the parameters $A_{\rm s}$ and $\tau_\mathrm{reio}$ are strongly degenerate, implying that temperature data can merely constrain the combination $10^9 A_{\rm s} e^{-2\tau}$ but not $A_{\rm s}$ and $\tau_\mathrm{reio}$ independently. Thus, the CMB temperature power spectrum can be made compatible with the $A_{\rm s}$ value that the SI$_\nu$ scenario requires to fit the galaxy power spectrum. However, the polarization power spectrum, which is sensitive to the reionization of the Universe, can break down this degeneracy, imposing important constraints on $\tau_\mathrm{reio}$ and $A_{\rm s}$. Such constraints exclude the lower values of $A_{\rm s}$ needed to compensate for the effect of delaying the neutrino free streaming, diminishing the statistical significance of the strongly self-interacting neutrino cosmology.

The role of CMB polarization data in the analysis is further illustrated by the color scheme of the circle and cross marks in Fig.~\ref{fig:profile_likelihood_all_alt}. From both panels we note that in the MI$_\nu$ regime, the best-fit values found for the CMB amplitude agree with the typical value expected in the standard paradigm, i.e., $10^9 A_{\rm s} e^{-2\tau} \approx 1.88$~\cite[see Ref.][for instance]{Planck:2018vyg}. Conversely, as one transitions towards the SI$_\nu$ regime, lower values of $10^9 A_{\rm s} e^{-2\tau}$ are obtained. While the temperature power spectrum allows such values (left panel), Planck polarization (right panel) can effectively constrain the reionization history, largely disfavouring scenarios with $10^9 A_{\rm s} e^{-2\tau} \lesssim 1.85$ and thereby diminishing the goodness of fit of the SI$_\nu$ mode.

We note that this preference for lower values of $A_{\rm s}$ when analyzing the power spectrum has been also found within the standard paradigm~\cite[see Refs.][for instance]{Philcox:2020vvt,Philcox:2021kcw} and, in general, follows the trend of the mild discrepancy found between the amount of clustering inferred from the LSS and CMB data, i.e., the $S_8$ tension~\cite{Abdalla:2022yfr}.

\subsection{\label{sub:main} Bayesian analysis}

\subsubsection{Cosmological constraints}

Using the NS algorithm (see Section~\ref{sub:nested}), we derive constraints on the neutrino self-interaction, $\Lambda$CDM, and $\Lambda$CDM + $N_\mathrm{eff}$ + $\sum m_\nu$ models considering the combination of $\mathrm{TT, EE, TE}\:+\:\mathrm{FS}$ and  $\mathrm{TT, EE, TE}\:+\:\mathrm{BAO}$ data.
Although the analysis presented in Sec.~\ref{sub:volu} reveals that Planck polarization power spectrum observations play a crucial role in the statistical significance of the SI$_\nu$ scenario, we have found that NS explorations disregarding the polarization data yield to unstable posterior inference. In Appendix~\ref{ap:NS}, we argue that such instabilities are related to the complex correlation among cosmological parameters when polarization data is not considered.

For completeness, we recall the constraints obtained in our previous CMB-independent analysis~\cite{Camarena:2023cku}, which uses FS data and Big Bang Nucleosynthesis priors on the primordial abundance of helium and deuterium coming from Ref.~\cite{Cooke:2017cwo} and Ref.~\cite{Aver:2015iza}, respectively. We label this result as $\mathrm{BBN\:+\:FS}$. Cosmological constraints for the self-interacting neutrino model are shown in Fig.~\ref{fig:triplot_no_modes}. We additionally presented cosmological constraints obtained when separating the SI$\nu$ and MI$\nu$ modes in Appendix~\ref{ap:modes_constraints}.

Figure~\ref{fig:triplot_no_modes} shows the constraints on the self-interacting neutrino scenario for the different combinations of data considered in this section. We observe that $\mathrm{TT, EE, TE}\:+\:\mathrm{FS}$ (red contours and lines) and $\mathrm{TT, EE, TE}\:+\:\mathrm{BAO}$ (dashed black contours and lines) data yield similar constraints for most cosmological parameters (see also Tables~\ref{tab:cons_CMB_PK}~and~\ref{tab:cons_CMB_BAO}). However, two notable differences arise. First, we note a moderately stronger constraint on $\sum m_\nu$ in the presence of BAO data. This moderate decrease in $\sum m_\nu$ uncertainty denotes that current geometrical measurements of the BAO can more effectively constrain the expansion history of the Universe and, consequently, the existence of non-relativistic species today. Furthermore, similar but weaker variations can be observed in $N_\mathrm{eff}$ and $H_0$ uncertainties.

Second, and more significantly, comparing both analyses reveals a notable reduction in the peak height of the SI$_\nu$ mode when analyzing the galaxy power spectrum data.  While our cross-check analysis provided by the heuristic MH pooling method partially attributes such a decrease to NS deficiencies in exploring the intricate parameter space featured by the combination of CMB and LSS data, see Appendix.~\ref{ap:pooling}, we emphasize that this reduction indicates a sizable depreciation in the statistical significance of the delayed neutrino free streaming scenario. 

Analysis presented in the previous section reveals that the challenge of simultaneously fitting the CMB and galaxy power spectra employing the simplest neutrino self-interaction scenario relates to the preference for lower values of $A_\mathrm{s}$ featured by the LSS data. As discussed in Sec.~\ref{sec:model}, a delay in the onset of neutrino free streaming induces a scale-dependent alteration of the linear power spectrum, including a suppression of its amplitude at scales crossing the horizon when neutrino are tightly-coupled. Ref.~\cite{Camarena:2023cku} shows that, in light of the galaxy power spectrum data, the SI$_\nu$ mode compensates for such a change by modifying the primordial curvature power spectrum, characterized by the parameters $A_{\rm s}$ and $n_{\rm s}$. Although the CMB data allows for a similar modification, the polarization data restrains the possible changes in $A_{\rm s}$, spoiling then the statistical significance of a cosmological scenario with strongly self-interacting neutrinos, see Sec.~\ref{sub:volu}.

\subsubsection{Model comparison}

The Bayes evidence serves as a metric for assessing the efficacy of a model in light of a specific data set. For a given model, $\mathcal{M}$, the Bayesian evidence, or marginalized likelihood, is computed as follows:
\begin{equation}
    \mathcal{Z} \equiv p(d|\mathcal{M}) = \int \mathcal{L}(d|\theta,\mathcal{M}) \pi(\theta|\mathcal{M})\: \mathrm{d}\theta\,, \label{eq:evidence}
\end{equation}
where $\mathcal{L}(d|\theta,\mathcal{M})$ represents the likelihood, $\pi(\theta|\mathcal{M})$ denotes the priors, $d$ is the provided data and $\theta$ is the set of model's parameters. Relying on the Bayesian evidence, a model comparison analysis can be performed by using the the Bayes factor, $\mathcal{B}_{ij}$, which contrasts the performance of two models through~\cite{Trotta:2004ty}:
\begin{equation}
    \mathcal{B}_{ij} = \frac{p(d|\mathcal{M}_i)}{p(d|\mathcal{M}_j)}\,. \label{eq:bayes_factor}
\end{equation}
A value $\mathcal{B}_{ij} > 1$  indicates that data increasingly prefer model $\mathcal{M}_i$ over model $\mathcal{M}_j$. If $\mathcal{B}_{ij} < 1$, the opposite is inferred. 

\begin{table*}[htb]
\caption{\label{tab:model_comparison} Results of the model comparison analysis when $\mathrm{TT, EE, TE}\:+\:\mathrm{FS}$ data is considered and the standard paradigm is used as reference model, i.e., $\mathcal{M}_0 = \Lambda$CDM.  We emphasize that $\Delta \chi^2 \equiv \Delta \chi^2_\mathrm{model} - \Delta \chi^2_{\Lambda\mathrm{CDM}}$. Values shown in parenthesis correspond to results obtained when analyzing the $\mathrm{TT, EE, TE}\:+\:\mathrm{BAO}$ combination of data.}
\begin{ruledtabular}
\begingroup
\renewcommand{\arraystretch}{1.25}
\begin{tabular}{lccc}
Data  &  $\Lambda$CDM + $N_\mathrm{eff}$ + $\sum m_\nu$ & MI$_\nu$ & SI$_\nu$ \\
\hline
$\Delta \chi^2_\mathrm{FS}$ & $-0.08$ & $-0.38$ & $0.20$ \\
$\Delta \chi^2_{\mathrm{high}\:\ell}$ & $-0.77$ & $-0.39$ & $-0.78$ \\ 
$\Delta \chi^2_{\mathrm{low}\:\ell}$ & $-0.53$ & $-0.41$ & $3.05$ \\
$\Delta \chi^2_{\mathrm{total}}$ & $-1.38$ & $-1.19$ & $2.46$ \\ 
\hline 
$\Delta \mathrm{AIC}_{i0}$ & $2.62$ & $4.81$ & $8.46$ \\
$B_{i0}$ & $0.031 \pm 0.0062$ ($0.025 \pm 0.0041$)  & $0.106 \pm 0.022$ ($0.053 \pm 0.0089$) & $0.0007 \pm 0.0005$ ($0.0027 \pm 0.0008$) \\
\end{tabular}
\endgroup
\end{ruledtabular}
\end{table*}

\begin{table}[htb]
\caption{\label{tab:model_comparison_extra} Results of the model comparison when analyzing $\mathrm{TT, EE, TE}\:+\:\mathrm{FS}$ data. Reference models, denoted here by $\mathcal{M}_0$, are compared to the SI$_\nu$ mode. We define $\Delta \chi^2 \equiv \Delta \chi^2_{\mathrm{SI}_\nu} - \Delta \chi^2_{\mathcal{M}_0}$.}
\begin{ruledtabular}
\begingroup
\renewcommand{\arraystretch}{1.25}
\begin{tabular}{lcc}
Data  &  $\mathcal{M}_0 = \Lambda$CDM + $N_\mathrm{eff}$ + $\sum m_\nu$ & $\mathcal{M}_0 =$ MI$_\nu$  \\
\hline
$\Delta \chi^2_\mathrm{total}$ & $3.84$ & $3.65$  \\
$\Delta \mathrm{AIC}_{i0}$ & $5.84$ & $3.65$  \\
$B_{i0}$ & $0.023 \pm 0.015$  & $0.0066 \pm 0.0045$ \\
\end{tabular}
\endgroup
\end{ruledtabular}
\end{table}

It is worth mentioning that the Bayes factor compares the marginalized likelihood of two distinct models, $p(d|\mathcal{M})$, rather than the posterior probability of the models, $p(\mathcal{M}|d)$. While one could argue for a direct model comparison by evaluating the posterior odds:
\begin{equation}
    \frac{p(\mathcal{M}_i|d)}{p(\mathcal{M}_j|d)} = \mathcal{B}_{ij} \frac{p(\mathcal{M}_i)}{p(\mathcal{M}_j)}\,, \label{eq:posterior_odds}
\end{equation}
we note that, overall, in agnostics analyses, there is no \textit{a priori} reason to assume $p(\mathcal{M}_i)/p(\mathcal{M}_j) \neq 1$. 

Complementary to the Bayes factor, whose values are computed using the NS algorithm, see Sec.~\ref{sub:nested}, we also consider the Akaike information criterion (AIC)~\cite{Akaike:1974ddd}:
\begin{equation}
    \mathrm{AIC} = \chi^2_{\mathrm{min}} + 2 n \,, \label{eq:AIC} 
\end{equation}
where $n$ is the number of parameters to fit, and $\chi^2_{\mathrm{min}}$ is the minimum chi-square fit to the data. The quality of fit between distinct models can then be evaluated by:
\begin{equation}
    \Delta \mathrm{AIC}_{ij} \equiv \mathrm{AIC}_i - \mathrm{AIC}_j =  \Delta \chi^2_{ij} + 2 \Delta n \,, \label{eq:DeltaAIC} 
\end{equation}
with $\Delta n$ being the difference in the number of parameters between $\mathcal{M}_i$ and $\mathcal{M}_j$ models. The AIC criterion favors the model with the lower value; thus, a positive (negative) $\Delta \mathrm{AIC}_{ij}$ indicates a preference for model $\mathrm{M}_j$~($\mathrm{M}_i$).

Unless otherwise stated, we use the $\Lambda$CDM model as the reference model, i.e., we compute $\mathcal{B}_{i0}$ and $\Delta \mathrm{AIC}_{i0}$ considering $\mathcal{M}_{0} = \Lambda$CDM. Table~\ref{tab:model_comparison} displays the values of $\Delta\mathrm{AIC}_{i0}$ and $\mathcal{B}_{i0}$ obtained from the analysis of $\mathrm{TT, EE, TE}\:+\:\mathrm{FS}$ data. For completeness, we individually list the minimum chi-square fit both for the different likelihoods used in the analyses and the total combination of them. We additionally display the Bayes factor corresponding to the analysis of $\mathrm{TT, EE, TE}\:+\:\mathrm{BAO}$ data. Results of this extra analysis are shown in parentheses in the last row of Table~\ref{tab:model_comparison}.   

As expected, both the $\Lambda$CDM + $N_\mathrm{eff}$ + $\sum m_\nu$ model and MI$_\nu$ mode yield comparable values of $\Delta \chi^2_\mathrm{total} < 0$, denoting that both cosmologies offer a better fit to the data than the standard paradigm. However, as indicated by the values of the $\Delta \mathrm{AIC}_{i0}$ and $\mathcal{B}_{i0}$ estimators, this enhancement in the goodness of fit is insufficient to compensate for the inclusion of the neutrino parameters.

A comparison between the values of $\mathcal{B}_{i0}$ and $\Delta \chi^2_\mathrm{total}$ obtained for the $\Lambda$CDM + $N_\mathrm{eff}$ + $\sum m_\nu$ model reveals that the preference for the standard paradigm, over the $\Lambda$CDM extension, is primarily due to Occam's razor principle --- the data suggest that introducing the $N_\mathrm{eff}$ and $\sum m_\nu$ parameters does not yield significant statistical improvements. In practice, this is related to the current inability of cosmological data to detect the mass of neutrinos; see Table~\ref{tab:cons_CMB_PK} in Appendix~\ref{ap:modes_constraints}, for instance. In addition, according to the Bayes factor, data slightly favors the MI$_\nu$ mode over the $\Lambda$CDM + $N_\mathrm{eff}$ + $\sum m_\nu$ model with odds of approximately $3 : 1$. Although this slight preference lacks statistical significance, the subtle inclination towards a scenario where the free-streaming phenomenology of neutrinos is constrained by the data could hint that cosmological data response better to $\logGeff$ rather than $\sum m_\nu$. 

Regarding the SI$_\nu$ mode, we observe a deterioration in the goodness of the fit, yielding $\Delta \chi^2_\mathrm{total} \approx 2.5 $. Such degradation is primarily attributed to a worsening of the fit in the CMB low $\ell$ power spectra data, which features $\Delta \chi^2_\mathrm{low\:\ell} = 3.05$. Furthermore, the combination of CMB and galaxy data displays $\Delta \chi^2_\mathrm{FS} = 0.2$, indicating that when CMB observations are considered in the analysis, the SI$_\nu$ mode provides a worse fit to the galaxy power spectrum compared to the $\Lambda$CDM model. It is interesting to highlight that this stands contrary to the CMB-independent analysis, where the strongly self-interacting neutrino cosmology fits the LSS data better than the standard paradigm with $\Delta \chi^2 \approx -2.5$~\cite{Camarena:2023cku}. As discussed before, the worsening of SI$_\nu$ galaxy power spectrum fit relates to the fact that Planck polarization data penalizes significantly lower values of~$A_{\rm s}$.

The preference displayed by the combination of $\mathrm{TT, EE, TE}\:+\:\mathrm{FS}$ data for the $\Lambda$CDM model over the SI$_\nu$ scenario is further noted in $\Delta \mathrm{AIC}_{i0} \sim 8.5$ and, remarkably, the Bayes factor, which indicates that the strongly self-interacting neutrino cosmology is disfavored with odds of $7:10000$. While this significant lower value calls attention to the difficulty of fitting the data with the simplest representation of a delayed neutrino free streaming scenario, we obtain more conservative values of the Bayes factor, $23:1000$ and $66:10000$, respectively, when comparing the SI$_\nu$ mode to the $\Lambda$CDM + $N_\mathrm{eff}$ + $\sum m_\nu$ and MI$_\nu$ models, see Table~\ref{tab:model_comparison_extra}. While these changes in the values of the Bayes factor does not revert the trend against the the simple self-interacting neutrino scenario, they reinforce the fact that the preference for the standard paradigm is partially due to Occam’s razor principle---data additionally penalize the inclusion $N_\mathrm{eff}$ and $\sum m_\nu$ as free parameters.

Compared to the analysis of $\mathrm{TT, EE, TE}\:+\:\mathrm{BAO}$ data, whose results are shown in parentheses in the last row of Table~\ref{tab:model_comparison}, analysis considering $\mathrm{TT, EE, TE}\:+\:\mathrm{FS}$ provides similar results for all models except for the SI$_\nu$ scenario. Indeed, a reduction of factor $\sim 4$ in the Bayes factor is observed when galaxy power spectrum data is considered instead of the geometrical distance provided by BAO measurements.  Furthermore, we notice that, in the analyses considering the galaxy power spectrum data, values found of the Bayes factor regarding the SI$_\nu$ mode feature substantial relative uncertainties. As discussed in Appendix~\ref{ap:pooling}, although such uncertainties hint at a sampling bias underestimating the statistical significance of the SI$_\nu$ mode, deficiencies in the NS sampling are unlikely to revert the evidence against this scenario.

\begin{figure*}[htb]
\includegraphics[width=0.975\textwidth]{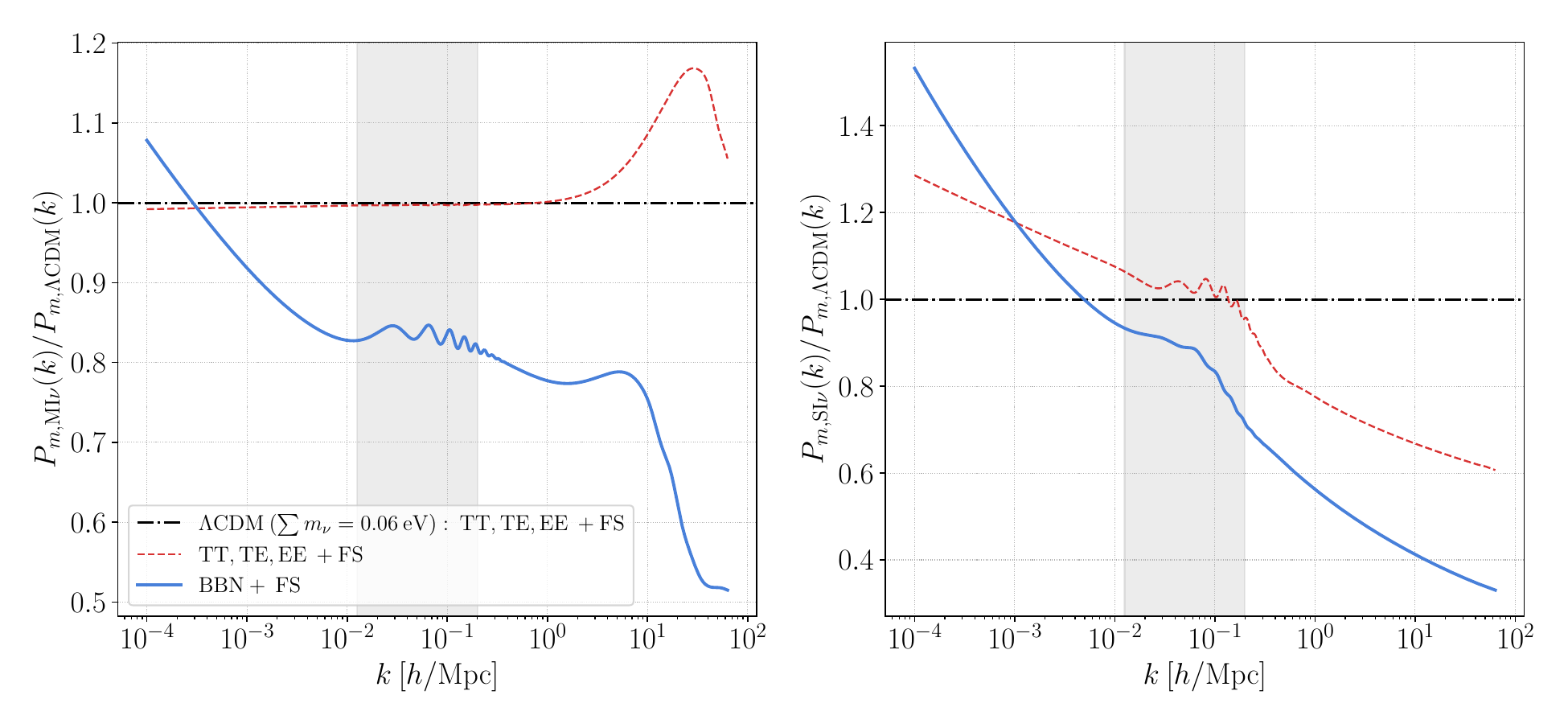}
\caption{\label{fig:Pk_lin_BF} Ratio between the best-fit prediction for the linear power spectrum of the standard $\Lambda$CDM model and the MI$_\nu$ mode (left panel) or SI$_\nu$ mode (right panel) for the different combination of data considered. The gray band represents the range of scales probed by the galaxy power spectrum data.}
\end{figure*}

\begin{figure*}[htb]
\includegraphics[width=0.975\textwidth]{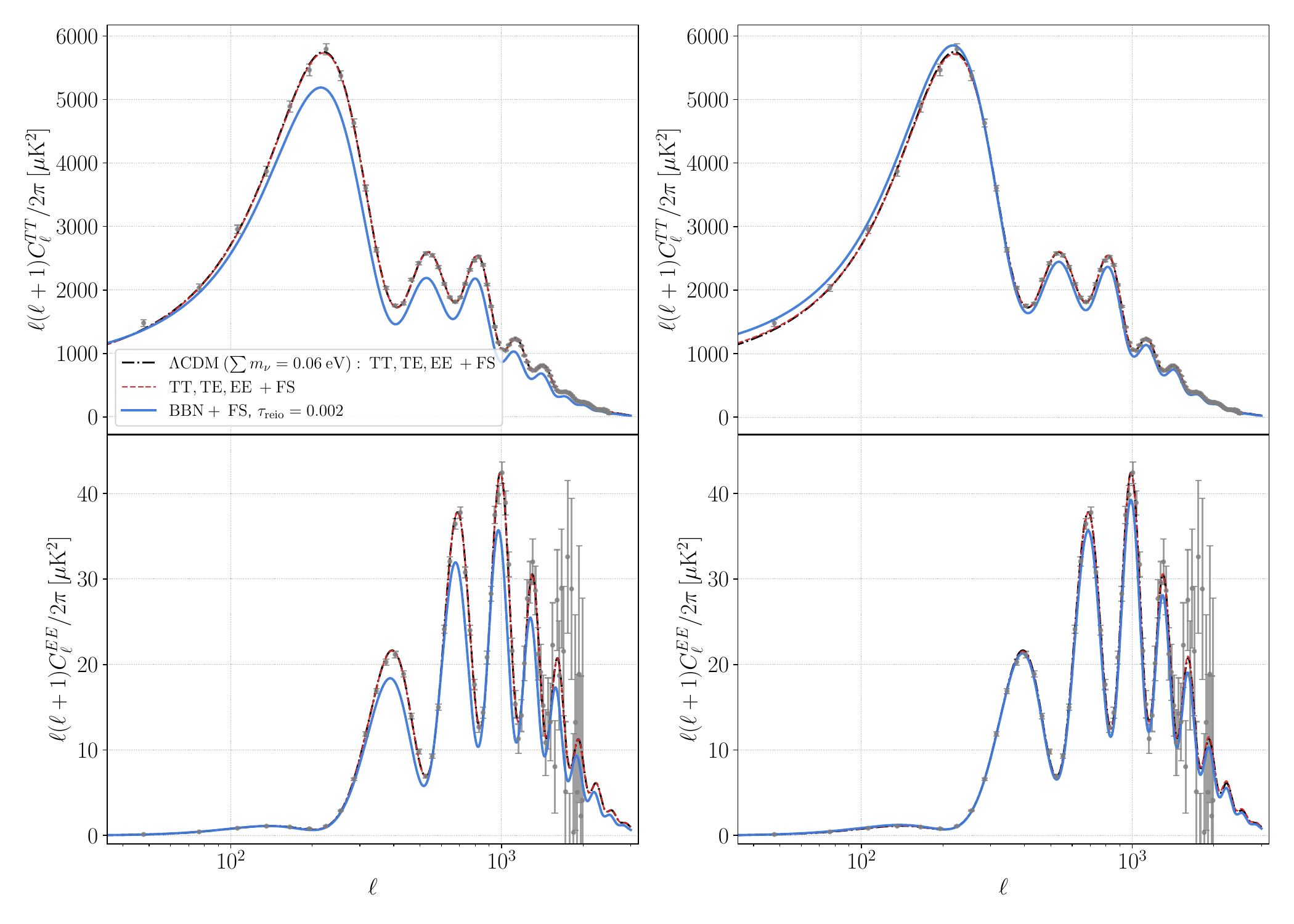}
\caption{\label{fig:CMB_TT_EE_BF} $\mathrm{TT}$ (top) and $\mathrm{EE}$ (bottom) CMB power spectra best-fit predictions for the $\Lambda$CDM (dot-dashed black lines), MI$_\nu$ (left panels), and SI$_\nu$ (right panels) cosmologies in light of different combination of data.}
\end{figure*}

\subsection{\label{sub:impli} Cosmological implications}

Figures~\ref{fig:Pk_lin_BF}~and~\ref{fig:CMB_TT_EE_BF} illustrate the linear matter and CMB power spectra, respectively, predicted by the best-fit of the SI$_\nu$ (right panel) and MI$_\nu$ (left panel) modes for the analyses of $\mathrm{TT, EE, TE}\:+\:\mathrm{FS}$ (dashed red lines) and $\mathrm{BBN}\:+\:\mathrm{FS}$ (solid blue lines) data compared to the predictions provided by the $\Lambda$CDM best-fit (dot-dashed black lines) when analyzing the CMB and galaxy power spectra. 

We first focus on the linear matter power spectrum. As shown by the dashed red line in the left panel of Fig.~\ref{fig:Pk_lin_BF}, in light of the same data, the $\Lambda$CDM model and MI$_\nu$ scenario provide comparable predictions across linear scales with a notable bump that peaks well inside nonlinear scales. On the other hand, the solid blue line shows notable differences across all scales when comparing the best-fit to the $\mathrm{TT, EE, TE}\:+\:\mathrm{FS}$ and $\mathrm{BBN}\:+\:\mathrm{FS}$ data obtained when assuming the $\Lambda$CDM and MI$_\nu$ models, respectively. We note that such differences are mainly driven by the lower values of the amplitude, $A_{\rm s}$, and tilt, $n_{\rm s}$, of the primordial curvature power spectrum preferred by the CMB-independent analysis. 

Contrary to this, the right panel of Fig.~\ref{fig:Pk_lin_BF} shows that both in the absence (solid blue line) or presence (dashed red line) of CMB data, the SI$_\nu$ matter power spectrum exhibits a unchanging general structure: the strongly self-interacting neutrinos feature a barely visible bump that peaks around $k \sim 0.1 \: h/\mathrm{Mpc}$ and predict an important suppression of structures on nonlinear scales. This distinct prediction is less pronounced when CMB power spectra are considered (dashed red line in the right panel), in which case the SI$_\nu$ best-fit predicts a $\gtrsim 20\%$ ($\gtrsim 30\%$) suppression of the power spectrum at galactic (sub-galactic) scales.

As discussed in Ref.~\cite{Kreisch:2019yzn}, the suppression induced by a delay of the onset of neutrino free streaming can add to the step-like suppression featured by massive neutrinos~\cite{Hu:1997mj,Lesgourgues:2006nd}. This characteristic could have potential implications for cosmological inference of the sum of masses of neutrinos. To exemplified this we point out that if taken at face value, strong self-interactions among neutrinos relax by a factor of $50\%$ the 95$\%$ CL upper limit obtained for $\sum m_\nu$ compared to the $\Lambda$CDM + $N_\mathrm{eff}$ + $\sum m_\nu$ model, see Table~\ref{tab:cons_CMB_PK} in Appendix~\ref{ap:modes_constraints}. While the interpretation of the constraints for the SI$_\nu$ mode should be careful, given the lack of support from the data, we find this effect worth of noting. We will investigate this in a future work.

Fig.~\ref{fig:CMB_TT_EE_BF} shows the that best-fit for the $\mathrm{BBN}\:+\:\mathrm{FS}$ (solid blue lines) data obtained within the SI$_\nu$ (right panels) and MI$_\nu$ (left panels) scenarios largely deviate from the $\Lambda$CDM analysis considering $\mathrm{TT, EE, TE}\:+\:\mathrm{FS}$ (dot-dashed black lines) data. Deviations between the baseline $\Lambda$CDM analysis and self-interacting neutrino models are less pronounced when simultaneously analyzing the CMB and galaxy power spectra (solid red lines).

\section{\label{sec:conclu} Conclusions}

During the last decade, several analyses have shown that certain CMB data exhibit a non-trivial preference for a cosmological scenario in which the onset of neutrino free streaming is delayed until close to the epoch of matter-radiation equality. More recently, additional analyses have revealed that LSS data also support this scenario, dismissing the hypothesis of an accidental feature in the CMB data and intensifying the puzzle surrounding the free-streaming nature of neutrinos~\cite{Camarena:2023cku,He:2023oke}. Here, by adopting the simplest cosmological representation for self-interacting neutrinos, we investigated whether this nonstandard cosmology can provide a self-consistent scenario for both the CMB power spectra and the large-scale distribution of galaxies simultaneously. We conducted an exhaustive analysis of the parameter space using three different approaches: profile likelihood, nested sampling, and heuristic Metropolis-Hastings approximation. We robustly demonstrated that the observed large-scale distribution of galaxies exacerbates the challenge already posed by the CMB Planck polarization data for the strongly self-interacting neutrino cosmology. This indicates that the simplest representation of a Universe with delayed neutrino free streaming fails to accommodate a consistent cosmological picture for the both CMB and galaxy power spectra.

The profile likelihood analysis points out that the difficulty of simultaneously fitting the CMB and galaxy power spectra employing the simplest neutrino self-interaction scenario relates to the fact that Planck polarization data disfavor the lower values of $A_{\rm s}$ required to compensate for the effects of neutrino interactions in the linear matter power spectrum. Based on the combination of CMB and galaxy power spectra data, our findings indicate that the SI$_\nu$ mode provides a mildly worse fit to the data when compared to the $\Lambda$CDM model, specifically, $\Delta \chi^2 \approx 2.5$. This degradation is mainly due to a worsening of the fit of the CMB low $\ell$ power spectra data, which yield $\Delta \chi^2_\mathrm{low\:\ell} \approx 3.1$. Additionally, we observe that the SI$_\nu$ mode offers a worse fit to the galaxy power spectrum compared to the $\Lambda$CDM model, with a quantified difference of $\Delta \chi^2_\mathrm{FS} \approx 0.2$. Interesting, we note that the preference displayed by galaxy power spectrum data of a smaller value of $A_s$ with respect to the CMB data follows the trend of the $S_8$ tension.

The challenge posed by the CMB and galaxy observations to the simple neutrino interaction model considered here is quantify through the Bayes factor. In particular, we have found that this nonstandard neutrino cosmology is disfavored over the standard paradigm with odds of $7:10000$. While this significantly lower value highlights the difficulty of providing a consistent cosmological picture for the data with a simple representation of delayed neutrino free streaming, we have noted that this is not solely induced by the presence of nonstandard interactions in the neutrino sector. Weaker results are found when accounting for the fact that current cosmological data cannot detect the sum of the masses of neutrinos, $\sum m_\nu$. In such cases, the analyses indicates that the data disfavor the SI$_\nu$ mode over the $\Lambda$CDM + $N_\mathrm{eff}$ + $\sum m_\nu$ and MI$_\nu$ cosmologies with odds of $23:1000$ and $66:10000$, respectively. The difficulty of simultaneously fitting the galaxy and CMB spectra with the SI$_\nu$ mode is further illustrated by the Bayes factor obtained when analyzing the CMB and BAO data, whose value increases by a factor of $\sim 4$ compared to the value of the galaxy power spectrum data analyses.

It is crucial to highlight that although our analysis robustly reveals strong statistical evidence against the simplest representation of a  cosmology with delayed neutrino free streaming, it does not dismiss or explain the presence of a rather unusual physical feature found in cosmological data (i.e.~the second local minima found in the $\Delta\chi^2$ surfaces shown in Fig.~\ref{fig:profile_likelihood_all_alt}). While this feature has so far been studied in the context of alternate neutrino streaming histories, it is possible that it is caused by a yet-to-be-discovered phenomenon deep in the radiation-dominated epoch that is not related at all to the neutrino sector. Shedding light on the nature of this data-driven feature will thus require considering a broader range of neutrino phenomenologies \cite[see Refs.][for instance]{Brinckmann:2022ajr,Venzor:2023aka,Aloni:2023tff} while also exploring other possible new-physics scenarios that could be responsible for it.
 
Although future analyses could help unveil the true nature of the non-standard feature currently captured by the SI$_\nu$ mode, we emphasize that significant challenges, for both the standard and nonstandard neutrino scenario, lie ahead. On the one hand, theoretical efforts will be needed to construct appropriate cosmological frameworks for more complex scenarios. In particular, new phenomenologies altering the evolution of the Universe at late times could affect the evolution of cosmological perturbations in the nonlinear regime, forcing us to reassess the way we model the LSS observables. On the other hand, given that the step-like suppression of the power spectrum produced by massive neutrinos adds to the effects of a delayed neutrino free streaming, analyses overlooking the underlying free-streaming nature of neutrinos could lead to biased estimations of the sum of the masses of neutrinos, $\sum m_\nu$. Understanding how our future analyses based on galaxy clustering data would change in light of this kind of effects will be then crucial to confirm the standard chronological evolution of the Universe or reveal the exact nature of the feature observed in the data. 

Lastly, even when we adopt the simplest representation for a delayed free streaming scenario, our analysis of the state-of-the-art cosmology data proves computationally challenging. While our conclusions remain robust across distinct approaches, we have found that exploration of the parameter space of the simplest framework adopted here can suffer from significant inefficiencies at several levels. In particular, when relying on the generally advised nested sampling method, we observe biased posterior estimation when analyzing the CMB and galaxy power spectra. Although the same inefficiencies are not necessarily expected to emerge in more complex realizations of the SI$_\nu$ mode, we emphasize that identifying such inefficiencies is neither a trivial task nor always computationally feasible. Even when specific-oriented tools are available for this task \cite[see][for instance]{Handley:2019mfs,Higson:2018cqj,Fowlie:2020mzs}, the complexity of modeling state-of-the-art cosmological data makes this problem inherently complex. Future observations are unlikely to ease this issue. Therefore, it is crucial to bear in mind the limitations of the statistical methods used in cosmological inference, especially when analyzing new physics that deviate from the typically proposed mild deformations of the $\Lambda$CDM model but can still provide a good fit to the cosmological observables. 

\begin{acknowledgments}
We are grateful to Kylar Greene for comments on an early version of this manuscript and John Houghteling for useful discussions. D. C. and F.-Y. C.-R. would like to thank the Robert E. Young Origins of the Universe Chair fund for its generous support. This work was supported in part by the National Science Foundation (NSF) under grant AST-2008696. We also would like to thank the UNM Center for Advanced Research Computing, supported in part by the National Science Foundation, for providing the research computing resources used in this work.
\end{acknowledgments}

\begin{figure*}[!htb]
\includegraphics[width=0.475\textwidth]{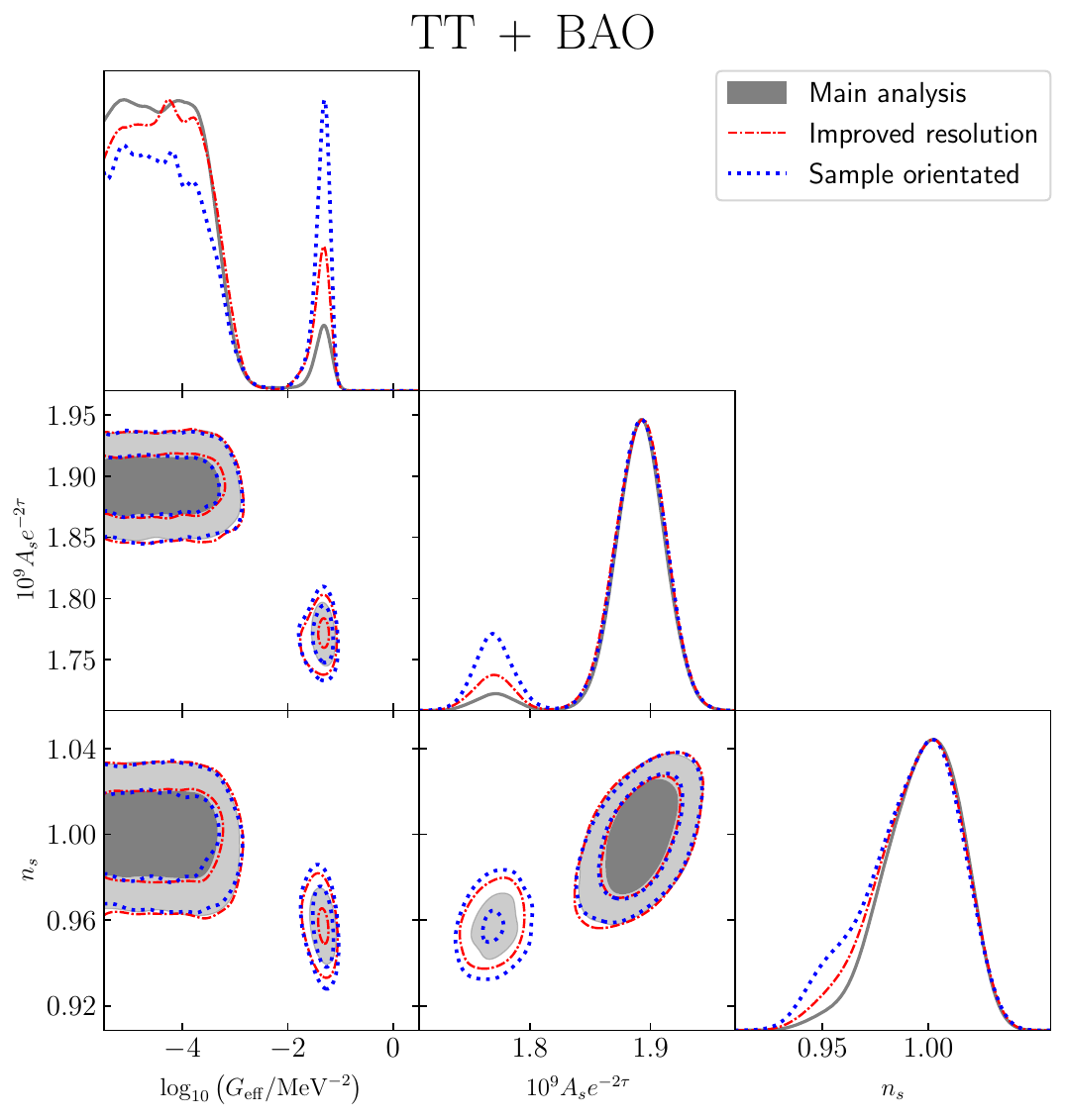}
\includegraphics[width=0.475\textwidth]{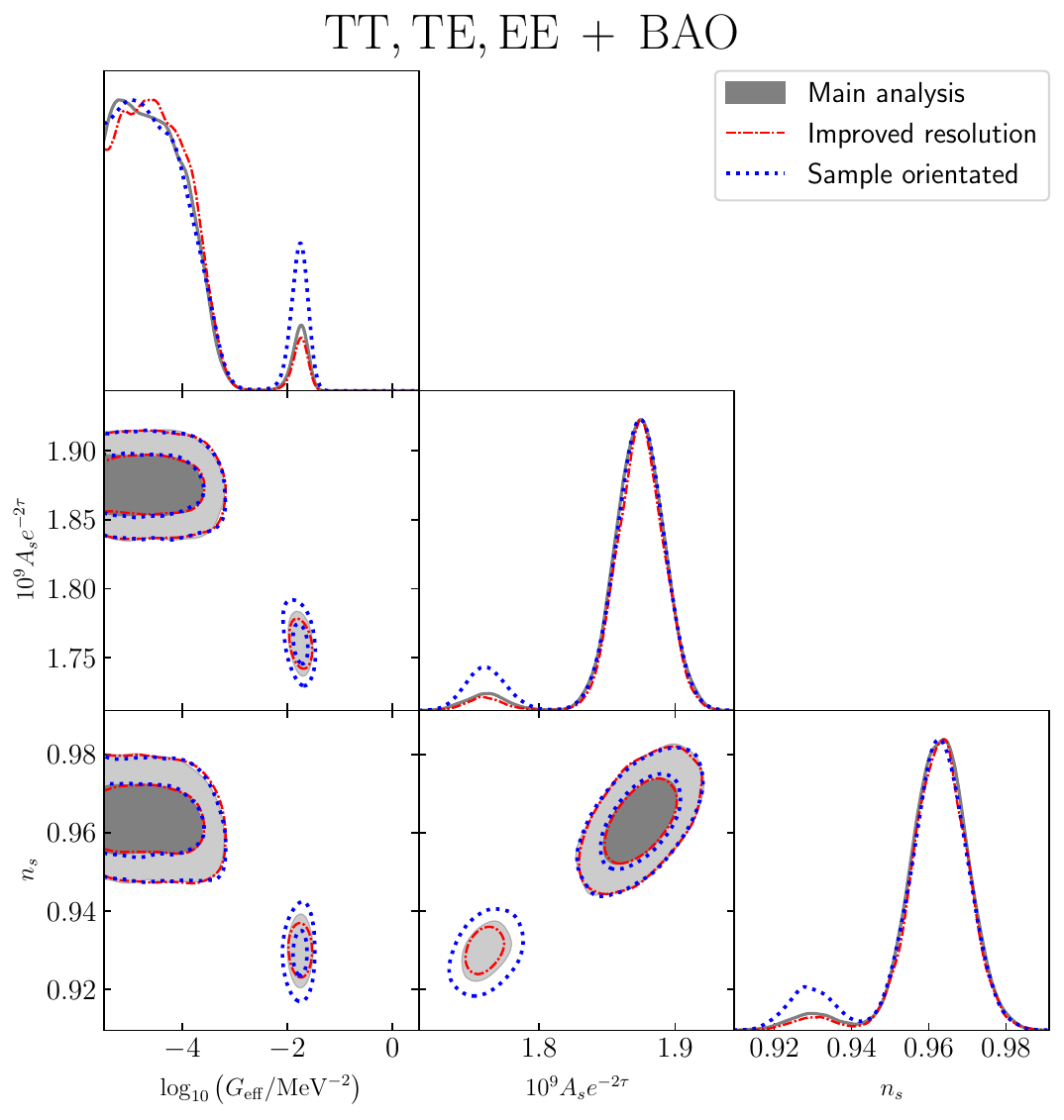}
\caption{\label{fig:triangle_NS_comparison} Comparison between the marginalized constraints, at $68\%$ and $95\%$ CL, on the relevant cosmological parameter of the self-interacting neutrino cosmology obtained when adopting three distinctive NS configurations and considering $\mathrm{TT}\:+\:\mathrm{BAO}$ (left plot) and $\mathrm{TT, EE, TE}\:+\:\mathrm{BAO}$ (right plot) data. }
\end{figure*}

\begin{figure*}[!htb]
\includegraphics[width=0.317\textwidth]{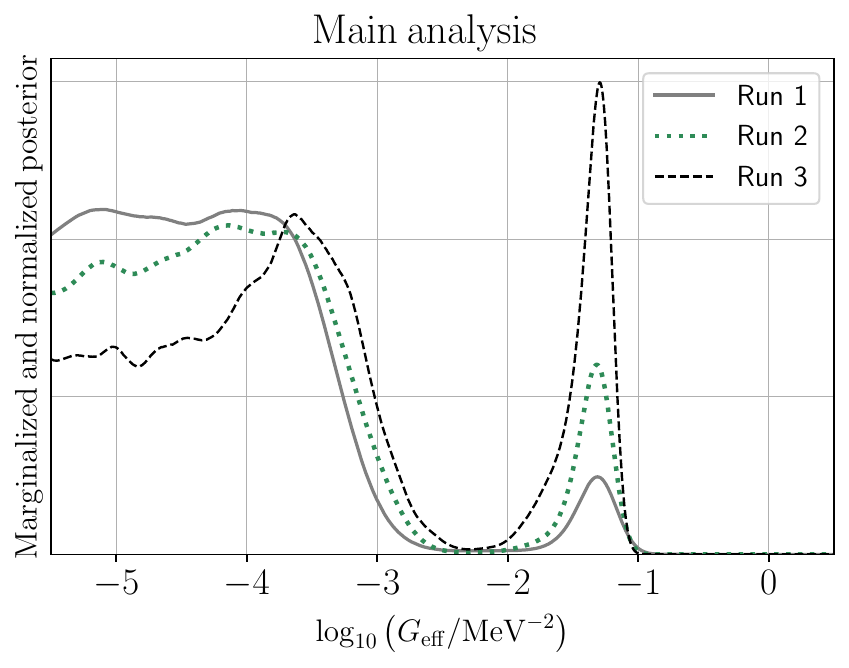}
\includegraphics[width=0.317\textwidth]{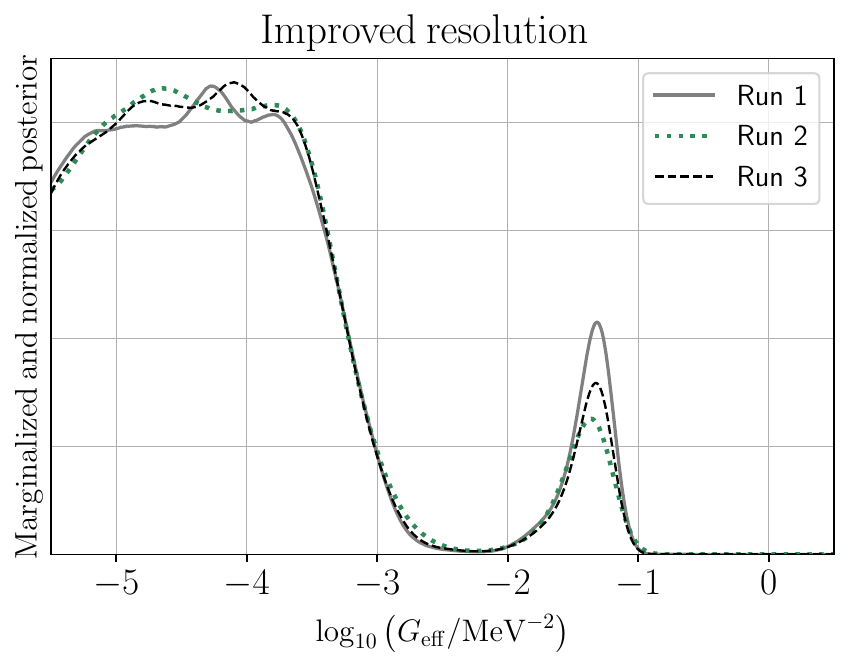}
\includegraphics[width=0.317\textwidth]{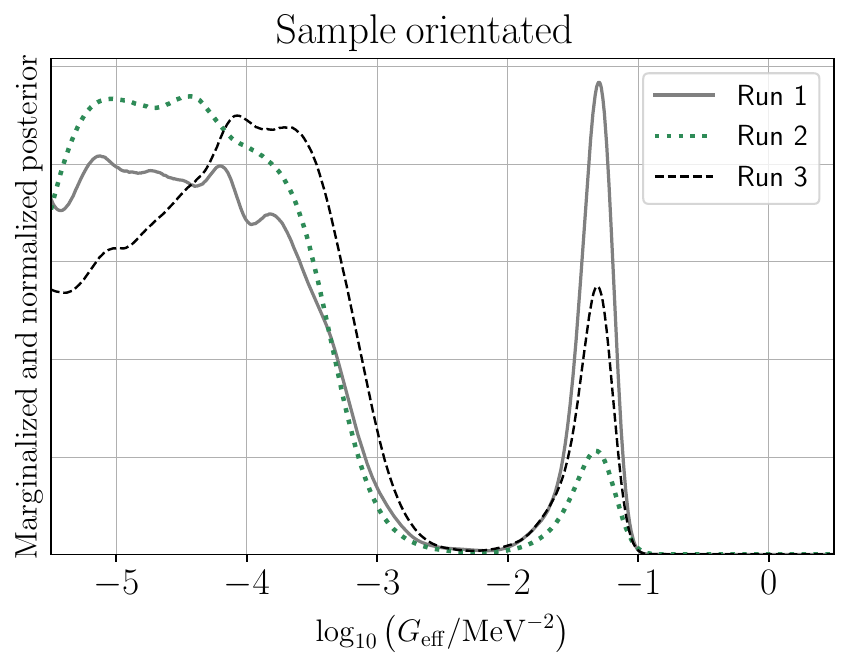}
\caption{\label{fig:TT_1d_NS_comparison} 1d marginalized (and normalized) $\logGeff$ posterior obtained when analyzing $\mathrm{TT}\:+\:\mathrm{BAO}$ data adopting the main analysis (left panel), improved resolution (central panel), and sample oriented (right panel) NS configurations. We conduct the analysis for each configuration three different times.}
\end{figure*}

\appendix

\section{Additional runs with different nested sampling configuration \label{ap:NS}}

\begin{figure*}[!htb]
\includegraphics[width=0.317\textwidth]{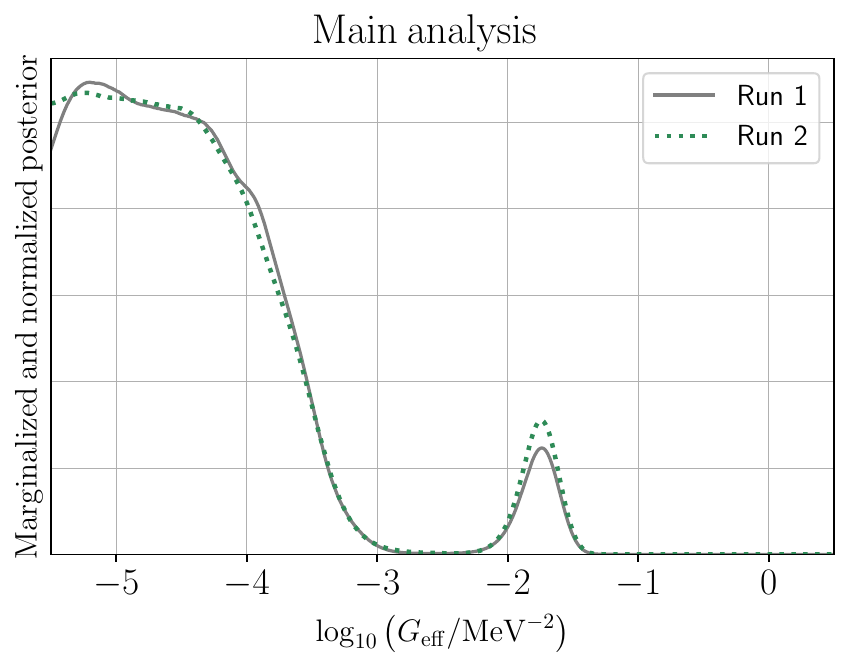}
\includegraphics[width=0.317\textwidth]{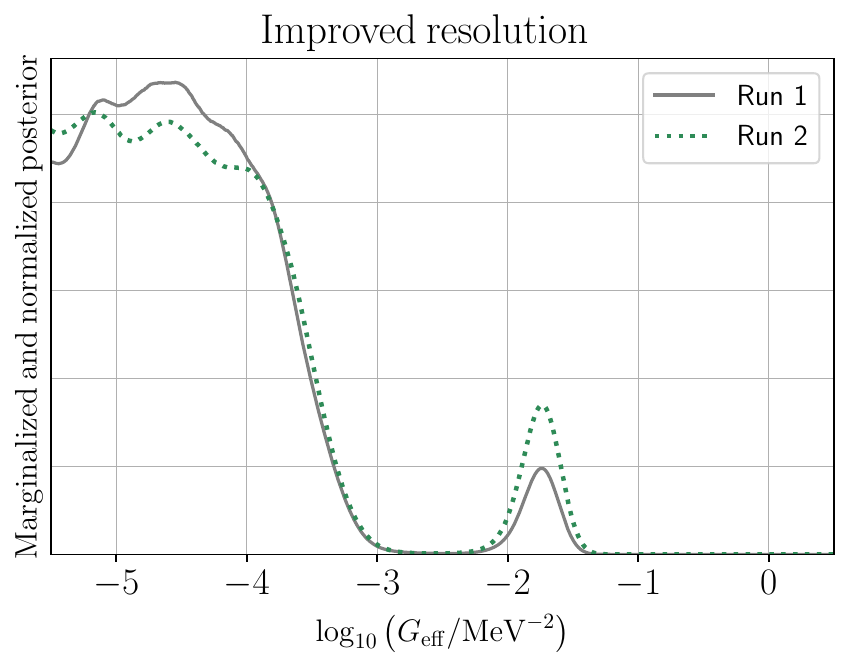}
\includegraphics[width=0.317\textwidth]{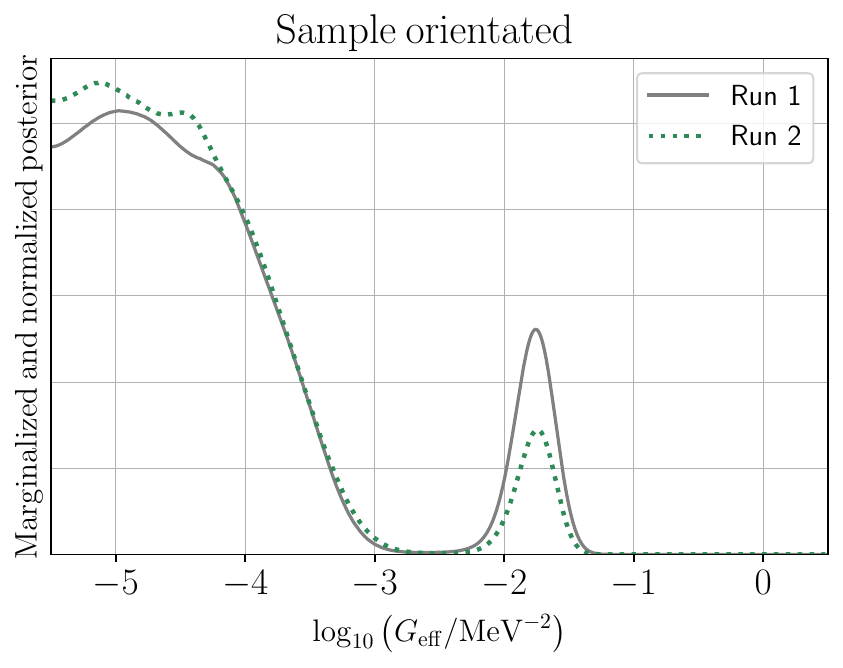}
\caption{\label{fig:TT_TE_EE_1d_NS_comparison} 1d marginalized (and normalized) $\logGeff$ posterior obtained when analyzing $\mathrm{TT,TE,EE}\:+\:\mathrm{BAO}$ data adopting the main analysis (left panel), improved resolution (central panel), and sample oriented (right panel) NS configurations. We conduct the analysis for each configuration two different times.}
\end{figure*}

Here, we present the results of additional analyses with varied NS configurations that could help to diagnose possible deficiencies in NS exploration. We consider the following configurations:
\begin{enumerate}
    \item Main analysis configuration: $2000$ live points and a sampling efficiency of $0.3$,
    \item Improved resolution configuration: same as previous configuration with improved resolution employing $3000$ live points,
    \item Sample oriented configuration: $2000$ live points and a sampling efficiency of $0.75$.
\end{enumerate}
In all these cases, we impose a minimum precision of $0.05$ in log-evidence. Due to the high-computational cost of analyzing the CMB and galaxy power spectra data, we only apply varied NS configuration to the combination of CMB and BAO data. We analyze $\mathrm{TT}\:+\:\mathrm{BAO}$ data performing three runs for each of the distinct configurations considered here. Furthermore, we perform two runs of each configuration when considering  $\mathrm{TT, EE, TE}\:+\:\mathrm{BAO}$ data. Results are shown in Fig.~\ref{fig:triangle_NS_comparison},~\ref{fig:TT_1d_NS_comparison},~and~\ref{fig:TT_TE_EE_1d_NS_comparison}. 

We find that the NS analysis disregarding Planck polarization data yields unstable constraints. Such constraints not only vary under different configurations of the NS framework, as shown by the left plot of Fig.~\ref{fig:triangle_NS_comparison}, but also under different runs of the same setup, as seen in Fig.~\ref{fig:TT_1d_NS_comparison}. We argue that this unstable behavior could be related to the intrinsic shape of the underlying posterior. As extensively discussed in the literature~\cite{Skilling:2006gxv,Higson:2018aaa,Higson:2018cqj,Fowlie:2020gfd,Buchner:2021,Ashton:2022grj}, NS can exhibit limitations depending on the nature of the likelihood under scrutiny, particularly affected by likelihoods with plateaus. The simplest representation of the self-interacting neutrino model intrinsically features a likelihood plateau for mildly (weakly) coupling strength, as scenarios with $\logGeff \lesssim -4$ are highly degenerate at (mildly) linear scale. Although this plateau is expected to impact the analysis independently of the combination of data considered, the absence of polarization data opens further degeneracies in the parameter space, complicating the problem. The CMB temperature power spectrum provides limited information about the reionization epoch, leading to loose constraints on the optical depth that are highly degenerate with the primordial amplitude $A_s$. This degeneration is reflected in the sum of the masses of neutrinos, whose parameter space widely opens up, allowing a large range of masses that modestly impact the value of the likelihood. The combination of the $\logGeff$ plateau and loose constraints on $\sum m_\nu$ leads to the unstable nature found for the constraints coming from the analysis of $\mathrm{TT}\:+\:\mathrm{BAO}$ data. 

Although more stable results are found when including the Planck polarization data in the analysis (see the right plot of Fig.~\ref{fig:triangle_NS_comparison}), we note that constraints become slightly unstable when adopting a sample-oriented configuration (see the right panel of Fig.~\ref{fig:TT_TE_EE_1d_NS_comparison}). Such slight instability could be related to the $\logGeff$ plateau. Additionally, it is important to emphasize that even when the analysis of $\mathrm{TT,TE,EE}\:+\:\mathrm{BAO}$ seems to be mainly unaffected by the aforementioned plateau, the impact of this could be more noticeable in a more complex parameter space, for instance, the one featured by the $\mathrm{TT,TE,EE}\:+\:\mathrm{FS}$ data. Additionally, further complications unrelated to the underlying shape of the posterior could bias the NS exploration. 

\begin{table*}[htb]
\caption{\label{tab:cons_CMB_PK}
68$\%$ CL intervals for the cosmological parameters (95$\%$ CL upper limit for the sum of the masses of neutrinos) obtained from the analysis of the $\mathrm{TT, EE, TE}\:+\:\mathrm{FS}$ data for the different cosmologies here considered.}
\begin{ruledtabular}
\begingroup
\renewcommand{\arraystretch}{1.35}
\begin{tabular} { l c c c c }
\multicolumn{5}{c}{TT, TE, EE + FS} \\ \hline 
 Parameter & $\Lambda$CDM & $\Lambda$CDM + $N_\mathrm{eff}$ + $\sum m_\nu$& MI$_\nu$ & SI$_\nu$ \\
\hline
$\log_{10} (G_\mathrm{eff}/\mathrm{MeV}^{-2})$ & ... & ... & $< -4.25$ & $-1.74^{+0.14}_{-0.11}$ \\
$N_\mathrm{eff}$ & $...$ & $2.97\pm 0.20$ & $2.99\pm 0.20$ & $2.77^{+0.19}_{-0.22}$ \\
$\sum m_\nu\, [\mathrm{eV}]$ & $...$ & $< 0.08 $ & $< 0.08$ & $< 0.12$ \\
$10^2 \omega_\mathrm{b}$ & $2.239\pm 0.013$ & $2.234\pm 0.018$ & $2.234\pm 0.018$ & $2.238^{+0.018}_{-0.020}$ \\
$\omega_\mathrm{cdm}$ & $0.1196\pm 0.001$ & $0.1185\pm 0.0033$ & $0.1187\pm 0.0032$ & $0.1165^{+0.0032}_{-0.0037}$ \\
$H_0\, [\mathrm{km\:s}^{-1}\:\mathrm{Mpc}^{-1}]$ & $67.53\pm 0.44$ & $67.1\pm 1.2$ & $67.2\pm 1.2$ & $66.6\pm 1.3$ \\
$\ln (10^{10}A_{\rm s})$ & $3.041\pm 0.016$ & $3.038\pm 0.018$ & $3.037\pm 0.018$ & $2.973\pm 0.017$ \\
$n_{\rm s}$ & $0.9660\pm 0.004$ & $0.9639\pm 0.007$ & $0.9631\pm 0.007$ & $0.9311\pm 0.006$ \\
$\tau_\mathrm{reio}$ & $0.054\pm 0.008$ & $0.053\pm 0.008$ & $0.053\pm 0.008$ & $0.052\pm 0.007$ \\
\hline
$\sigma_8$ & $0.8086\pm 0.007$ & $0.805^{+0.016}_{-0.013}$ & $0.806^{+0.016}_{-0.013}$ & $0.800^{+0.020}_{-0.014}$ \\
$S_8$ & $0.826\pm 0.012$ & $0.824\pm 0.015$ & $0.825\pm 0.014$ & $0.820^{+0.017}_{-0.014}$ \\
$10^{9} A_{\rm s} e^{-2\tau}$ & $1.881\pm 0.011$ & $1.876\pm 0.018$ & $1.875\pm 0.017$ & $1.761\pm 0.015$ \\
\end{tabular}
\endgroup
\end{ruledtabular}
\end{table*}

\begin{table*}[htb]
\caption{\label{tab:cons_CMB_BAO}
68$\%$ CL intervals for the cosmological parameters (95$\%$ CL upper limit for the sum of the masses of neutrinos) obtained from the analysis of the $\mathrm{TT, EE, TE}\:+\:\mathrm{BAO}$ data for the different cosmologies here considered.}
\begin{ruledtabular}
\begingroup
\renewcommand{\arraystretch}{1.35}
\begin{tabular}{lcccc}
\multicolumn{5}{c}{TT, TE, EE + BAO} \\ \hline 
 Parameter & $\Lambda$CDM & $\Lambda$CDM + $N_\mathrm{eff}$ + $\sum m_\nu$& MI$_\nu$ & SI$_\nu$ \\
\hline
$\log_{10} (G_\mathrm{eff}/\mathrm{MeV}^{-2})$ & ... & ... & $< -4.25$ & $-1.77^{+0.16}_{-0.10}$ \\
$N_\mathrm{eff}$ & $...$ & $2.97\pm 0.18$ & $2.99\pm 0.18$ & $2.78\pm 0.17$ \\
$\sum m_\nu\, [\mathrm{eV}]$ & ... & $< 0.051$ & $< 0.05$ & $< 0.08$ \\
$10^2 \omega_\mathrm{b}$ & $2.240\pm 0.014$ & $2.234\pm 0.019$ & $2.234\pm 0.018$ & $2.236\pm 0.019$ \\
$\omega_\mathrm{cdm}$ & $0.1193\pm 0.001$ & $0.1183\pm 0.003$ & $0.1185\pm 0.003$ & $0.1165\pm 0.0031$ \\
$H_0\, [\mathrm{km\:s}^{-1}\:\mathrm{Mpc}^{-1}]$ & $67.65\pm 0.46$ & $67.3\pm 1.1$ & $67.4\pm 1.1$ & $67.0\pm 1.2$ \\
$\ln (10^{10}A_{\rm s})$ & $3.045\pm 0.016$ & $3.041\pm 0.018$ & $3.041\pm 0.018$ & $2.975\pm 0.017$ \\
$n_{\rm s}$ & $0.9664\pm 0.004$ & $0.9637\pm 0.007$ & $0.9630\pm 0.007$ & $0.9298\pm 0.006$ \\
$\tau_\mathrm{reio}$ & $0.056\pm 0.008$ & $0.055\pm 0.008$ & $0.055\pm 0.008$ & $0.053\pm 0.007$ \\
\hline
$\sigma_8$ & $0.809\pm 0.007$ & $0.810^{+0.015}_{-0.012}$ & $0.811^{+0.014}_{-0.013}$ & $0.809^{+0.017}_{-0.014}$ \\
$S_8$ & $0.824\pm 0.013$ & $0.825\pm 0.015$ & $0.826\pm 0.014$ & $0.824^{+0.016}_{-0.015}$ \\
$10^{9} A_{\rm s} e^{-2\tau}$ & $1.881\pm 0.011$ & $1.875\pm 0.018$ & $1.874\pm 0.017$ & $1.761\pm 0.016$ \\
\end{tabular}
\endgroup
\end{ruledtabular}
\end{table*}

\section{Modes separation: cosmological constraints \label{ap:modes_constraints}}

We separate the modes obtained from the NS analysis using the Hierarchical Density-Based Spatial Clustering of Applications with Noise (\texttt{HDBSCAN}) algorithm~\cite{Campello:2013aaa,Campello:2015aaa,McInnes:2017aaa}. We assessed the robustness of the \texttt{HDBSCAN} algorithm in accurately separating modes by repeating the clustering analysis using an alternative approach. In such an approach, we first identified the value of $\logGeff$ that yields the lowest value of its corresponding one-dimensional marginalized posterior. This value is then employed to partition the NS chains into two regions corresponding to the SI$_\nu$ and MI$_\nu$ modes. We found that this alternative method and the \texttt{HDBSCAN} algorithm produce identical results.

Tables~\ref{tab:cons_CMB_PK}~and~\ref{tab:cons_CMB_BAO}, show the constraints for the $\Lambda$CDM and $\Lambda$CDM + $N_\mathrm{eff}$ + $\sum m_\nu$ models and the MI$\nu$ and SI$\nu$ modes of the self-interacting neutrino cosmology obtained when analyzing $\mathrm{TT, EE, TE}\:+\:\mathrm{FS}$ and  $\mathrm{TT, EE, TE}\:+\:\mathrm{BAO}$ data, respectively. 

Given that current cosmological data is insensitive to models with $\logGeff \lesssim -4.5$, we observe in Tables~\ref{tab:cons_CMB_PK}~and~\ref{tab:cons_CMB_BAO} that constraints for the MI$_\nu$ mode and $\Lambda$CDM + $N_\mathrm{eff}$ + $\sum m_\nu$ model are indistinguishable. In agreement to this, constraints obtained for the mildly self-interacting neutrinos concur with the $\Lambda$CDM constraints. Conversely, constraints on the strongly self-interacting neutrino cosmology highly deviate from the standard scenario, specially for the $A_{\rm s} e^{-2\tau}$ and $n_{\rm s}$ parameters. Additionally, we note a relaxation of $\gtrsim 50\%$ on the upper limit on the sum of neutrino masses for SI$_\nu$ mode in comparison with the $\Lambda$CDM + $N_\mathrm{eff}$ + $\sum m_\nu$ and the MI$_\nu$ cosmologies when analyzing either the $\mathrm{TT, EE, TE}\:+\:\mathrm{FS}$ of $\mathrm{TT, EE, TE}\:+\:\mathrm{BAO}$ data.

\section{Metropolis-Hasting pooling \label{ap:pooling}}

It is well-known that Metropolis-Hasting methods (MH) exhibit inefficiencies in sampling multi-modal posteriors. These inefficiencies primarily arise from their intrinsic incapacity to traverse low-probability regions separating different modes. While some adaptations of the MH algorithm, such as parallel tempering~\cite{Swendsen:1986}, seek to address this limitation, they often come at the cost of more likelihood evaluations or require fine-tuned configuration. 

Here, we introduce a heuristic method to sample the anticipated multi-modal posterior. Our method relies on the fact that low-probability barriers can be overcome by independently sampling disconnected portions of the parameter space where modes are expected. While directly pooling these independent samples would lead to a spurious realization of the target posterior, we note that is possible to recover the underlying posterior with an appropriate re-weighting process. We have found that an adequate re-weight can be achieved by exploiting the fact that, under uniform priors, for each mode $i$, the point that maximizes the posterior and likelihood, $\hat{\theta}^{i}$, leads to a constant ratio across the different modes,
\begin{equation}
    \mathcal{R} = \frac{\mathcal{P}(\hat{\theta}^{i})}{\mathcal{L}(d|\hat{\theta}^{i})}\,,
\end{equation}
where $\mathcal{P}(\theta)$ is the posterior, $\mathcal{L}(d|\theta)$ the likelihood, $d$ a given dataset, and  $\theta$ the set of cosmological parameters, such that $\theta = \left\lbrace \theta_0, \theta_1, ..., \theta_n \right\rbrace$, with $n$ being the number of free parameters considered into the analysis.

While computing $\mathcal{R}$ requires an \textit{a priori} knowledge of the posterior, in practice, MH enables an approximation to the posterior-likelihood ratio. Based on the fact that the number of effective samples in the neighborhood of the peak of the likelihood should be proportional to $\mathcal{P}(\hat{\theta}^{i})$, we can compute such a ratio for each mode:
\begin{equation}
    \tilde{\mathcal{R}}^{i} = \sum_{\hat{\theta}^{i} - \epsilon^i < \theta^i < \hat{\theta}^{i} + \epsilon^i} \frac{\alpha_i w(\theta^i) }{\mathcal{L}(d|\hat{\theta}^{i})} \,, \label{eq:R_mode}
\end{equation}
where $[\hat{\theta}^{i} - \epsilon^i, \hat{\theta}^{i} + \epsilon^i]$ defines a region that approximates the neighborhood of the peak of the likelihood, $w(\theta^i)$ is the weight of the MH sample, and $\alpha_i$ is a re-weighting factor to be determined. The appropriate re-weight of samples can then be obtained by arbitrarily fixing one of re-weighting factors, $\alpha_j = 1$, to later demand $\tilde{\mathcal{R}}^{i} = \tilde{\mathcal{R}}^j$ for all the remaining modes $i \neq j$.  

We highlight that, although our method does not require fine-tuning extra parameters, Eq~\eqref{eq:R_mode} exhibits some arbitrariness owing to the 
loose definition of the region $\smash{[\hat{\theta}^{i} - \epsilon^i, \hat{\theta}^{i} + \epsilon^i]}$. Such arbitrariness, however, can be overcome by pragmatically defining $\epsilon^i$. We adopt the Freedman-Diaconis rule~\cite{Freedman1981OnTH} and define:
\begin{equation}
    \epsilon^i_j = 2\frac{\mathrm{IQR}(\theta^i_j)}{\sqrt[3]{N}} \,, \label{eq:epsilon}
\end{equation}
where $N$ is the number of samples obtained for the corresponding mode and $\mathrm{IQR}$ is the so-called interquartile range. We apply Eq.~\eqref{eq:epsilon} for all the pertinent parameters, but the one that drives the multi-modality, $\theta_k$, which is left unbounded, $\epsilon_k \rightarrow \infty$.\footnote{In the presence of modes with significantly different standard deviations, a bounded $\epsilon_k$ can lead to substantial variations in the sizes of neighborhood ranges among the modes. Owing to volume biases, these differences tend to spoil the re-weighting process.} 

We present a proof of concept in the following by analyzing two toy models mimicking the expected bi-modal behaviour of $\logGeff$ posterior. The analyses of such models show that the MH pooling method can recover the target posterior with considerable precision.

\subsection{Proof of concept \label{ap:pooling_proof}}

\begin{figure*}[htb]
\includegraphics[width=0.475\textwidth]{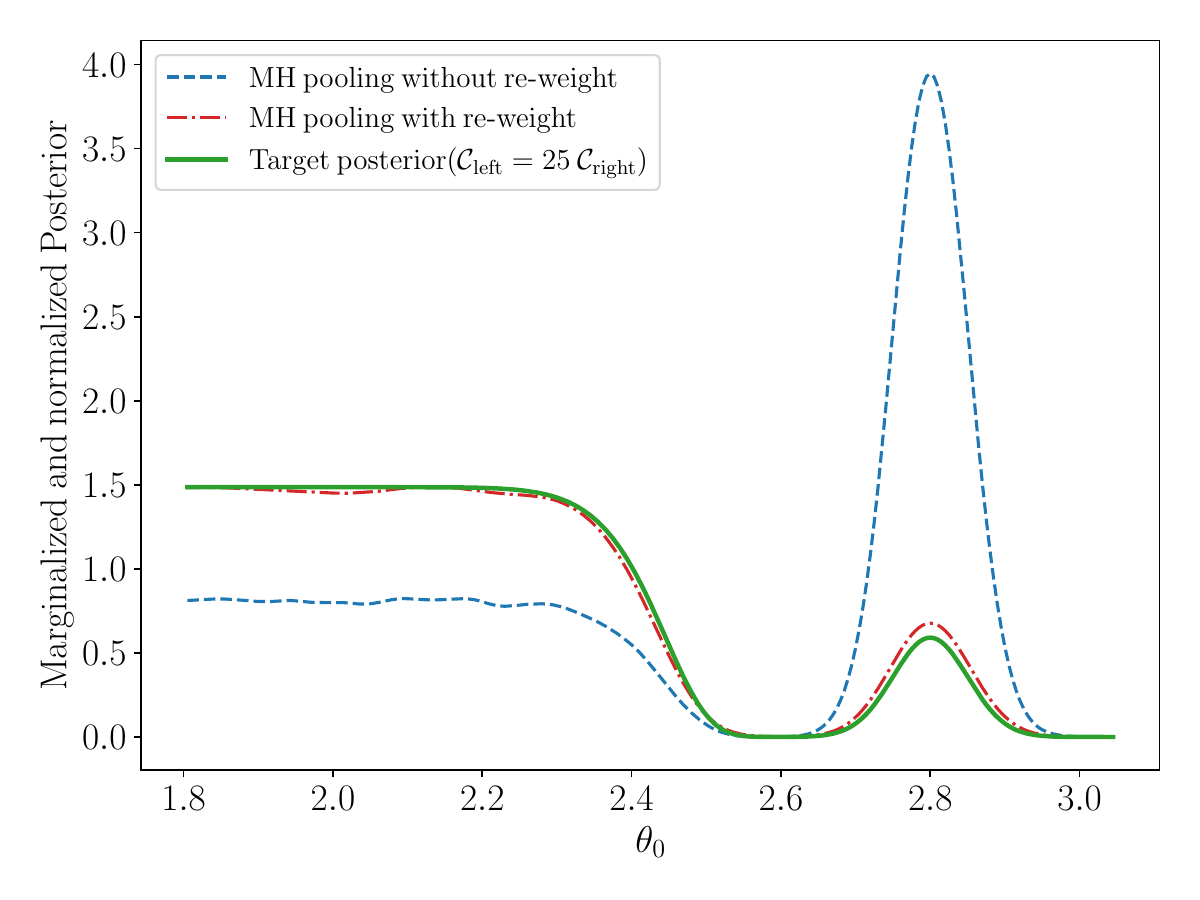}
\includegraphics[width=0.475\textwidth]{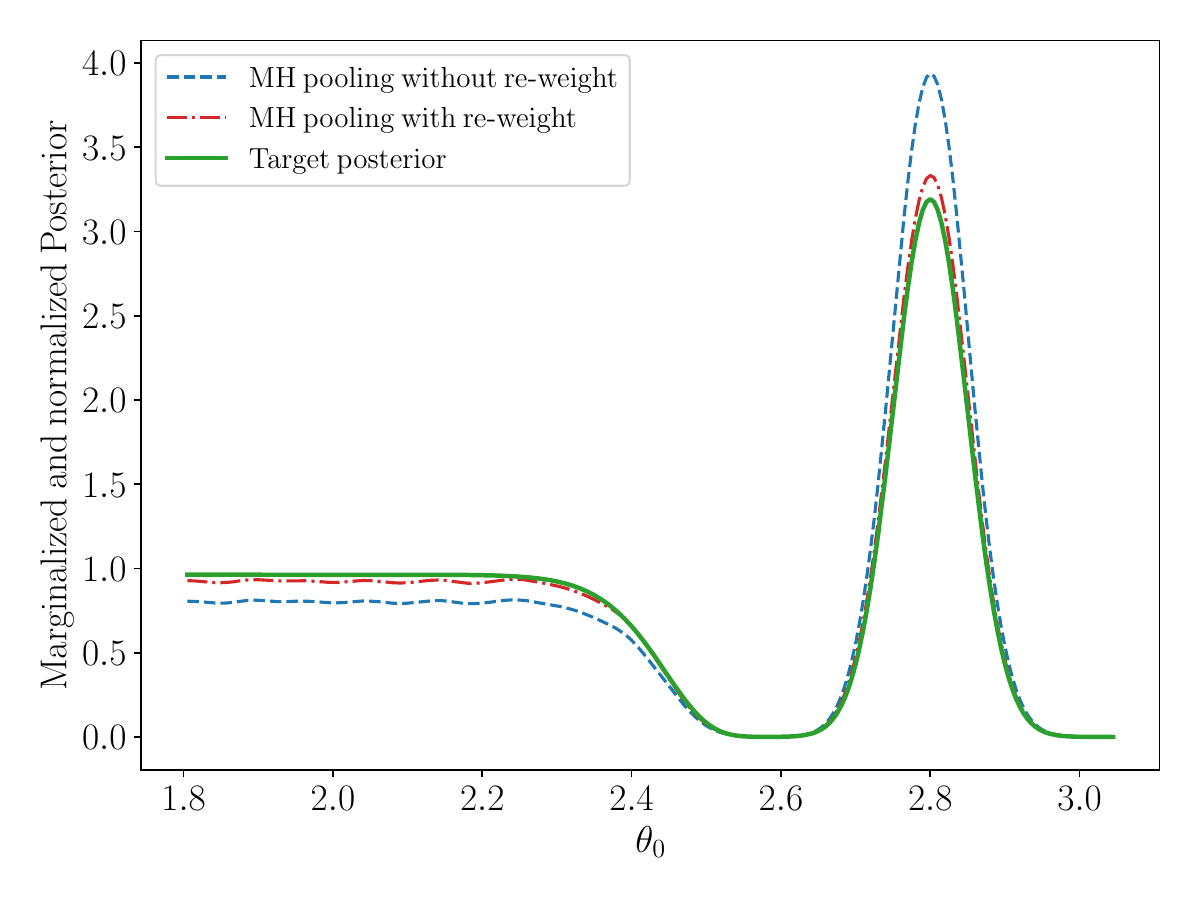}
\caption{\label{fig:toy_models} Results obtained from the MH pooling method when applied to two toys models mimicking,  Eq.~\ref{eq:toys}, the $\logGeff$ posterior.}
\end{figure*}

As a proof of concept for the heuristic Metropolis-Hasting framework, we analyze two toy models mimicking the expected $\logGeff$ posterior. The toy models are built assuming a two-dimensional parameter space, $\hat{\theta} =\left\lbrace \theta_0, \theta_1 \right\rbrace$, with a marked bi-modality in $\theta_0$. Assuming $\theta_0$ and $\theta_1$ as uncorrelated variables, we model the bi-modality by imposing a likelihood encompassing a normal distribution and a generalized normal distribution that peak at different regions of the parameter space. More specifically, we adopt
\begin{eqnarray} \label{eq:toys}
    \nonumber \mathcal{L}(\theta_0)  = && \mathcal{C}_\mathrm{right} \; \frac{1}{\sqrt{2\pi \sigma^2}} \exp{\left[-\frac{1}{2}\left(\frac{\theta_0-\mu_1}{\sigma}\right)^2\right]} \\ + && \frac{\mathcal{C}_\mathrm{left}}{2\alpha} \: \frac{\beta}{\Gamma(1/\beta)} \exp{\left[-\left(\frac{\theta_0-\mu_2}{\alpha}\right)^\beta\right]}\,, 
\end{eqnarray}
where $\Gamma$ denotes the Gamma function, $\mu_1 = 2.8, \sigma=0.05, \mu_2 = 1.8, \alpha = 0.65$, and $\beta = 12$. The first term of this equation mimics the SI$_\nu$ mode, while the second term introduces a plateau-like behavior similar to the MI$_\nu$ mode. The ratio between $\mathcal{C}_\mathrm{right}$ and $\mathcal{C}_\mathrm{left}$ determines whether the plateau, expected at $\theta_0 \rightarrow \mu_2$, dominates the posterior. We model two different scenarios assuming:
\begin{enumerate}
    \item $\mathcal{C}_\mathrm{left} = 25\: \mathcal{C}_\mathrm{right}$, emulating a posterior dominated by the plateau and saturating at $\mu_2$,
    \item $\mathcal{C}_\mathrm{left} = 3\: \mathcal{C}_\mathrm{right}$, emulating a posterior that peaks at~$\mu_1$. 
\end{enumerate}
Considering a uniform prior $\theta_0 = [1.8,3.0]$, we explore these models using the heuristic MH pooling method. Results of our analyses and the underlying real posterior are illustrated in Fig.~\ref{fig:toy_models}. 

 We observe that, unlike a pooling approach where no re-weight is applied (dashed blue lines), our heuristic method (dot-dashed red lines) --- following Eq.~\eqref{eq:R_mode} to obtain the appropriate re-weighting factors --- enables precise estimation of the target posterior (solid green lines). As shown by both panels of Fig.~\ref{fig:toy_models}, the underlying posterior is recovered regardless of the relative height between the plateau at $\mu_2$ and peak at $\mu_1$, demonstrating the efficiency of the MH pooling approximation. The results obtained for analyses of the toy models validate, at first approximation, the use of MH pooling as a method capable of exposing potential deficiencies in the posterior estimation provided by the NS exploration.

 \begin{figure*}[htb]
\includegraphics[width=0.475\textwidth]{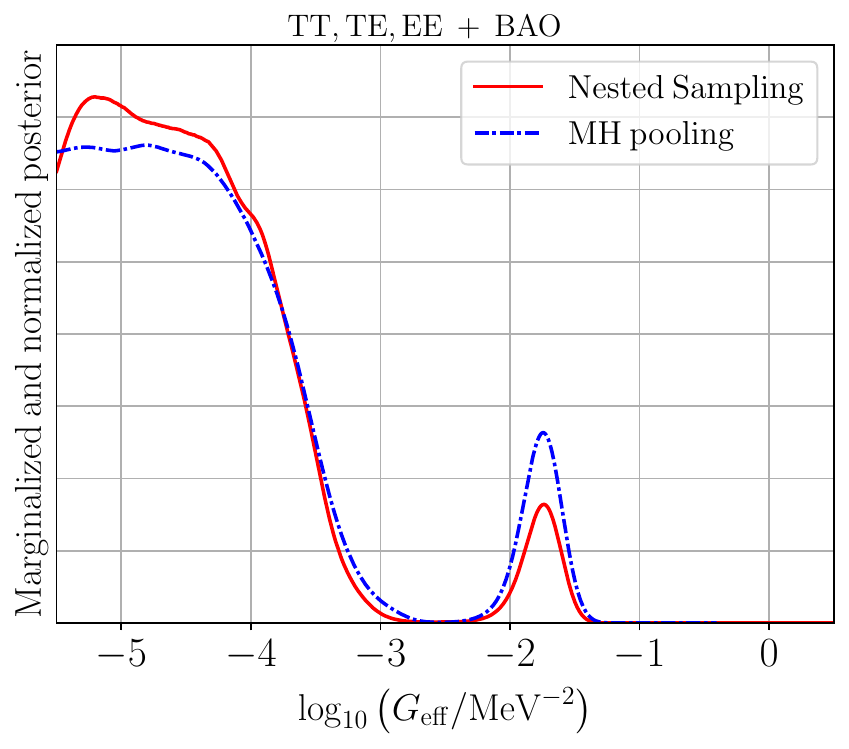}
\includegraphics[width=0.475\textwidth]{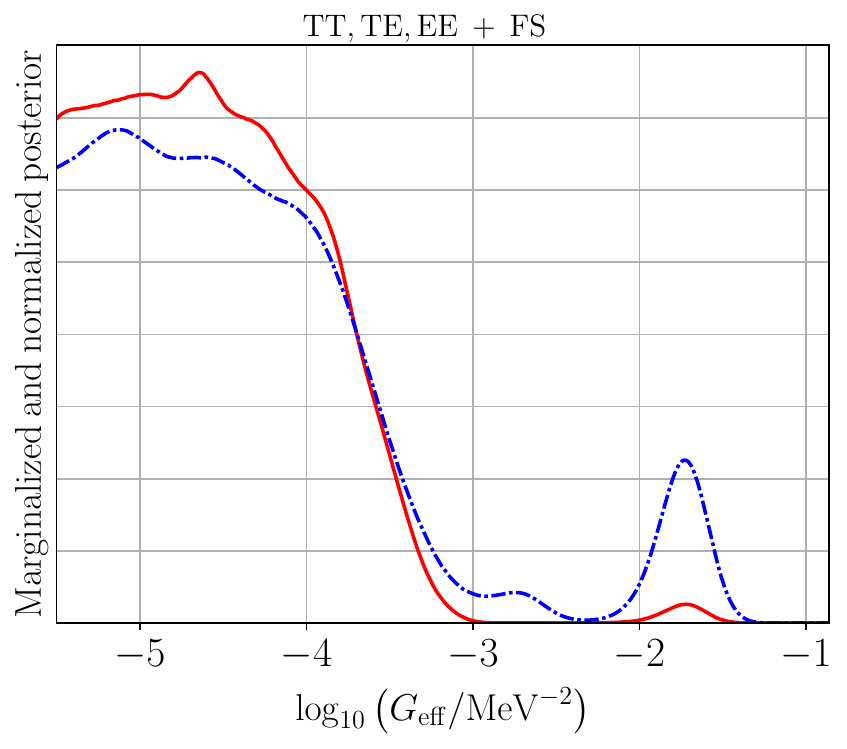}
\caption{\label{fig:NS_vs_MH} 1d marginalized (and normalized) $\logGeff$ posterior obtained when analyzing $\mathrm{TT, EE, TE}\:+\:\mathrm{FS}$ (right) and $\mathrm{TT, EE, TE}\:+\:\mathrm{BAO}$ (left) data employing the MH pooling (dashed blue lines) and NS (red solid lines) techniques.}
\end{figure*}

 Lastly, we highlight that the MH pooling method can be improved. For instance, we have assumed $\theta_0$ and $\theta_1$ are uncorrelated, yielding results insensitive to any assumption about $\theta_1$. However, this could change drastically in the presence of non-trivial correlations. Given that $\theta_1$ plays a crucial role in determining the region that approximates the neighborhood of the peak of the likelihood via Eq.~\eqref{eq:epsilon}, correlations with the bi-modal parameter could introduce volume bias affecting the posterior estimation. To overcome this issue, one could propose more sophisticated methods to delineate the region in which Eq.~\ref{eq:R_mode} is applied. Clustering algorithms or differentiable versions of the likelihood, allowing fast computation of their gradient, for instance, could enable more precise approximations of the neighborhood of a likelihood peak. These enhancements could significantly improve the re-weighting process, promoting the MH pooling method to a higher level beyond a complementary tool.

\subsection{\label{sub:pooling_cross} Cross-checking NS results}

Anticipating a bi-modality in $\logGeff$, we apply the MH pooling method considering two separate regions of the parameter space, i.e., we sample the SI$_\nu$ and MI$_\nu$ modes using the priors $\logGeff = [-2.5,0.5]$ and $\logGeff = [-5.5,-2.5]$, respectively considering the combination of $\mathrm{TT, EE, TE}\:+\:\mathrm{FS}$ and  $\mathrm{TT, EE, TE}\:+\:\mathrm{BAO}$ data. The remaining cosmological parameters follow the priors described in  Table~\ref{tab:priors}. We evaluate the convergence on the corresponding chains demanding $R - 1 \lesssim 0.005$ for all cosmological parameters. To avoid biases due to differences on the effective number of samples, we run chains with approximately the same length and consider normalized weights on Eq.~\eqref{eq:R_mode}. The marginalized $\logGeff$ posterior distributions obtained from both this method and the NS algorithm are presented in Fig.~\ref{fig:NS_vs_MH}.

We observe that, for both combinations of data, the marginalized posteriors provided by the MH pooling method (dashed blue lines) show higher SI$_\nu$ peaks than the NS posteriors (solid red line). In the case of the analysis considering BAO data (left panel), we note a slight difference between the MH pooling and NS method. This mismatch can be attributed to the intrinsic sampling error induced by both methods in the final kernel density estimation of the posterior. Conversely, the notable difference observed in the case of the analysis of CMB and galaxy power spectra (right panel) data could hint to a deficient NS exploration that fails to accurately portray the relative height between the MI$_\nu$ and SI$_\nu$ modes. It is noteworthy to mention that a comparison of the goodness of fit of the MI$_\nu$ and SI$_\nu$ modes yields $\Delta \chi^2_\mathrm{total} = 3.65$ (see Table~\ref{tab:model_comparison_extra}) implying a maximum likelihood ratio of $[\mathcal{L}^{\mathrm{MI}_\nu} / \mathcal{L}^{\mathrm{SI}_\nu}]_\mathrm{max} \sim 6$. Right panel of Fig.~\ref{fig:NS_vs_MH} shows that the NS algorithm (solid red line) overestimates this ratio.

An examination of the final NS product reveals a sampling bias favoring the MI$\nu$ mode. Specifically, we found that our NS exploration produces few samples for the SI$\nu$ mode within the $68\%$ confidence level. While pinpointing the precise cause of this bias is beyond the scope of this paper, it likely relates to the specific resolution assumed in our analyses, i.e., the number of NS live points. In this context, it is important to highlight that the likelihoods used here encompass a complex parameter space, expanding from ten to twenty-two dimensions when considering FS data instead of BAO measurements in the analysis. Thus, while one could theoretically increase the NS resolution, in practice, the complexity of the parameter space suggests that overcoming this bias might be challenging without significantly increasing computational time.

The sampling bias is also noticeable in the uncertainties of the evidence of the SI$_\nu$ mode when $\mathrm{TT, EE, TE}\:+\:\mathrm{FS}$ data is analyzed. Employing this data, NS algorithm provides a $15\%$ estimation of evidence for the MI$_\nu$ mode while a $70\%$ estimation for the SI$_\nu$ mode. As indicated by Table~\ref{tab:model_comparison_extra} and the last column of Table~\ref{tab:model_comparison}, these uncertainties play a dominant role in the errors associated with the Bayes factor, $\mathcal{B}_{i0}$, whose values are attained with approximately $70\%$ accuracy for the strongly self-interacting neutrino scenario. While significant, it is crucial to highlight that, at first approximation, these errors would enlarge (reduce) the value obtained for $\mathcal{B}_{i0}$ by a factor of $\sim 2$ ($\sim 0.5$), marginally affecting the level of support that data display for the simplest representation of delayed neutrino free streaming cosmology.

We argue that the challenges encountered in exploring the parameter space are related to the complexity of modeling the state-of-the-art cosmological data. Acknowledging that upcoming observations are unlikely to be less complex to model within a Bayesian likelihood-based framework, we emphasize the importance of recognizing and assessing the potential limitations of the diverse statistical methods employed in cosmological inference. Even when the challenges encountered in exploring the parameter space in this study may not be directly applicable to other analyses, discarding potential deficiencies will be fundamental in exploring new physics, particularly when analyzing nonstandard scenarios that, despite deviating significantly from the standard paradigm, can provide a good fit for cosmological observables.

\bibliography{apssamp}

\begin{thebibliography}{146}%
\makeatletter
\providecommand \@ifxundefined [1]{%
 \@ifx{#1\undefined}
}%
\providecommand \@ifnum [1]{%
 \ifnum #1\expandafter \@firstoftwo
 \else \expandafter \@secondoftwo
 \fi
}%
\providecommand \@ifx [1]{%
 \ifx #1\expandafter \@firstoftwo
 \else \expandafter \@secondoftwo
 \fi
}%
\providecommand \natexlab [1]{#1}%
\providecommand \enquote  [1]{``#1''}%
\providecommand \bibnamefont  [1]{#1}%
\providecommand \bibfnamefont [1]{#1}%
\providecommand \citenamefont [1]{#1}%
\providecommand \href@noop [0]{\@secondoftwo}%
\providecommand \href [0]{\begingroup \@sanitize@url \@href}%
\providecommand \@href[1]{\@@startlink{#1}\@@href}%
\providecommand \@@href[1]{\endgroup#1\@@endlink}%
\providecommand \@sanitize@url [0]{\catcode `\\12\catcode `\$12\catcode
  `\&12\catcode `\#12\catcode `\^12\catcode `\_12\catcode `\%12\relax}%
\providecommand \@@startlink[1]{}%
\providecommand \@@endlink[0]{}%
\providecommand \url  [0]{\begingroup\@sanitize@url \@url }%
\providecommand \@url [1]{\endgroup\@href {#1}{\urlprefix }}%
\providecommand \urlprefix  [0]{URL }%
\providecommand \Eprint [0]{\href }%
\providecommand \doibase [0]{https://doi.org/}%
\providecommand \selectlanguage [0]{\@gobble}%
\providecommand \bibinfo  [0]{\@secondoftwo}%
\providecommand \bibfield  [0]{\@secondoftwo}%
\providecommand \translation [1]{[#1]}%
\providecommand \BibitemOpen [0]{}%
\providecommand \bibitemStop [0]{}%
\providecommand \bibitemNoStop [0]{.\EOS\space}%
\providecommand \EOS [0]{\spacefactor3000\relax}%
\providecommand \BibitemShut  [1]{\csname bibitem#1\endcsname}%
\let\auto@bib@innerbib\@empty
\bibitem [{\citenamefont {Aghanim}\ \emph
  {et~al.}(2020{\natexlab{a}})\citenamefont {Aghanim} \emph
  {et~al.}}]{Planck:2018vyg}%
  \BibitemOpen
  \bibfield  {author} {\bibinfo {author} {\bibfnamefont {N.}~\bibnamefont
  {Aghanim}} \emph {et~al.} (\bibinfo {collaboration} {Planck}),\ }\bibfield
  {title} {\bibinfo {title} {{Planck 2018 results. VI. Cosmological
  parameters}},\ }\href {https://doi.org/10.1051/0004-6361/201833910}
  {\bibfield  {journal} {\bibinfo  {journal} {Astron. Astrophys.}\ }\textbf
  {\bibinfo {volume} {641}},\ \bibinfo {pages} {A6} (\bibinfo {year}
  {2020}{\natexlab{a}})},\ \bibinfo {note} {[Erratum: Astron.Astrophys. 652, C4
  (2021)]},\ \Eprint {https://arxiv.org/abs/1807.06209} {arXiv:1807.06209
  [astro-ph.CO]} \BibitemShut {NoStop}%
\bibitem [{\citenamefont {Lattanzi}\ and\ \citenamefont
  {Gerbino}(2018)}]{Lattanzi:2017ubx}%
  \BibitemOpen
  \bibfield  {author} {\bibinfo {author} {\bibfnamefont {M.}~\bibnamefont
  {Lattanzi}}\ and\ \bibinfo {author} {\bibfnamefont {M.}~\bibnamefont
  {Gerbino}},\ }\bibfield  {title} {\bibinfo {title} {{Status of neutrino
  properties and future prospects - Cosmological and astrophysical
  constraints}},\ }\href {https://doi.org/10.3389/fphy.2017.00070} {\bibfield
  {journal} {\bibinfo  {journal} {Front. in Phys.}\ }\textbf {\bibinfo {volume}
  {5}},\ \bibinfo {pages} {70} (\bibinfo {year} {2018})},\ \Eprint
  {https://arxiv.org/abs/1712.07109} {arXiv:1712.07109 [astro-ph.CO]}
  \BibitemShut {NoStop}%
\bibitem [{\citenamefont {Sakr}(2022)}]{Sakr:2022ans}%
  \BibitemOpen
  \bibfield  {author} {\bibinfo {author} {\bibfnamefont {Z.}~\bibnamefont
  {Sakr}},\ }\bibfield  {title} {\bibinfo {title} {{A Short Review on the
  Latest Neutrinos Mass and Number Constraints from Cosmological
  Observables}},\ }\href {https://doi.org/10.3390/universe8050284} {\bibfield
  {journal} {\bibinfo  {journal} {Universe}\ }\textbf {\bibinfo {volume} {8}},\
  \bibinfo {pages} {284} (\bibinfo {year} {2022})}\BibitemShut {NoStop}%
\bibitem [{\citenamefont {Bashinsky}\ and\ \citenamefont
  {Seljak}(2004)}]{Bashinsky:2003tk}%
  \BibitemOpen
  \bibfield  {author} {\bibinfo {author} {\bibfnamefont {S.}~\bibnamefont
  {Bashinsky}}\ and\ \bibinfo {author} {\bibfnamefont {U.}~\bibnamefont
  {Seljak}},\ }\bibfield  {title} {\bibinfo {title} {{Neutrino perturbations in
  CMB anisotropy and matter clustering}},\ }\href
  {https://doi.org/10.1103/PhysRevD.69.083002} {\bibfield  {journal} {\bibinfo
  {journal} {Phys. Rev. D}\ }\textbf {\bibinfo {volume} {69}},\ \bibinfo
  {pages} {083002} (\bibinfo {year} {2004})},\ \Eprint
  {https://arxiv.org/abs/astro-ph/0310198} {arXiv:astro-ph/0310198}
  \BibitemShut {NoStop}%
\bibitem [{\citenamefont {Baumann}\ \emph {et~al.}(2016)\citenamefont
  {Baumann}, \citenamefont {Green}, \citenamefont {Meyers},\ and\ \citenamefont
  {Wallisch}}]{Baumann:2015rya}%
  \BibitemOpen
  \bibfield  {author} {\bibinfo {author} {\bibfnamefont {D.}~\bibnamefont
  {Baumann}}, \bibinfo {author} {\bibfnamefont {D.}~\bibnamefont {Green}},
  \bibinfo {author} {\bibfnamefont {J.}~\bibnamefont {Meyers}},\ and\ \bibinfo
  {author} {\bibfnamefont {B.}~\bibnamefont {Wallisch}},\ }\bibfield  {title}
  {\bibinfo {title} {{Phases of New Physics in the CMB}},\ }\href
  {https://doi.org/10.1088/1475-7516/2016/01/007} {\bibfield  {journal}
  {\bibinfo  {journal} {JCAP}\ }\textbf {\bibinfo {volume} {01}},\ \bibinfo
  {pages} {007}},\ \Eprint {https://arxiv.org/abs/1508.06342} {arXiv:1508.06342
  [astro-ph.CO]} \BibitemShut {NoStop}%
\bibitem [{\citenamefont {Konoplich}\ and\ \citenamefont
  {Khlopov}(1988)}]{Konoplich:1988mj}%
  \BibitemOpen
  \bibfield  {author} {\bibinfo {author} {\bibfnamefont {R.~V.}\ \bibnamefont
  {Konoplich}}\ and\ \bibinfo {author} {\bibfnamefont {M.~Y.}\ \bibnamefont
  {Khlopov}},\ }\bibfield  {title} {\bibinfo {title} {{Constraints on triplet
  Majoron model due to observations of neutrinos from stellar collapse}},\
  }\href@noop {} {\bibfield  {journal} {\bibinfo  {journal} {Sov. J. Nucl.
  Phys.}\ }\textbf {\bibinfo {volume} {47}},\ \bibinfo {pages} {565} (\bibinfo
  {year} {1988})}\BibitemShut {NoStop}%
\bibitem [{\citenamefont {Berkov}\ \emph {et~al.}(1988)\citenamefont {Berkov},
  \citenamefont {Nikitin}, \citenamefont {Sudarikov},\ and\ \citenamefont
  {Khlopov}}]{Berkov:1988sd}%
  \BibitemOpen
  \bibfield  {author} {\bibinfo {author} {\bibfnamefont {A.~V.}\ \bibnamefont
  {Berkov}}, \bibinfo {author} {\bibfnamefont {Y.~P.}\ \bibnamefont {Nikitin}},
  \bibinfo {author} {\bibfnamefont {A.~L.}\ \bibnamefont {Sudarikov}},\ and\
  \bibinfo {author} {\bibfnamefont {M.~Y.}\ \bibnamefont {Khlopov}},\
  }\bibfield  {title} {\bibinfo {title} {{POSSIBLE MANIFESTATIONS OF ANOMALOUS
  4 NEUTRINO INTERACTION. (IN RUSSIAN)}},\ }\href@noop {} {\bibfield  {journal}
  {\bibinfo  {journal} {Sov. J. Nucl. Phys.}\ }\textbf {\bibinfo {volume}
  {48}},\ \bibinfo {pages} {497} (\bibinfo {year} {1988})}\BibitemShut
  {NoStop}%
\bibitem [{\citenamefont {Belotsky}\ \emph {et~al.}(2001)\citenamefont
  {Belotsky}, \citenamefont {Sudarikov},\ and\ \citenamefont
  {Khlopov}}]{Belotsky:2001fb}%
  \BibitemOpen
  \bibfield  {author} {\bibinfo {author} {\bibfnamefont {K.~M.}\ \bibnamefont
  {Belotsky}}, \bibinfo {author} {\bibfnamefont {A.~L.}\ \bibnamefont
  {Sudarikov}},\ and\ \bibinfo {author} {\bibfnamefont {M.~Y.}\ \bibnamefont
  {Khlopov}},\ }\bibfield  {title} {\bibinfo {title} {{Constraint on anomalous
  4nu interaction}},\ }\href {https://doi.org/10.1134/1.1409505} {\bibfield
  {journal} {\bibinfo  {journal} {Phys. Atom. Nucl.}\ }\textbf {\bibinfo
  {volume} {64}},\ \bibinfo {pages} {1637} (\bibinfo {year}
  {2001})}\BibitemShut {NoStop}%
\bibitem [{\citenamefont {Cyr-Racine}\ and\ \citenamefont
  {Sigurdson}(2014)}]{Cyr-Racine:2013jua}%
  \BibitemOpen
  \bibfield  {author} {\bibinfo {author} {\bibfnamefont {F.-Y.}\ \bibnamefont
  {Cyr-Racine}}\ and\ \bibinfo {author} {\bibfnamefont {K.}~\bibnamefont
  {Sigurdson}},\ }\bibfield  {title} {\bibinfo {title} {{Limits on
  Neutrino-Neutrino Scattering in the Early Universe}},\ }\href
  {https://doi.org/10.1103/PhysRevD.90.123533} {\bibfield  {journal} {\bibinfo
  {journal} {Phys. Rev. D}\ }\textbf {\bibinfo {volume} {90}},\ \bibinfo
  {pages} {123533} (\bibinfo {year} {2014})},\ \Eprint
  {https://arxiv.org/abs/1306.1536} {arXiv:1306.1536 [astro-ph.CO]}
  \BibitemShut {NoStop}%
\bibitem [{\citenamefont {Archidiacono}\ and\ \citenamefont
  {Hannestad}(2014)}]{Archidiacono:2013dua}%
  \BibitemOpen
  \bibfield  {author} {\bibinfo {author} {\bibfnamefont {M.}~\bibnamefont
  {Archidiacono}}\ and\ \bibinfo {author} {\bibfnamefont {S.}~\bibnamefont
  {Hannestad}},\ }\bibfield  {title} {\bibinfo {title} {{Updated constraints on
  non-standard neutrino interactions from Planck}},\ }\href
  {https://doi.org/10.1088/1475-7516/2014/07/046} {\bibfield  {journal}
  {\bibinfo  {journal} {JCAP}\ }\textbf {\bibinfo {volume} {07}},\ \bibinfo
  {pages} {046}},\ \Eprint {https://arxiv.org/abs/1311.3873} {arXiv:1311.3873
  [astro-ph.CO]} \BibitemShut {NoStop}%
\bibitem [{\citenamefont {Lancaster}\ \emph {et~al.}(2017)\citenamefont
  {Lancaster}, \citenamefont {Cyr-Racine}, \citenamefont {Knox},\ and\
  \citenamefont {Pan}}]{Lancaster:2017ksf}%
  \BibitemOpen
  \bibfield  {author} {\bibinfo {author} {\bibfnamefont {L.}~\bibnamefont
  {Lancaster}}, \bibinfo {author} {\bibfnamefont {F.-Y.}\ \bibnamefont
  {Cyr-Racine}}, \bibinfo {author} {\bibfnamefont {L.}~\bibnamefont {Knox}},\
  and\ \bibinfo {author} {\bibfnamefont {Z.}~\bibnamefont {Pan}},\ }\bibfield
  {title} {\bibinfo {title} {{A tale of two modes: Neutrino free-streaming in
  the early universe}},\ }\href {https://doi.org/10.1088/1475-7516/2017/07/033}
  {\bibfield  {journal} {\bibinfo  {journal} {JCAP}\ }\textbf {\bibinfo
  {volume} {07}},\ \bibinfo {pages} {033}},\ \Eprint
  {https://arxiv.org/abs/1704.06657} {arXiv:1704.06657 [astro-ph.CO]}
  \BibitemShut {NoStop}%
\bibitem [{\citenamefont {Oldengott}\ \emph {et~al.}(2017)\citenamefont
  {Oldengott}, \citenamefont {Tram}, \citenamefont {Rampf},\ and\ \citenamefont
  {Wong}}]{Oldengott:2017fhy}%
  \BibitemOpen
  \bibfield  {author} {\bibinfo {author} {\bibfnamefont {I.~M.}\ \bibnamefont
  {Oldengott}}, \bibinfo {author} {\bibfnamefont {T.}~\bibnamefont {Tram}},
  \bibinfo {author} {\bibfnamefont {C.}~\bibnamefont {Rampf}},\ and\ \bibinfo
  {author} {\bibfnamefont {Y.~Y.~Y.}\ \bibnamefont {Wong}},\ }\bibfield
  {title} {\bibinfo {title} {{Interacting neutrinos in cosmology: exact
  description and constraints}},\ }\href
  {https://doi.org/10.1088/1475-7516/2017/11/027} {\bibfield  {journal}
  {\bibinfo  {journal} {JCAP}\ }\textbf {\bibinfo {volume} {11}},\ \bibinfo
  {pages} {027}},\ \Eprint {https://arxiv.org/abs/1706.02123} {arXiv:1706.02123
  [astro-ph.CO]} \BibitemShut {NoStop}%
\bibitem [{\citenamefont {Choi}\ \emph {et~al.}(2018)\citenamefont {Choi},
  \citenamefont {Chiang},\ and\ \citenamefont {LoVerde}}]{Choi:2018gho}%
  \BibitemOpen
  \bibfield  {author} {\bibinfo {author} {\bibfnamefont {G.}~\bibnamefont
  {Choi}}, \bibinfo {author} {\bibfnamefont {C.-T.}\ \bibnamefont {Chiang}},\
  and\ \bibinfo {author} {\bibfnamefont {M.}~\bibnamefont {LoVerde}},\
  }\bibfield  {title} {\bibinfo {title} {{Probing Decoupling in Dark Sectors
  with the Cosmic Microwave Background}},\ }\href
  {https://doi.org/10.1088/1475-7516/2018/06/044} {\bibfield  {journal}
  {\bibinfo  {journal} {JCAP}\ }\textbf {\bibinfo {volume} {06}},\ \bibinfo
  {pages} {044}},\ \Eprint {https://arxiv.org/abs/1804.10180} {arXiv:1804.10180
  [astro-ph.CO]} \BibitemShut {NoStop}%
\bibitem [{\citenamefont {Song}\ \emph {et~al.}(2018)\citenamefont {Song},
  \citenamefont {Gonzalez-Garcia},\ and\ \citenamefont
  {Salvado}}]{Song:2018zyl}%
  \BibitemOpen
  \bibfield  {author} {\bibinfo {author} {\bibfnamefont {N.}~\bibnamefont
  {Song}}, \bibinfo {author} {\bibfnamefont {M.~C.}\ \bibnamefont
  {Gonzalez-Garcia}},\ and\ \bibinfo {author} {\bibfnamefont {J.}~\bibnamefont
  {Salvado}},\ }\bibfield  {title} {\bibinfo {title} {{Cosmological constraints
  with self-interacting sterile neutrinos}},\ }\href
  {https://doi.org/10.1088/1475-7516/2018/10/055} {\bibfield  {journal}
  {\bibinfo  {journal} {JCAP}\ }\textbf {\bibinfo {volume} {10}},\ \bibinfo
  {pages} {055}},\ \Eprint {https://arxiv.org/abs/1805.08218} {arXiv:1805.08218
  [astro-ph.CO]} \BibitemShut {NoStop}%
\bibitem [{\citenamefont {Lorenz}\ \emph {et~al.}(2019)\citenamefont {Lorenz},
  \citenamefont {Funcke}, \citenamefont {Calabrese},\ and\ \citenamefont
  {Hannestad}}]{Lorenz:2018fzb}%
  \BibitemOpen
  \bibfield  {author} {\bibinfo {author} {\bibfnamefont {C.~S.}\ \bibnamefont
  {Lorenz}}, \bibinfo {author} {\bibfnamefont {L.}~\bibnamefont {Funcke}},
  \bibinfo {author} {\bibfnamefont {E.}~\bibnamefont {Calabrese}},\ and\
  \bibinfo {author} {\bibfnamefont {S.}~\bibnamefont {Hannestad}},\ }\bibfield
  {title} {\bibinfo {title} {{Time-varying neutrino mass from a supercooled
  phase transition: current cosmological constraints and impact on the
  $\Omega_m$-$\sigma_8$ plane}},\ }\href
  {https://doi.org/10.1103/PhysRevD.99.023501} {\bibfield  {journal} {\bibinfo
  {journal} {Phys. Rev. D}\ }\textbf {\bibinfo {volume} {99}},\ \bibinfo
  {pages} {023501} (\bibinfo {year} {2019})},\ \Eprint
  {https://arxiv.org/abs/1811.01991} {arXiv:1811.01991 [astro-ph.CO]}
  \BibitemShut {NoStop}%
\bibitem [{\citenamefont {Barenboim}\ \emph {et~al.}(2019)\citenamefont
  {Barenboim}, \citenamefont {Denton},\ and\ \citenamefont
  {Oldengott}}]{Barenboim:2019tux}%
  \BibitemOpen
  \bibfield  {author} {\bibinfo {author} {\bibfnamefont {G.}~\bibnamefont
  {Barenboim}}, \bibinfo {author} {\bibfnamefont {P.~B.}\ \bibnamefont
  {Denton}},\ and\ \bibinfo {author} {\bibfnamefont {I.~M.}\ \bibnamefont
  {Oldengott}},\ }\bibfield  {title} {\bibinfo {title} {{Constraints on
  inflation with an extended neutrino sector}},\ }\href
  {https://doi.org/10.1103/PhysRevD.99.083515} {\bibfield  {journal} {\bibinfo
  {journal} {Phys. Rev. D}\ }\textbf {\bibinfo {volume} {99}},\ \bibinfo
  {pages} {083515} (\bibinfo {year} {2019})},\ \Eprint
  {https://arxiv.org/abs/1903.02036} {arXiv:1903.02036 [astro-ph.CO]}
  \BibitemShut {NoStop}%
\bibitem [{\citenamefont {Forastieri}\ \emph {et~al.}(2019)\citenamefont
  {Forastieri}, \citenamefont {Lattanzi},\ and\ \citenamefont
  {Natoli}}]{Forastieri:2019cuf}%
  \BibitemOpen
  \bibfield  {author} {\bibinfo {author} {\bibfnamefont {F.}~\bibnamefont
  {Forastieri}}, \bibinfo {author} {\bibfnamefont {M.}~\bibnamefont
  {Lattanzi}},\ and\ \bibinfo {author} {\bibfnamefont {P.}~\bibnamefont
  {Natoli}},\ }\bibfield  {title} {\bibinfo {title} {{Cosmological constraints
  on neutrino self-interactions with a light mediator}},\ }\href
  {https://doi.org/10.1103/PhysRevD.100.103526} {\bibfield  {journal} {\bibinfo
   {journal} {Phys. Rev. D}\ }\textbf {\bibinfo {volume} {100}},\ \bibinfo
  {pages} {103526} (\bibinfo {year} {2019})},\ \Eprint
  {https://arxiv.org/abs/1904.07810} {arXiv:1904.07810 [astro-ph.CO]}
  \BibitemShut {NoStop}%
\bibitem [{\citenamefont {Smirnov}\ and\ \citenamefont
  {Xu}(2019)}]{Smirnov:2019cae}%
  \BibitemOpen
  \bibfield  {author} {\bibinfo {author} {\bibfnamefont {A.~Y.}\ \bibnamefont
  {Smirnov}}\ and\ \bibinfo {author} {\bibfnamefont {X.-J.}\ \bibnamefont
  {Xu}},\ }\bibfield  {title} {\bibinfo {title} {{Wolfenstein potentials for
  neutrinos induced by ultra-light mediators}},\ }\href
  {https://doi.org/10.1007/JHEP12(2019)046} {\bibfield  {journal} {\bibinfo
  {journal} {JHEP}\ }\textbf {\bibinfo {volume} {12}},\ \bibinfo {pages}
  {046}},\ \Eprint {https://arxiv.org/abs/1909.07505} {arXiv:1909.07505
  [hep-ph]} \BibitemShut {NoStop}%
\bibitem [{\citenamefont {Escudero}\ and\ \citenamefont
  {Witte}(2020)}]{Escudero:2019gvw}%
  \BibitemOpen
  \bibfield  {author} {\bibinfo {author} {\bibfnamefont {M.}~\bibnamefont
  {Escudero}}\ and\ \bibinfo {author} {\bibfnamefont {S.~J.}\ \bibnamefont
  {Witte}},\ }\bibfield  {title} {\bibinfo {title} {{A CMB search for the
  neutrino mass mechanism and its relation to the Hubble tension}},\ }\href
  {https://doi.org/10.1140/epjc/s10052-020-7854-5} {\bibfield  {journal}
  {\bibinfo  {journal} {Eur. Phys. J. C}\ }\textbf {\bibinfo {volume} {80}},\
  \bibinfo {pages} {294} (\bibinfo {year} {2020})},\ \Eprint
  {https://arxiv.org/abs/1909.04044} {arXiv:1909.04044 [astro-ph.CO]}
  \BibitemShut {NoStop}%
\bibitem [{\citenamefont {Ghosh}\ \emph {et~al.}(2020)\citenamefont {Ghosh},
  \citenamefont {Khatri},\ and\ \citenamefont {Roy}}]{Ghosh:2019tab}%
  \BibitemOpen
  \bibfield  {author} {\bibinfo {author} {\bibfnamefont {S.}~\bibnamefont
  {Ghosh}}, \bibinfo {author} {\bibfnamefont {R.}~\bibnamefont {Khatri}},\ and\
  \bibinfo {author} {\bibfnamefont {T.~S.}\ \bibnamefont {Roy}},\ }\bibfield
  {title} {\bibinfo {title} {{Can dark neutrino interactions phase out the
  Hubble tension?}},\ }\href {https://doi.org/10.1103/PhysRevD.102.123544}
  {\bibfield  {journal} {\bibinfo  {journal} {Phys. Rev. D}\ }\textbf {\bibinfo
  {volume} {102}},\ \bibinfo {pages} {123544} (\bibinfo {year} {2020})},\
  \Eprint {https://arxiv.org/abs/1908.09843} {arXiv:1908.09843 [hep-ph]}
  \BibitemShut {NoStop}%
\bibitem [{\citenamefont {Funcke}\ \emph {et~al.}(2020)\citenamefont {Funcke},
  \citenamefont {Raffelt},\ and\ \citenamefont {Vitagliano}}]{Funcke:2019grs}%
  \BibitemOpen
  \bibfield  {author} {\bibinfo {author} {\bibfnamefont {L.}~\bibnamefont
  {Funcke}}, \bibinfo {author} {\bibfnamefont {G.}~\bibnamefont {Raffelt}},\
  and\ \bibinfo {author} {\bibfnamefont {E.}~\bibnamefont {Vitagliano}},\
  }\bibfield  {title} {\bibinfo {title} {{Distinguishing Dirac and Majorana
  neutrinos by their decays via Nambu-Goldstone bosons in the
  gravitational-anomaly model of neutrino masses}},\ }\href
  {https://doi.org/10.1103/PhysRevD.101.015025} {\bibfield  {journal} {\bibinfo
   {journal} {Phys. Rev. D}\ }\textbf {\bibinfo {volume} {101}},\ \bibinfo
  {pages} {015025} (\bibinfo {year} {2020})},\ \Eprint
  {https://arxiv.org/abs/1905.01264} {arXiv:1905.01264 [hep-ph]} \BibitemShut
  {NoStop}%
\bibitem [{\citenamefont {Sakstein}\ and\ \citenamefont
  {Trodden}(2020)}]{Sakstein:2019fmf}%
  \BibitemOpen
  \bibfield  {author} {\bibinfo {author} {\bibfnamefont {J.}~\bibnamefont
  {Sakstein}}\ and\ \bibinfo {author} {\bibfnamefont {M.}~\bibnamefont
  {Trodden}},\ }\bibfield  {title} {\bibinfo {title} {{Early Dark Energy from
  Massive Neutrinos as a Natural Resolution of the Hubble Tension}},\ }\href
  {https://doi.org/10.1103/PhysRevLett.124.161301} {\bibfield  {journal}
  {\bibinfo  {journal} {Phys. Rev. Lett.}\ }\textbf {\bibinfo {volume} {124}},\
  \bibinfo {pages} {161301} (\bibinfo {year} {2020})},\ \Eprint
  {https://arxiv.org/abs/1911.11760} {arXiv:1911.11760 [astro-ph.CO]}
  \BibitemShut {NoStop}%
\bibitem [{\citenamefont {Mazumdar}\ \emph {et~al.}(2020)\citenamefont
  {Mazumdar}, \citenamefont {Mohanty},\ and\ \citenamefont
  {Parashari}}]{Mazumdar:2019tbm}%
  \BibitemOpen
  \bibfield  {author} {\bibinfo {author} {\bibfnamefont {A.}~\bibnamefont
  {Mazumdar}}, \bibinfo {author} {\bibfnamefont {S.}~\bibnamefont {Mohanty}},\
  and\ \bibinfo {author} {\bibfnamefont {P.}~\bibnamefont {Parashari}},\
  }\bibfield  {title} {\bibinfo {title} {{Inflation models in the light of
  self-interacting sterile neutrinos}},\ }\href
  {https://doi.org/10.1103/PhysRevD.101.083521} {\bibfield  {journal} {\bibinfo
   {journal} {Phys. Rev. D}\ }\textbf {\bibinfo {volume} {101}},\ \bibinfo
  {pages} {083521} (\bibinfo {year} {2020})},\ \Eprint
  {https://arxiv.org/abs/1911.08512} {arXiv:1911.08512 [astro-ph.CO]}
  \BibitemShut {NoStop}%
\bibitem [{\citenamefont {Blinov}\ and\ \citenamefont
  {Marques-Tavares}(2020)}]{Blinov:2020hmc}%
  \BibitemOpen
  \bibfield  {author} {\bibinfo {author} {\bibfnamefont {N.}~\bibnamefont
  {Blinov}}\ and\ \bibinfo {author} {\bibfnamefont {G.}~\bibnamefont
  {Marques-Tavares}},\ }\bibfield  {title} {\bibinfo {title} {{Interacting
  radiation after Planck and its implications for the Hubble Tension}},\ }\href
  {https://doi.org/10.1088/1475-7516/2020/09/029} {\bibfield  {journal}
  {\bibinfo  {journal} {JCAP}\ }\textbf {\bibinfo {volume} {09}},\ \bibinfo
  {pages} {029}},\ \Eprint {https://arxiv.org/abs/2003.08387} {arXiv:2003.08387
  [astro-ph.CO]} \BibitemShut {NoStop}%
\bibitem [{\citenamefont {de~Gouv\^ea}\ \emph {et~al.}(2020)\citenamefont
  {de~Gouv\^ea}, \citenamefont {Dev}, \citenamefont {Dutta}, \citenamefont
  {Ghosh}, \citenamefont {Han},\ and\ \citenamefont
  {Zhang}}]{deGouvea:2019qaz}%
  \BibitemOpen
  \bibfield  {author} {\bibinfo {author} {\bibfnamefont {A.}~\bibnamefont
  {de~Gouv\^ea}}, \bibinfo {author} {\bibfnamefont {P.~S.~B.}\ \bibnamefont
  {Dev}}, \bibinfo {author} {\bibfnamefont {B.}~\bibnamefont {Dutta}}, \bibinfo
  {author} {\bibfnamefont {T.}~\bibnamefont {Ghosh}}, \bibinfo {author}
  {\bibfnamefont {T.}~\bibnamefont {Han}},\ and\ \bibinfo {author}
  {\bibfnamefont {Y.}~\bibnamefont {Zhang}},\ }\bibfield  {title} {\bibinfo
  {title} {{Leptonic Scalars at the LHC}},\ }\href
  {https://doi.org/10.1007/JHEP07(2020)142} {\bibfield  {journal} {\bibinfo
  {journal} {JHEP}\ }\textbf {\bibinfo {volume} {07}},\ \bibinfo {pages}
  {142}},\ \Eprint {https://arxiv.org/abs/1910.01132} {arXiv:1910.01132
  [hep-ph]} \BibitemShut {NoStop}%
\bibitem [{\citenamefont {Froustey}\ \emph {et~al.}(2020)\citenamefont
  {Froustey}, \citenamefont {Pitrou},\ and\ \citenamefont
  {Volpe}}]{Froustey:2020mcq}%
  \BibitemOpen
  \bibfield  {author} {\bibinfo {author} {\bibfnamefont {J.}~\bibnamefont
  {Froustey}}, \bibinfo {author} {\bibfnamefont {C.}~\bibnamefont {Pitrou}},\
  and\ \bibinfo {author} {\bibfnamefont {M.~C.}\ \bibnamefont {Volpe}},\
  }\bibfield  {title} {\bibinfo {title} {{Neutrino decoupling including flavour
  oscillations and primordial nucleosynthesis}},\ }\href
  {https://doi.org/10.1088/1475-7516/2020/12/015} {\bibfield  {journal}
  {\bibinfo  {journal} {JCAP}\ }\textbf {\bibinfo {volume} {12}},\ \bibinfo
  {pages} {015}},\ \Eprint {https://arxiv.org/abs/2008.01074} {arXiv:2008.01074
  [hep-ph]} \BibitemShut {NoStop}%
\bibitem [{\citenamefont {Babu}\ \emph {et~al.}(2020)\citenamefont {Babu},
  \citenamefont {Chauhan},\ and\ \citenamefont {Dev}}]{Babu:2019iml}%
  \BibitemOpen
  \bibfield  {author} {\bibinfo {author} {\bibfnamefont {K.~S.}\ \bibnamefont
  {Babu}}, \bibinfo {author} {\bibfnamefont {G.}~\bibnamefont {Chauhan}},\ and\
  \bibinfo {author} {\bibfnamefont {P.~S.~B.}\ \bibnamefont {Dev}},\ }\bibfield
   {title} {\bibinfo {title} {{Neutrino nonstandard interactions via light
  scalars in the Earth, Sun, supernovae, and the early Universe}},\ }\href
  {https://doi.org/10.1103/PhysRevD.101.095029} {\bibfield  {journal} {\bibinfo
   {journal} {Phys. Rev. D}\ }\textbf {\bibinfo {volume} {101}},\ \bibinfo
  {pages} {095029} (\bibinfo {year} {2020})},\ \Eprint
  {https://arxiv.org/abs/1912.13488} {arXiv:1912.13488 [hep-ph]} \BibitemShut
  {NoStop}%
\bibitem [{\citenamefont {Kreisch}\ \emph {et~al.}(2020)\citenamefont
  {Kreisch}, \citenamefont {Cyr-Racine},\ and\ \citenamefont
  {Dor\'e}}]{Kreisch:2019yzn}%
  \BibitemOpen
  \bibfield  {author} {\bibinfo {author} {\bibfnamefont {C.~D.}\ \bibnamefont
  {Kreisch}}, \bibinfo {author} {\bibfnamefont {F.-Y.}\ \bibnamefont
  {Cyr-Racine}},\ and\ \bibinfo {author} {\bibfnamefont {O.}~\bibnamefont
  {Dor\'e}},\ }\bibfield  {title} {\bibinfo {title} {{Neutrino puzzle:
  Anomalies, interactions, and cosmological tensions}},\ }\href
  {https://doi.org/10.1103/PhysRevD.101.123505} {\bibfield  {journal} {\bibinfo
   {journal} {Phys. Rev. D}\ }\textbf {\bibinfo {volume} {101}},\ \bibinfo
  {pages} {123505} (\bibinfo {year} {2020})},\ \Eprint
  {https://arxiv.org/abs/1902.00534} {arXiv:1902.00534 [astro-ph.CO]}
  \BibitemShut {NoStop}%
\bibitem [{\citenamefont {Park}\ \emph {et~al.}(2019)\citenamefont {Park},
  \citenamefont {Kreisch}, \citenamefont {Dunkley}, \citenamefont
  {Hadzhiyska},\ and\ \citenamefont {Cyr-Racine}}]{Park:2019ibn}%
  \BibitemOpen
  \bibfield  {author} {\bibinfo {author} {\bibfnamefont {M.}~\bibnamefont
  {Park}}, \bibinfo {author} {\bibfnamefont {C.~D.}\ \bibnamefont {Kreisch}},
  \bibinfo {author} {\bibfnamefont {J.}~\bibnamefont {Dunkley}}, \bibinfo
  {author} {\bibfnamefont {B.}~\bibnamefont {Hadzhiyska}},\ and\ \bibinfo
  {author} {\bibfnamefont {F.-Y.}\ \bibnamefont {Cyr-Racine}},\ }\bibfield
  {title} {\bibinfo {title} {{$\Lambda$CDM or self-interacting neutrinos: How
  CMB data can tell the two models apart}},\ }\href
  {https://doi.org/10.1103/PhysRevD.100.063524} {\bibfield  {journal} {\bibinfo
   {journal} {Phys. Rev. D}\ }\textbf {\bibinfo {volume} {100}},\ \bibinfo
  {pages} {063524} (\bibinfo {year} {2019})},\ \Eprint
  {https://arxiv.org/abs/1904.02625} {arXiv:1904.02625 [astro-ph.CO]}
  \BibitemShut {NoStop}%
\bibitem [{\citenamefont {Deppisch}\ \emph {et~al.}(2020)\citenamefont
  {Deppisch}, \citenamefont {Graf}, \citenamefont {Rodejohann},\ and\
  \citenamefont {Xu}}]{Deppisch:2020sqh}%
  \BibitemOpen
  \bibfield  {author} {\bibinfo {author} {\bibfnamefont {F.~F.}\ \bibnamefont
  {Deppisch}}, \bibinfo {author} {\bibfnamefont {L.}~\bibnamefont {Graf}},
  \bibinfo {author} {\bibfnamefont {W.}~\bibnamefont {Rodejohann}},\ and\
  \bibinfo {author} {\bibfnamefont {X.-J.}\ \bibnamefont {Xu}},\ }\bibfield
  {title} {\bibinfo {title} {Neutrino self-interactions and double beta
  decay},\ }\href {https://doi.org/10.1103/PhysRevD.102.051701} {\bibfield
  {journal} {\bibinfo  {journal} {Phys. Rev. D}\ }\textbf {\bibinfo {volume}
  {102}},\ \bibinfo {pages} {051701} (\bibinfo {year} {2020})},\ \Eprint
  {https://arxiv.org/abs/2004.11919} {arXiv:2004.11919 [hep-ph]} \BibitemShut
  {NoStop}%
\bibitem [{\citenamefont {Kelly}\ \emph {et~al.}(2020)\citenamefont {Kelly},
  \citenamefont {Sen}, \citenamefont {Tangarife},\ and\ \citenamefont
  {Zhang}}]{Kelly:2020pcy}%
  \BibitemOpen
  \bibfield  {author} {\bibinfo {author} {\bibfnamefont {K.~J.}\ \bibnamefont
  {Kelly}}, \bibinfo {author} {\bibfnamefont {M.}~\bibnamefont {Sen}}, \bibinfo
  {author} {\bibfnamefont {W.}~\bibnamefont {Tangarife}},\ and\ \bibinfo
  {author} {\bibfnamefont {Y.}~\bibnamefont {Zhang}},\ }\bibfield  {title}
  {\bibinfo {title} {{Origin of sterile neutrino dark matter via secret
  neutrino interactions with vector bosons}},\ }\href
  {https://doi.org/10.1103/PhysRevD.101.115031} {\bibfield  {journal} {\bibinfo
   {journal} {Phys. Rev. D}\ }\textbf {\bibinfo {volume} {101}},\ \bibinfo
  {pages} {115031} (\bibinfo {year} {2020})},\ \Eprint
  {https://arxiv.org/abs/2005.03681} {arXiv:2005.03681 [hep-ph]} \BibitemShut
  {NoStop}%
\bibitem [{\citenamefont {Escudero~Abenza}(2020)}]{EscuderoAbenza:2020cmq}%
  \BibitemOpen
  \bibfield  {author} {\bibinfo {author} {\bibfnamefont {M.}~\bibnamefont
  {Escudero~Abenza}},\ }\bibfield  {title} {\bibinfo {title} {{Precision early
  universe thermodynamics made simple: $N_{\rm eff}$ and neutrino decoupling in
  the Standard Model and beyond}},\ }\href
  {https://doi.org/10.1088/1475-7516/2020/05/048} {\bibfield  {journal}
  {\bibinfo  {journal} {JCAP}\ }\textbf {\bibinfo {volume} {05}},\ \bibinfo
  {pages} {048}},\ \Eprint {https://arxiv.org/abs/2001.04466} {arXiv:2001.04466
  [hep-ph]} \BibitemShut {NoStop}%
\bibitem [{\citenamefont {He}\ \emph {et~al.}(2020)\citenamefont {He},
  \citenamefont {Ma},\ and\ \citenamefont {Zheng}}]{He:2020zns}%
  \BibitemOpen
  \bibfield  {author} {\bibinfo {author} {\bibfnamefont {H.-J.}\ \bibnamefont
  {He}}, \bibinfo {author} {\bibfnamefont {Y.-Z.}\ \bibnamefont {Ma}},\ and\
  \bibinfo {author} {\bibfnamefont {J.}~\bibnamefont {Zheng}},\ }\bibfield
  {title} {\bibinfo {title} {{Resolving Hubble Tension by Self-Interacting
  Neutrinos with Dirac Seesaw}},\ }\href
  {https://doi.org/10.1088/1475-7516/2020/11/003} {\bibfield  {journal}
  {\bibinfo  {journal} {JCAP}\ }\textbf {\bibinfo {volume} {11}},\ \bibinfo
  {pages} {003}},\ \Eprint {https://arxiv.org/abs/2003.12057} {arXiv:2003.12057
  [hep-ph]} \BibitemShut {NoStop}%
\bibitem [{\citenamefont {Ding}\ and\ \citenamefont
  {Feruglio}(2020)}]{Ding:2020yen}%
  \BibitemOpen
  \bibfield  {author} {\bibinfo {author} {\bibfnamefont {G.-J.}\ \bibnamefont
  {Ding}}\ and\ \bibinfo {author} {\bibfnamefont {F.}~\bibnamefont
  {Feruglio}},\ }\bibfield  {title} {\bibinfo {title} {{Testing Moduli and
  Flavon Dynamics with Neutrino Oscillations}},\ }\href
  {https://doi.org/10.1007/JHEP06(2020)134} {\bibfield  {journal} {\bibinfo
  {journal} {JHEP}\ }\textbf {\bibinfo {volume} {06}},\ \bibinfo {pages}
  {134}},\ \Eprint {https://arxiv.org/abs/2003.13448} {arXiv:2003.13448
  [hep-ph]} \BibitemShut {NoStop}%
\bibitem [{\citenamefont {Berbig}\ \emph {et~al.}(2020)\citenamefont {Berbig},
  \citenamefont {Jana},\ and\ \citenamefont {Trautner}}]{Berbig:2020wve}%
  \BibitemOpen
  \bibfield  {author} {\bibinfo {author} {\bibfnamefont {M.}~\bibnamefont
  {Berbig}}, \bibinfo {author} {\bibfnamefont {S.}~\bibnamefont {Jana}},\ and\
  \bibinfo {author} {\bibfnamefont {A.}~\bibnamefont {Trautner}},\ }\bibfield
  {title} {\bibinfo {title} {{The Hubble tension and a renormalizable model of
  gauged neutrino self-interactions}},\ }\href
  {https://doi.org/10.1103/PhysRevD.102.115008} {\bibfield  {journal} {\bibinfo
   {journal} {Phys. Rev. D}\ }\textbf {\bibinfo {volume} {102}},\ \bibinfo
  {pages} {115008} (\bibinfo {year} {2020})},\ \Eprint
  {https://arxiv.org/abs/2004.13039} {arXiv:2004.13039 [hep-ph]} \BibitemShut
  {NoStop}%
\bibitem [{\citenamefont {Gogoi}\ \emph {et~al.}(2021)\citenamefont {Gogoi},
  \citenamefont {Sharma}, \citenamefont {Chanda},\ and\ \citenamefont
  {Das}}]{Gogoi:2020qif}%
  \BibitemOpen
  \bibfield  {author} {\bibinfo {author} {\bibfnamefont {A.}~\bibnamefont
  {Gogoi}}, \bibinfo {author} {\bibfnamefont {R.~K.}\ \bibnamefont {Sharma}},
  \bibinfo {author} {\bibfnamefont {P.}~\bibnamefont {Chanda}},\ and\ \bibinfo
  {author} {\bibfnamefont {S.}~\bibnamefont {Das}},\ }\bibfield  {title}
  {\bibinfo {title} {{Early Mass-varying Neutrino Dark Energy: Nugget Formation
  and Hubble Anomaly}},\ }\href {https://doi.org/10.3847/1538-4357/abfe5b}
  {\bibfield  {journal} {\bibinfo  {journal} {Astrophys. J.}\ }\textbf
  {\bibinfo {volume} {915}},\ \bibinfo {pages} {132} (\bibinfo {year}
  {2021})},\ \Eprint {https://arxiv.org/abs/2005.11889} {arXiv:2005.11889
  [astro-ph.CO]} \BibitemShut {NoStop}%
\bibitem [{\citenamefont {Barenboim}\ and\ \citenamefont
  {Nierste}(2021)}]{Barenboim:2020dmg}%
  \BibitemOpen
  \bibfield  {author} {\bibinfo {author} {\bibfnamefont {G.}~\bibnamefont
  {Barenboim}}\ and\ \bibinfo {author} {\bibfnamefont {U.}~\bibnamefont
  {Nierste}},\ }\bibfield  {title} {\bibinfo {title} {{Modified majoron model
  for cosmological anomalies}},\ }\href
  {https://doi.org/10.1103/PhysRevD.104.023013} {\bibfield  {journal} {\bibinfo
   {journal} {Phys. Rev. D}\ }\textbf {\bibinfo {volume} {104}},\ \bibinfo
  {pages} {023013} (\bibinfo {year} {2021})},\ \Eprint
  {https://arxiv.org/abs/2005.13280} {arXiv:2005.13280 [hep-ph]} \BibitemShut
  {NoStop}%
\bibitem [{\citenamefont {Das}\ and\ \citenamefont
  {Ghosh}(2021)}]{Das:2020xke}%
  \BibitemOpen
  \bibfield  {author} {\bibinfo {author} {\bibfnamefont {A.}~\bibnamefont
  {Das}}\ and\ \bibinfo {author} {\bibfnamefont {S.}~\bibnamefont {Ghosh}},\
  }\bibfield  {title} {\bibinfo {title} {{Flavor-specific interaction favors
  strong neutrino self-coupling in the early universe}},\ }\href
  {https://doi.org/10.1088/1475-7516/2021/07/038} {\bibfield  {journal}
  {\bibinfo  {journal} {JCAP}\ }\textbf {\bibinfo {volume} {07}},\ \bibinfo
  {pages} {038}},\ \Eprint {https://arxiv.org/abs/2011.12315} {arXiv:2011.12315
  [astro-ph.CO]} \BibitemShut {NoStop}%
\bibitem [{\citenamefont {Mazumdar}\ \emph {et~al.}(2022)\citenamefont
  {Mazumdar}, \citenamefont {Mohanty},\ and\ \citenamefont
  {Parashari}}]{Mazumdar:2020ibx}%
  \BibitemOpen
  \bibfield  {author} {\bibinfo {author} {\bibfnamefont {A.}~\bibnamefont
  {Mazumdar}}, \bibinfo {author} {\bibfnamefont {S.}~\bibnamefont {Mohanty}},\
  and\ \bibinfo {author} {\bibfnamefont {P.}~\bibnamefont {Parashari}},\
  }\bibfield  {title} {\bibinfo {title} {{Flavour specific neutrino
  self-interaction: H $_{0}$ tension and IceCube}},\ }\href
  {https://doi.org/10.1088/1475-7516/2022/10/011} {\bibfield  {journal}
  {\bibinfo  {journal} {JCAP}\ }\textbf {\bibinfo {volume} {10}},\ \bibinfo
  {pages} {011}},\ \Eprint {https://arxiv.org/abs/2011.13685} {arXiv:2011.13685
  [hep-ph]} \BibitemShut {NoStop}%
\bibitem [{\citenamefont {Brinckmann}\ \emph {et~al.}(2021)\citenamefont
  {Brinckmann}, \citenamefont {Chang},\ and\ \citenamefont
  {LoVerde}}]{Brinckmann:2020bcn}%
  \BibitemOpen
  \bibfield  {author} {\bibinfo {author} {\bibfnamefont {T.}~\bibnamefont
  {Brinckmann}}, \bibinfo {author} {\bibfnamefont {J.~H.}\ \bibnamefont
  {Chang}},\ and\ \bibinfo {author} {\bibfnamefont {M.}~\bibnamefont
  {LoVerde}},\ }\bibfield  {title} {\bibinfo {title} {{Self-interacting
  neutrinos, the Hubble parameter tension, and the cosmic microwave
  background}},\ }\href {https://doi.org/10.1103/PhysRevD.104.063523}
  {\bibfield  {journal} {\bibinfo  {journal} {Phys. Rev. D}\ }\textbf {\bibinfo
  {volume} {104}},\ \bibinfo {pages} {063523} (\bibinfo {year} {2021})},\
  \Eprint {https://arxiv.org/abs/2012.11830} {arXiv:2012.11830 [astro-ph.CO]}
  \BibitemShut {NoStop}%
\bibitem [{\citenamefont {Kelly}\ \emph {et~al.}(2021)\citenamefont {Kelly},
  \citenamefont {Sen},\ and\ \citenamefont {Zhang}}]{Kelly:2020aks}%
  \BibitemOpen
  \bibfield  {author} {\bibinfo {author} {\bibfnamefont {K.~J.}\ \bibnamefont
  {Kelly}}, \bibinfo {author} {\bibfnamefont {M.}~\bibnamefont {Sen}},\ and\
  \bibinfo {author} {\bibfnamefont {Y.}~\bibnamefont {Zhang}},\ }\bibfield
  {title} {\bibinfo {title} {{Intimate Relationship between Sterile Neutrino
  Dark Matter and \ensuremath{\Delta}Neff}},\ }\href
  {https://doi.org/10.1103/PhysRevLett.127.041101} {\bibfield  {journal}
  {\bibinfo  {journal} {Phys. Rev. Lett.}\ }\textbf {\bibinfo {volume} {127}},\
  \bibinfo {pages} {041101} (\bibinfo {year} {2021})},\ \Eprint
  {https://arxiv.org/abs/2011.02487} {arXiv:2011.02487 [hep-ph]} \BibitemShut
  {NoStop}%
\bibitem [{\citenamefont {Esteban}\ and\ \citenamefont
  {Salvado}(2021)}]{Esteban:2021ozz}%
  \BibitemOpen
  \bibfield  {author} {\bibinfo {author} {\bibfnamefont {I.}~\bibnamefont
  {Esteban}}\ and\ \bibinfo {author} {\bibfnamefont {J.}~\bibnamefont
  {Salvado}},\ }\bibfield  {title} {\bibinfo {title} {{Long Range Interactions
  in Cosmology: Implications for Neutrinos}},\ }\href
  {https://doi.org/10.1088/1475-7516/2021/05/036} {\bibfield  {journal}
  {\bibinfo  {journal} {JCAP}\ }\textbf {\bibinfo {volume} {05}},\ \bibinfo
  {pages} {036}},\ \Eprint {https://arxiv.org/abs/2101.05804} {arXiv:2101.05804
  [hep-ph]} \BibitemShut {NoStop}%
\bibitem [{\citenamefont {Arias-Aragon}\ \emph {et~al.}(2021)\citenamefont
  {Arias-Aragon}, \citenamefont {Fernandez-Martinez}, \citenamefont
  {Gonzalez-Lopez},\ and\ \citenamefont {Merlo}}]{Arias-Aragon:2020qip}%
  \BibitemOpen
  \bibfield  {author} {\bibinfo {author} {\bibfnamefont {F.}~\bibnamefont
  {Arias-Aragon}}, \bibinfo {author} {\bibfnamefont {E.}~\bibnamefont
  {Fernandez-Martinez}}, \bibinfo {author} {\bibfnamefont {M.}~\bibnamefont
  {Gonzalez-Lopez}},\ and\ \bibinfo {author} {\bibfnamefont {L.}~\bibnamefont
  {Merlo}},\ }\bibfield  {title} {\bibinfo {title} {{Neutrino Masses and Hubble
  Tension via a Majoron in MFV}},\ }\href
  {https://doi.org/10.1140/epjc/s10052-020-08825-8} {\bibfield  {journal}
  {\bibinfo  {journal} {Eur. Phys. J. C}\ }\textbf {\bibinfo {volume} {81}},\
  \bibinfo {pages} {28} (\bibinfo {year} {2021})},\ \Eprint
  {https://arxiv.org/abs/2009.01848} {arXiv:2009.01848 [hep-ph]} \BibitemShut
  {NoStop}%
\bibitem [{\citenamefont {Du}\ and\ \citenamefont {Yu}(2021)}]{Du:2021idh}%
  \BibitemOpen
  \bibfield  {author} {\bibinfo {author} {\bibfnamefont {Y.}~\bibnamefont
  {Du}}\ and\ \bibinfo {author} {\bibfnamefont {J.-H.}\ \bibnamefont {Yu}},\
  }\bibfield  {title} {\bibinfo {title} {{Neutrino non-standard interactions
  meet precision measurements of N$_{eff}$}},\ }\href
  {https://doi.org/10.1007/JHEP05(2021)058} {\bibfield  {journal} {\bibinfo
  {journal} {JHEP}\ }\textbf {\bibinfo {volume} {05}},\ \bibinfo {pages}
  {058}},\ \Eprint {https://arxiv.org/abs/2101.10475} {arXiv:2101.10475
  [hep-ph]} \BibitemShut {NoStop}%
\bibitem [{\citenamefont {Carrillo~Gonz\'alez}\ \emph
  {et~al.}(2021)\citenamefont {Carrillo~Gonz\'alez}, \citenamefont {Liang},
  \citenamefont {Sakstein},\ and\ \citenamefont
  {Trodden}}]{CarrilloGonzalez:2020oac}%
  \BibitemOpen
  \bibfield  {author} {\bibinfo {author} {\bibfnamefont {M.}~\bibnamefont
  {Carrillo~Gonz\'alez}}, \bibinfo {author} {\bibfnamefont {Q.}~\bibnamefont
  {Liang}}, \bibinfo {author} {\bibfnamefont {J.}~\bibnamefont {Sakstein}},\
  and\ \bibinfo {author} {\bibfnamefont {M.}~\bibnamefont {Trodden}},\
  }\bibfield  {title} {\bibinfo {title} {{Neutrino-Assisted Early Dark Energy:
  Theory and Cosmology}},\ }\href
  {https://doi.org/10.1088/1475-7516/2021/04/063} {\bibfield  {journal}
  {\bibinfo  {journal} {JCAP}\ }\textbf {\bibinfo {volume} {04}},\ \bibinfo
  {pages} {063}},\ \Eprint {https://arxiv.org/abs/2011.09895} {arXiv:2011.09895
  [astro-ph.CO]} \BibitemShut {NoStop}%
\bibitem [{\citenamefont {Huang}\ and\ \citenamefont
  {Rodejohann}(2021)}]{Huang:2021dba}%
  \BibitemOpen
  \bibfield  {author} {\bibinfo {author} {\bibfnamefont {G.-y.}\ \bibnamefont
  {Huang}}\ and\ \bibinfo {author} {\bibfnamefont {W.}~\bibnamefont
  {Rodejohann}},\ }\bibfield  {title} {\bibinfo {title} {{Solving the Hubble
  tension without spoiling Big Bang Nucleosynthesis}},\ }\href
  {https://doi.org/10.1103/PhysRevD.103.123007} {\bibfield  {journal} {\bibinfo
   {journal} {Phys. Rev. D}\ }\textbf {\bibinfo {volume} {103}},\ \bibinfo
  {pages} {123007} (\bibinfo {year} {2021})},\ \Eprint
  {https://arxiv.org/abs/2102.04280} {arXiv:2102.04280 [hep-ph]} \BibitemShut
  {NoStop}%
\bibitem [{\citenamefont {Sung}\ \emph {et~al.}(2021)\citenamefont {Sung},
  \citenamefont {Guo},\ and\ \citenamefont {Wu}}]{Sung:2021swd}%
  \BibitemOpen
  \bibfield  {author} {\bibinfo {author} {\bibfnamefont {A.}~\bibnamefont
  {Sung}}, \bibinfo {author} {\bibfnamefont {G.}~\bibnamefont {Guo}},\ and\
  \bibinfo {author} {\bibfnamefont {M.-R.}\ \bibnamefont {Wu}},\ }\bibfield
  {title} {\bibinfo {title} {{Supernova Constraint on Self-Interacting Dark
  Sector Particles}},\ }\href {https://doi.org/10.1103/PhysRevD.103.103005}
  {\bibfield  {journal} {\bibinfo  {journal} {Phys. Rev. D}\ }\textbf {\bibinfo
  {volume} {103}},\ \bibinfo {pages} {103005} (\bibinfo {year} {2021})},\
  \Eprint {https://arxiv.org/abs/2102.04601} {arXiv:2102.04601 [hep-ph]}
  \BibitemShut {NoStop}%
\bibitem [{\citenamefont {Escudero}\ and\ \citenamefont
  {Witte}(2021)}]{Escudero:2021rfi}%
  \BibitemOpen
  \bibfield  {author} {\bibinfo {author} {\bibfnamefont {M.}~\bibnamefont
  {Escudero}}\ and\ \bibinfo {author} {\bibfnamefont {S.~J.}\ \bibnamefont
  {Witte}},\ }\bibfield  {title} {\bibinfo {title} {{The hubble tension as a
  hint of leptogenesis and neutrino mass generation}},\ }\href
  {https://doi.org/10.1140/epjc/s10052-021-09276-5} {\bibfield  {journal}
  {\bibinfo  {journal} {Eur. Phys. J. C}\ }\textbf {\bibinfo {volume} {81}},\
  \bibinfo {pages} {515} (\bibinfo {year} {2021})},\ \Eprint
  {https://arxiv.org/abs/2103.03249} {arXiv:2103.03249 [hep-ph]} \BibitemShut
  {NoStop}%
\bibitem [{\citenamefont {Roy~Choudhury}\ \emph {et~al.}(2021)\citenamefont
  {Roy~Choudhury}, \citenamefont {Hannestad},\ and\ \citenamefont
  {Tram}}]{RoyChoudhury:2020dmd}%
  \BibitemOpen
  \bibfield  {author} {\bibinfo {author} {\bibfnamefont {S.}~\bibnamefont
  {Roy~Choudhury}}, \bibinfo {author} {\bibfnamefont {S.}~\bibnamefont
  {Hannestad}},\ and\ \bibinfo {author} {\bibfnamefont {T.}~\bibnamefont
  {Tram}},\ }\bibfield  {title} {\bibinfo {title} {{Updated constraints on
  massive neutrino self-interactions from cosmology in light of the $H_0$
  tension}},\ }\href {https://doi.org/10.1088/1475-7516/2021/03/084} {\bibfield
   {journal} {\bibinfo  {journal} {JCAP}\ }\textbf {\bibinfo {volume} {03}},\
  \bibinfo {pages} {084}},\ \Eprint {https://arxiv.org/abs/2012.07519}
  {arXiv:2012.07519 [astro-ph.CO]} \BibitemShut {NoStop}%
\bibitem [{\citenamefont {Carpio}\ \emph {et~al.}(2023)\citenamefont {Carpio},
  \citenamefont {Murase}, \citenamefont {Shoemaker},\ and\ \citenamefont
  {Tabrizi}}]{Carpio:2021jhu}%
  \BibitemOpen
  \bibfield  {author} {\bibinfo {author} {\bibfnamefont {J.~A.}\ \bibnamefont
  {Carpio}}, \bibinfo {author} {\bibfnamefont {K.}~\bibnamefont {Murase}},
  \bibinfo {author} {\bibfnamefont {I.~M.}\ \bibnamefont {Shoemaker}},\ and\
  \bibinfo {author} {\bibfnamefont {Z.}~\bibnamefont {Tabrizi}},\ }\bibfield
  {title} {\bibinfo {title} {{High-energy cosmic neutrinos as a probe of the
  vector mediator scenario in light of the muon g-2 anomaly and Hubble
  tension}},\ }\href {https://doi.org/10.1103/PhysRevD.107.103057} {\bibfield
  {journal} {\bibinfo  {journal} {Phys. Rev. D}\ }\textbf {\bibinfo {volume}
  {107}},\ \bibinfo {pages} {103057} (\bibinfo {year} {2023})},\ \Eprint
  {https://arxiv.org/abs/2104.15136} {arXiv:2104.15136 [hep-ph]} \BibitemShut
  {NoStop}%
\bibitem [{\citenamefont {Orlofsky}\ and\ \citenamefont
  {Zhang}(2021)}]{Orlofsky:2021mmy}%
  \BibitemOpen
  \bibfield  {author} {\bibinfo {author} {\bibfnamefont {N.}~\bibnamefont
  {Orlofsky}}\ and\ \bibinfo {author} {\bibfnamefont {Y.}~\bibnamefont
  {Zhang}},\ }\bibfield  {title} {\bibinfo {title} {{Neutrino as the dark
  force}},\ }\href {https://doi.org/10.1103/PhysRevD.104.075010} {\bibfield
  {journal} {\bibinfo  {journal} {Phys. Rev. D}\ }\textbf {\bibinfo {volume}
  {104}},\ \bibinfo {pages} {075010} (\bibinfo {year} {2021})},\ \Eprint
  {https://arxiv.org/abs/2106.08339} {arXiv:2106.08339 [hep-ph]} \BibitemShut
  {NoStop}%
\bibitem [{\citenamefont {Green}\ \emph {et~al.}(2021)\citenamefont {Green},
  \citenamefont {Kaplan},\ and\ \citenamefont {Rajendran}}]{Green:2021gdc}%
  \BibitemOpen
  \bibfield  {author} {\bibinfo {author} {\bibfnamefont {D.}~\bibnamefont
  {Green}}, \bibinfo {author} {\bibfnamefont {D.~E.}\ \bibnamefont {Kaplan}},\
  and\ \bibinfo {author} {\bibfnamefont {S.}~\bibnamefont {Rajendran}},\
  }\bibfield  {title} {\bibinfo {title} {{Neutrino interactions in the late
  universe}},\ }\href {https://doi.org/10.1007/JHEP11(2021)162} {\bibfield
  {journal} {\bibinfo  {journal} {JHEP}\ }\textbf {\bibinfo {volume} {11}},\
  \bibinfo {pages} {162}},\ \Eprint {https://arxiv.org/abs/2108.06928}
  {arXiv:2108.06928 [hep-ph]} \BibitemShut {NoStop}%
\bibitem [{\citenamefont {Esteban}\ \emph {et~al.}(2021)\citenamefont
  {Esteban}, \citenamefont {Pandey}, \citenamefont {Brdar},\ and\ \citenamefont
  {Beacom}}]{Esteban:2021tub}%
  \BibitemOpen
  \bibfield  {author} {\bibinfo {author} {\bibfnamefont {I.}~\bibnamefont
  {Esteban}}, \bibinfo {author} {\bibfnamefont {S.}~\bibnamefont {Pandey}},
  \bibinfo {author} {\bibfnamefont {V.}~\bibnamefont {Brdar}},\ and\ \bibinfo
  {author} {\bibfnamefont {J.~F.}\ \bibnamefont {Beacom}},\ }\bibfield  {title}
  {\bibinfo {title} {{Probing secret interactions of astrophysical neutrinos in
  the high-statistics era}},\ }\href
  {https://doi.org/10.1103/PhysRevD.104.123014} {\bibfield  {journal} {\bibinfo
   {journal} {Phys. Rev. D}\ }\textbf {\bibinfo {volume} {104}},\ \bibinfo
  {pages} {123014} (\bibinfo {year} {2021})},\ \Eprint
  {https://arxiv.org/abs/2107.13568} {arXiv:2107.13568 [hep-ph]} \BibitemShut
  {NoStop}%
\bibitem [{\citenamefont {Venzor}\ \emph {et~al.}(2022)\citenamefont {Venzor},
  \citenamefont {Garcia-Arroyo}, \citenamefont {P\'erez-Lorenzana},\ and\
  \citenamefont {De-Santiago}}]{Venzor:2022hql}%
  \BibitemOpen
  \bibfield  {author} {\bibinfo {author} {\bibfnamefont {J.}~\bibnamefont
  {Venzor}}, \bibinfo {author} {\bibfnamefont {G.}~\bibnamefont
  {Garcia-Arroyo}}, \bibinfo {author} {\bibfnamefont {A.}~\bibnamefont
  {P\'erez-Lorenzana}},\ and\ \bibinfo {author} {\bibfnamefont
  {J.}~\bibnamefont {De-Santiago}},\ }\bibfield  {title} {\bibinfo {title}
  {{Massive neutrino self-interactions with a light mediator in cosmology}},\
  }\href {https://doi.org/10.1103/PhysRevD.105.123539} {\bibfield  {journal}
  {\bibinfo  {journal} {Phys. Rev. D}\ }\textbf {\bibinfo {volume} {105}},\
  \bibinfo {pages} {123539} (\bibinfo {year} {2022})},\ \Eprint
  {https://arxiv.org/abs/2202.09310} {arXiv:2202.09310 [astro-ph.CO]}
  \BibitemShut {NoStop}%
\bibitem [{\citenamefont {Taule}\ \emph {et~al.}(2022)\citenamefont {Taule},
  \citenamefont {Escudero},\ and\ \citenamefont {Garny}}]{Taule:2022jrz}%
  \BibitemOpen
  \bibfield  {author} {\bibinfo {author} {\bibfnamefont {P.}~\bibnamefont
  {Taule}}, \bibinfo {author} {\bibfnamefont {M.}~\bibnamefont {Escudero}},\
  and\ \bibinfo {author} {\bibfnamefont {M.}~\bibnamefont {Garny}},\ }\bibfield
   {title} {\bibinfo {title} {{Global view of neutrino interactions in
  cosmology: The free streaming window as seen by Planck}},\ }\href
  {https://doi.org/10.1103/PhysRevD.106.063539} {\bibfield  {journal} {\bibinfo
   {journal} {Phys. Rev. D}\ }\textbf {\bibinfo {volume} {106}},\ \bibinfo
  {pages} {063539} (\bibinfo {year} {2022})},\ \Eprint
  {https://arxiv.org/abs/2207.04062} {arXiv:2207.04062 [astro-ph.CO]}
  \BibitemShut {NoStop}%
\bibitem [{\citenamefont {Roy~Choudhury}\ \emph {et~al.}(2022)\citenamefont
  {Roy~Choudhury}, \citenamefont {Hannestad},\ and\ \citenamefont
  {Tram}}]{RoyChoudhury:2022rva}%
  \BibitemOpen
  \bibfield  {author} {\bibinfo {author} {\bibfnamefont {S.}~\bibnamefont
  {Roy~Choudhury}}, \bibinfo {author} {\bibfnamefont {S.}~\bibnamefont
  {Hannestad}},\ and\ \bibinfo {author} {\bibfnamefont {T.}~\bibnamefont
  {Tram}},\ }\bibfield  {title} {\bibinfo {title} {{Massive neutrino
  self-interactions and inflation}},\ }\href
  {https://doi.org/10.1088/1475-7516/2022/10/018} {\bibfield  {journal}
  {\bibinfo  {journal} {JCAP}\ }\textbf {\bibinfo {volume} {10}},\ \bibinfo
  {pages} {018}},\ \Eprint {https://arxiv.org/abs/2207.07142} {arXiv:2207.07142
  [astro-ph.CO]} \BibitemShut {NoStop}%
\bibitem [{\citenamefont {Loverde}\ and\ \citenamefont
  {Weiner}(2023)}]{Loverde:2022wih}%
  \BibitemOpen
  \bibfield  {author} {\bibinfo {author} {\bibfnamefont {M.}~\bibnamefont
  {Loverde}}\ and\ \bibinfo {author} {\bibfnamefont {Z.~J.}\ \bibnamefont
  {Weiner}},\ }\bibfield  {title} {\bibinfo {title} {{Probing neutrino
  interactions and dark radiation with gravitational waves}},\ }\href
  {https://doi.org/10.1088/1475-7516/2023/02/064} {\bibfield  {journal}
  {\bibinfo  {journal} {JCAP}\ }\textbf {\bibinfo {volume} {02}},\ \bibinfo
  {pages} {064}},\ \Eprint {https://arxiv.org/abs/2208.11714} {arXiv:2208.11714
  [astro-ph.CO]} \BibitemShut {NoStop}%
\bibitem [{\citenamefont {Kreisch}\ \emph {et~al.}(2022)\citenamefont {Kreisch}
  \emph {et~al.}}]{Kreisch:2022zxp}%
  \BibitemOpen
  \bibfield  {author} {\bibinfo {author} {\bibfnamefont {C.~D.}\ \bibnamefont
  {Kreisch}} \emph {et~al.},\ }\href@noop {} {\bibinfo {title} {{The Atacama
  Cosmology Telescope: The Persistence of Neutrino Self-Interaction in
  Cosmological Measurements}}} (\bibinfo {year} {2022}),\ \Eprint
  {https://arxiv.org/abs/2207.03164} {arXiv:2207.03164 [astro-ph.CO]}
  \BibitemShut {NoStop}%
\bibitem [{\citenamefont {Das}\ and\ \citenamefont
  {Ghosh}(2023)}]{Das:2023npl}%
  \BibitemOpen
  \bibfield  {author} {\bibinfo {author} {\bibfnamefont {A.}~\bibnamefont
  {Das}}\ and\ \bibinfo {author} {\bibfnamefont {S.}~\bibnamefont {Ghosh}},\
  }\href@noop {} {\bibinfo {title} {{The magnificent ACT of flavor-specific
  neutrino self-interaction}}} (\bibinfo {year} {2023}),\ \Eprint
  {https://arxiv.org/abs/2303.08843} {arXiv:2303.08843 [astro-ph.CO]}
  \BibitemShut {NoStop}%
\bibitem [{\citenamefont {Venzor}\ \emph {et~al.}(2023)\citenamefont {Venzor},
  \citenamefont {Garcia-Arroyo}, \citenamefont {P\'erez-Lorenzana},\ and\
  \citenamefont {De-Santiago}}]{Venzor:2023aka}%
  \BibitemOpen
  \bibfield  {author} {\bibinfo {author} {\bibfnamefont {J.}~\bibnamefont
  {Venzor}}, \bibinfo {author} {\bibfnamefont {G.}~\bibnamefont
  {Garcia-Arroyo}}, \bibinfo {author} {\bibfnamefont {A.}~\bibnamefont
  {P\'erez-Lorenzana}},\ and\ \bibinfo {author} {\bibfnamefont
  {J.}~\bibnamefont {De-Santiago}},\ }\href@noop {} {\bibinfo {title}
  {{Resonant neutrino self-interactions and the $H_0$ tension}}} (\bibinfo
  {year} {2023}),\ \Eprint {https://arxiv.org/abs/2303.12792} {arXiv:2303.12792
  [astro-ph.CO]} \BibitemShut {NoStop}%
\bibitem [{\citenamefont {Sandner}\ \emph {et~al.}(2023)\citenamefont
  {Sandner}, \citenamefont {Escudero},\ and\ \citenamefont
  {Witte}}]{Sandner:2023ptm}%
  \BibitemOpen
  \bibfield  {author} {\bibinfo {author} {\bibfnamefont {S.}~\bibnamefont
  {Sandner}}, \bibinfo {author} {\bibfnamefont {M.}~\bibnamefont {Escudero}},\
  and\ \bibinfo {author} {\bibfnamefont {S.~J.}\ \bibnamefont {Witte}},\
  }\bibfield  {title} {\bibinfo {title} {{Precision CMB constraints on eV-scale
  bosons coupled to neutrinos}},\ }\href
  {https://doi.org/10.1140/epjc/s10052-023-11864-6} {\bibfield  {journal}
  {\bibinfo  {journal} {Eur. Phys. J. C}\ }\textbf {\bibinfo {volume} {83}},\
  \bibinfo {pages} {709} (\bibinfo {year} {2023})},\ \Eprint
  {https://arxiv.org/abs/2305.01692} {arXiv:2305.01692 [hep-ph]} \BibitemShut
  {NoStop}%
\bibitem [{\citenamefont {Camarena}\ \emph {et~al.}(2023)\citenamefont
  {Camarena}, \citenamefont {Cyr-Racine},\ and\ \citenamefont
  {Houghteling}}]{Camarena:2023cku}%
  \BibitemOpen
  \bibfield  {author} {\bibinfo {author} {\bibfnamefont {D.}~\bibnamefont
  {Camarena}}, \bibinfo {author} {\bibfnamefont {F.-Y.}\ \bibnamefont
  {Cyr-Racine}},\ and\ \bibinfo {author} {\bibfnamefont {J.}~\bibnamefont
  {Houghteling}},\ }\bibfield  {title} {\bibinfo {title} {{The two-mode puzzle:
  Confronting self-interacting neutrinos with the full shape of the galaxy
  power spectrum}},\ }\href@noop {} {\  (\bibinfo {year} {2023})},\ \Eprint
  {https://arxiv.org/abs/2309.03941} {arXiv:2309.03941 [astro-ph.CO]}
  \BibitemShut {NoStop}%
\bibitem [{\citenamefont {He}\ \emph {et~al.}(2023)\citenamefont {He},
  \citenamefont {An}, \citenamefont {Ivanov},\ and\ \citenamefont
  {Gluscevic}}]{He:2023oke}%
  \BibitemOpen
  \bibfield  {author} {\bibinfo {author} {\bibfnamefont {A.}~\bibnamefont
  {He}}, \bibinfo {author} {\bibfnamefont {R.}~\bibnamefont {An}}, \bibinfo
  {author} {\bibfnamefont {M.~M.}\ \bibnamefont {Ivanov}},\ and\ \bibinfo
  {author} {\bibfnamefont {V.}~\bibnamefont {Gluscevic}},\ }\href@noop {}
  {\bibinfo {title} {{Self-Interacting Neutrinos in Light of Large-Scale
  Structure Data}}} (\bibinfo {year} {2023}),\ \Eprint
  {https://arxiv.org/abs/2309.03956} {arXiv:2309.03956 [astro-ph.CO]}
  \BibitemShut {NoStop}%
\bibitem [{\citenamefont {Follin}\ \emph {et~al.}(2015)\citenamefont {Follin},
  \citenamefont {Knox}, \citenamefont {Millea},\ and\ \citenamefont
  {Pan}}]{Follin:2015hya}%
  \BibitemOpen
  \bibfield  {author} {\bibinfo {author} {\bibfnamefont {B.}~\bibnamefont
  {Follin}}, \bibinfo {author} {\bibfnamefont {L.}~\bibnamefont {Knox}},
  \bibinfo {author} {\bibfnamefont {M.}~\bibnamefont {Millea}},\ and\ \bibinfo
  {author} {\bibfnamefont {Z.}~\bibnamefont {Pan}},\ }\bibfield  {title}
  {\bibinfo {title} {{First Detection of the Acoustic Oscillation Phase Shift
  Expected from the Cosmic Neutrino Background}},\ }\href
  {https://doi.org/10.1103/PhysRevLett.115.091301} {\bibfield  {journal}
  {\bibinfo  {journal} {Phys. Rev. Lett.}\ }\textbf {\bibinfo {volume} {115}},\
  \bibinfo {pages} {091301} (\bibinfo {year} {2015})},\ \Eprint
  {https://arxiv.org/abs/1503.07863} {arXiv:1503.07863 [astro-ph.CO]}
  \BibitemShut {NoStop}%
\bibitem [{\citenamefont {Baumann}\ \emph {et~al.}(2017)\citenamefont
  {Baumann}, \citenamefont {Green},\ and\ \citenamefont
  {Zaldarriaga}}]{Baumann:2017lmt}%
  \BibitemOpen
  \bibfield  {author} {\bibinfo {author} {\bibfnamefont {D.}~\bibnamefont
  {Baumann}}, \bibinfo {author} {\bibfnamefont {D.}~\bibnamefont {Green}},\
  and\ \bibinfo {author} {\bibfnamefont {M.}~\bibnamefont {Zaldarriaga}},\
  }\bibfield  {title} {\bibinfo {title} {{Phases of New Physics in the BAO
  Spectrum}},\ }\href {https://doi.org/10.1088/1475-7516/2017/11/007}
  {\bibfield  {journal} {\bibinfo  {journal} {JCAP}\ }\textbf {\bibinfo
  {volume} {11}},\ \bibinfo {pages} {007}},\ \Eprint
  {https://arxiv.org/abs/1703.00894} {arXiv:1703.00894 [astro-ph.CO]}
  \BibitemShut {NoStop}%
\bibitem [{\citenamefont {Baumann}\ \emph {et~al.}(2019)\citenamefont
  {Baumann}, \citenamefont {Beutler}, \citenamefont {Flauger}, \citenamefont
  {Green}, \citenamefont {Slosar}, \citenamefont {Vargas-Maga\~na},
  \citenamefont {Wallisch},\ and\ \citenamefont {Y\`eche}}]{Baumann:2019keh}%
  \BibitemOpen
  \bibfield  {author} {\bibinfo {author} {\bibfnamefont {D.}~\bibnamefont
  {Baumann}}, \bibinfo {author} {\bibfnamefont {F.}~\bibnamefont {Beutler}},
  \bibinfo {author} {\bibfnamefont {R.}~\bibnamefont {Flauger}}, \bibinfo
  {author} {\bibfnamefont {D.}~\bibnamefont {Green}}, \bibinfo {author}
  {\bibfnamefont {A.}~\bibnamefont {Slosar}}, \bibinfo {author} {\bibfnamefont
  {M.}~\bibnamefont {Vargas-Maga\~na}}, \bibinfo {author} {\bibfnamefont
  {B.}~\bibnamefont {Wallisch}},\ and\ \bibinfo {author} {\bibfnamefont
  {C.}~\bibnamefont {Y\`eche}},\ }\bibfield  {title} {\bibinfo {title} {{First
  constraint on the neutrino-induced phase shift in the spectrum of baryon
  acoustic oscillations}},\ }\href {https://doi.org/10.1038/s41567-019-0435-6}
  {\bibfield  {journal} {\bibinfo  {journal} {Nature Phys.}\ }\textbf {\bibinfo
  {volume} {15}},\ \bibinfo {pages} {465} (\bibinfo {year} {2019})},\ \Eprint
  {https://arxiv.org/abs/1803.10741} {arXiv:1803.10741 [astro-ph.CO]}
  \BibitemShut {NoStop}%
\bibitem [{\citenamefont {Aiola}\ \emph {et~al.}(2020)\citenamefont {Aiola}
  \emph {et~al.}}]{ACT:2020gnv}%
  \BibitemOpen
  \bibfield  {author} {\bibinfo {author} {\bibfnamefont {S.}~\bibnamefont
  {Aiola}} \emph {et~al.} (\bibinfo {collaboration} {ACT}),\ }\bibfield
  {title} {\bibinfo {title} {{The Atacama Cosmology Telescope: DR4 Maps and
  Cosmological Parameters}},\ }\href
  {https://doi.org/10.1088/1475-7516/2020/12/047} {\bibfield  {journal}
  {\bibinfo  {journal} {JCAP}\ }\textbf {\bibinfo {volume} {12}},\ \bibinfo
  {pages} {047}},\ \Eprint {https://arxiv.org/abs/2007.07288} {arXiv:2007.07288
  [astro-ph.CO]} \BibitemShut {NoStop}%
\bibitem [{\citenamefont {Chabanier}\ \emph {et~al.}(2019)\citenamefont
  {Chabanier} \emph {et~al.}}]{Chabanier:2018rga}%
  \BibitemOpen
  \bibfield  {author} {\bibinfo {author} {\bibfnamefont {S.}~\bibnamefont
  {Chabanier}} \emph {et~al.},\ }\bibfield  {title} {\bibinfo {title} {{The
  one-dimensional power spectrum from the SDSS DR14 Ly$\alpha$ forests}},\
  }\href {https://doi.org/10.1088/1475-7516/2019/07/017} {\bibfield  {journal}
  {\bibinfo  {journal} {JCAP}\ }\textbf {\bibinfo {volume} {07}},\ \bibinfo
  {pages} {017}},\ \Eprint {https://arxiv.org/abs/1812.03554} {arXiv:1812.03554
  [astro-ph.CO]} \BibitemShut {NoStop}%
\bibitem [{\citenamefont {Berryman}\ \emph {et~al.}(2023)\citenamefont
  {Berryman} \emph {et~al.}}]{Berryman:2022hds}%
  \BibitemOpen
  \bibfield  {author} {\bibinfo {author} {\bibfnamefont {J.~M.}\ \bibnamefont
  {Berryman}} \emph {et~al.},\ }\bibfield  {title} {\bibinfo {title} {{Neutrino
  self-interactions: A white paper}},\ }\href
  {https://doi.org/10.1016/j.dark.2023.101267} {\bibfield  {journal} {\bibinfo
  {journal} {Phys. Dark Univ.}\ }\textbf {\bibinfo {volume} {42}},\ \bibinfo
  {pages} {101267} (\bibinfo {year} {2023})},\ \Eprint
  {https://arxiv.org/abs/2203.01955} {arXiv:2203.01955 [hep-ph]} \BibitemShut
  {NoStop}%
\bibitem [{\citenamefont {Kolb}\ and\ \citenamefont
  {Turner}(1987)}]{Kolb:1987qy}%
  \BibitemOpen
  \bibfield  {author} {\bibinfo {author} {\bibfnamefont {E.~W.}\ \bibnamefont
  {Kolb}}\ and\ \bibinfo {author} {\bibfnamefont {M.~S.}\ \bibnamefont
  {Turner}},\ }\bibfield  {title} {\bibinfo {title} {{Supernova SN 1987a and
  the Secret Interactions of Neutrinos}},\ }\href
  {https://doi.org/10.1103/PhysRevD.36.2895} {\bibfield  {journal} {\bibinfo
  {journal} {Phys. Rev. D}\ }\textbf {\bibinfo {volume} {36}},\ \bibinfo
  {pages} {2895} (\bibinfo {year} {1987})}\BibitemShut {NoStop}%
\bibitem [{\citenamefont {Manohar}(1987)}]{Manohar:1987ec}%
  \BibitemOpen
  \bibfield  {author} {\bibinfo {author} {\bibfnamefont {A.}~\bibnamefont
  {Manohar}},\ }\bibfield  {title} {\bibinfo {title} {{A Limit on the
  Neutrino-neutrino Scattering Cross-section From the Supernova}},\ }\href
  {https://doi.org/10.1016/0370-2693(87)91171-3} {\bibfield  {journal}
  {\bibinfo  {journal} {Phys. Lett. B}\ }\textbf {\bibinfo {volume} {192}},\
  \bibinfo {pages} {217} (\bibinfo {year} {1987})}\BibitemShut {NoStop}%
\bibitem [{\citenamefont {Dicus}\ \emph {et~al.}(1989)\citenamefont {Dicus},
  \citenamefont {Nussinov}, \citenamefont {Pal},\ and\ \citenamefont
  {Teplitz}}]{Dicus:1988jh}%
  \BibitemOpen
  \bibfield  {author} {\bibinfo {author} {\bibfnamefont {D.~A.}\ \bibnamefont
  {Dicus}}, \bibinfo {author} {\bibfnamefont {S.}~\bibnamefont {Nussinov}},
  \bibinfo {author} {\bibfnamefont {P.~B.}\ \bibnamefont {Pal}},\ and\ \bibinfo
  {author} {\bibfnamefont {V.~L.}\ \bibnamefont {Teplitz}},\ }\bibfield
  {title} {\bibinfo {title} {{Implications of Relativistic Gas Dynamics for
  Neutrino-neutrino Cross-sections}},\ }\href
  {https://doi.org/10.1016/0370-2693(89)90480-2} {\bibfield  {journal}
  {\bibinfo  {journal} {Phys. Lett. B}\ }\textbf {\bibinfo {volume} {218}},\
  \bibinfo {pages} {84} (\bibinfo {year} {1989})}\BibitemShut {NoStop}%
\bibitem [{\citenamefont {Davoudiasl}\ and\ \citenamefont
  {Huber}(2005)}]{Davoudiasl:2005fd}%
  \BibitemOpen
  \bibfield  {author} {\bibinfo {author} {\bibfnamefont {H.}~\bibnamefont
  {Davoudiasl}}\ and\ \bibinfo {author} {\bibfnamefont {P.}~\bibnamefont
  {Huber}},\ }\bibfield  {title} {\bibinfo {title} {{Probing the origins of
  neutrino mass with supernova data}},\ }\href
  {https://doi.org/10.1103/PhysRevLett.95.191302} {\bibfield  {journal}
  {\bibinfo  {journal} {Phys. Rev. Lett.}\ }\textbf {\bibinfo {volume} {95}},\
  \bibinfo {pages} {191302} (\bibinfo {year} {2005})},\ \Eprint
  {https://arxiv.org/abs/hep-ph/0504265} {arXiv:hep-ph/0504265} \BibitemShut
  {NoStop}%
\bibitem [{\citenamefont {Sher}\ and\ \citenamefont
  {Triola}(2011)}]{Sher:2011mx}%
  \BibitemOpen
  \bibfield  {author} {\bibinfo {author} {\bibfnamefont {M.}~\bibnamefont
  {Sher}}\ and\ \bibinfo {author} {\bibfnamefont {C.}~\bibnamefont {Triola}},\
  }\bibfield  {title} {\bibinfo {title} {{Astrophysical Consequences of a
  Neutrinophilic Two-Higgs-Doublet Model}},\ }\href
  {https://doi.org/10.1103/PhysRevD.83.117702} {\bibfield  {journal} {\bibinfo
  {journal} {Phys. Rev. D}\ }\textbf {\bibinfo {volume} {83}},\ \bibinfo
  {pages} {117702} (\bibinfo {year} {2011})},\ \Eprint
  {https://arxiv.org/abs/1105.4844} {arXiv:1105.4844 [hep-ph]} \BibitemShut
  {NoStop}%
\bibitem [{\citenamefont {Fayet}\ \emph {et~al.}(2006)\citenamefont {Fayet},
  \citenamefont {Hooper},\ and\ \citenamefont {Sigl}}]{Fayet:2006sa}%
  \BibitemOpen
  \bibfield  {author} {\bibinfo {author} {\bibfnamefont {P.}~\bibnamefont
  {Fayet}}, \bibinfo {author} {\bibfnamefont {D.}~\bibnamefont {Hooper}},\ and\
  \bibinfo {author} {\bibfnamefont {G.}~\bibnamefont {Sigl}},\ }\bibfield
  {title} {\bibinfo {title} {{Constraints on light dark matter from
  core-collapse supernovae}},\ }\href
  {https://doi.org/10.1103/PhysRevLett.96.211302} {\bibfield  {journal}
  {\bibinfo  {journal} {Phys. Rev. Lett.}\ }\textbf {\bibinfo {volume} {96}},\
  \bibinfo {pages} {211302} (\bibinfo {year} {2006})},\ \Eprint
  {https://arxiv.org/abs/hep-ph/0602169} {arXiv:hep-ph/0602169} \BibitemShut
  {NoStop}%
\bibitem [{\citenamefont {Choi}\ and\ \citenamefont
  {Santamaria}(1990)}]{Choi:1989hi}%
  \BibitemOpen
  \bibfield  {author} {\bibinfo {author} {\bibfnamefont {K.}~\bibnamefont
  {Choi}}\ and\ \bibinfo {author} {\bibfnamefont {A.}~\bibnamefont
  {Santamaria}},\ }\bibfield  {title} {\bibinfo {title} {{Majorons and
  Supernova Cooling}},\ }\href {https://doi.org/10.1103/PhysRevD.42.293}
  {\bibfield  {journal} {\bibinfo  {journal} {Phys. Rev. D}\ }\textbf {\bibinfo
  {volume} {42}},\ \bibinfo {pages} {293} (\bibinfo {year} {1990})}\BibitemShut
  {NoStop}%
\bibitem [{\citenamefont {Blennow}\ \emph {et~al.}(2008)\citenamefont
  {Blennow}, \citenamefont {Mirizzi},\ and\ \citenamefont
  {Serpico}}]{Blennow:2008er}%
  \BibitemOpen
  \bibfield  {author} {\bibinfo {author} {\bibfnamefont {M.}~\bibnamefont
  {Blennow}}, \bibinfo {author} {\bibfnamefont {A.}~\bibnamefont {Mirizzi}},\
  and\ \bibinfo {author} {\bibfnamefont {P.~D.}\ \bibnamefont {Serpico}},\
  }\bibfield  {title} {\bibinfo {title} {{Nonstandard neutrino-neutrino
  refractive effects in dense neutrino gases}},\ }\href
  {https://doi.org/10.1103/PhysRevD.78.113004} {\bibfield  {journal} {\bibinfo
  {journal} {Phys. Rev. D}\ }\textbf {\bibinfo {volume} {78}},\ \bibinfo
  {pages} {113004} (\bibinfo {year} {2008})},\ \Eprint
  {https://arxiv.org/abs/0810.2297} {arXiv:0810.2297 [hep-ph]} \BibitemShut
  {NoStop}%
\bibitem [{\citenamefont {Galais}\ \emph {et~al.}(2012)\citenamefont {Galais},
  \citenamefont {Kneller},\ and\ \citenamefont {Volpe}}]{Galais:2011jh}%
  \BibitemOpen
  \bibfield  {author} {\bibinfo {author} {\bibfnamefont {S.}~\bibnamefont
  {Galais}}, \bibinfo {author} {\bibfnamefont {J.}~\bibnamefont {Kneller}},\
  and\ \bibinfo {author} {\bibfnamefont {C.}~\bibnamefont {Volpe}},\ }\bibfield
   {title} {\bibinfo {title} {{The neutrino-neutrino interaction effects in
  supernovae: the point of view from the matter basis}},\ }\href
  {https://doi.org/10.1088/0954-3899/39/3/035201} {\bibfield  {journal}
  {\bibinfo  {journal} {J. Phys. G}\ }\textbf {\bibinfo {volume} {39}},\
  \bibinfo {pages} {035201} (\bibinfo {year} {2012})},\ \Eprint
  {https://arxiv.org/abs/1102.1471} {arXiv:1102.1471 [astro-ph.SR]}
  \BibitemShut {NoStop}%
\bibitem [{\citenamefont {Kachelriess}\ \emph {et~al.}(2000)\citenamefont
  {Kachelriess}, \citenamefont {Tomas},\ and\ \citenamefont
  {Valle}}]{Kachelriess:2000qc}%
  \BibitemOpen
  \bibfield  {author} {\bibinfo {author} {\bibfnamefont {M.}~\bibnamefont
  {Kachelriess}}, \bibinfo {author} {\bibfnamefont {R.}~\bibnamefont {Tomas}},\
  and\ \bibinfo {author} {\bibfnamefont {J.~W.~F.}\ \bibnamefont {Valle}},\
  }\bibfield  {title} {\bibinfo {title} {{Supernova bounds on Majoron emitting
  decays of light neutrinos}},\ }\href
  {https://doi.org/10.1103/PhysRevD.62.023004} {\bibfield  {journal} {\bibinfo
  {journal} {Phys. Rev. D}\ }\textbf {\bibinfo {volume} {62}},\ \bibinfo
  {pages} {023004} (\bibinfo {year} {2000})},\ \Eprint
  {https://arxiv.org/abs/hep-ph/0001039} {arXiv:hep-ph/0001039} \BibitemShut
  {NoStop}%
\bibitem [{\citenamefont {Farzan}(2003)}]{Farzan:2002wx}%
  \BibitemOpen
  \bibfield  {author} {\bibinfo {author} {\bibfnamefont {Y.}~\bibnamefont
  {Farzan}},\ }\bibfield  {title} {\bibinfo {title} {{Bounds on the coupling of
  the Majoron to light neutrinos from supernova cooling}},\ }\href
  {https://doi.org/10.1103/PhysRevD.67.073015} {\bibfield  {journal} {\bibinfo
  {journal} {Phys. Rev. D}\ }\textbf {\bibinfo {volume} {67}},\ \bibinfo
  {pages} {073015} (\bibinfo {year} {2003})},\ \Eprint
  {https://arxiv.org/abs/hep-ph/0211375} {arXiv:hep-ph/0211375} \BibitemShut
  {NoStop}%
\bibitem [{\citenamefont {Zhou}(2011)}]{Zhou:2011rc}%
  \BibitemOpen
  \bibfield  {author} {\bibinfo {author} {\bibfnamefont {S.}~\bibnamefont
  {Zhou}},\ }\bibfield  {title} {\bibinfo {title} {{Comment on astrophysical
  consequences of a neutrinophilic 2HDM}},\ }\href
  {https://doi.org/10.1103/PhysRevD.84.038701} {\bibfield  {journal} {\bibinfo
  {journal} {Phys. Rev. D}\ }\textbf {\bibinfo {volume} {84}},\ \bibinfo
  {pages} {038701} (\bibinfo {year} {2011})},\ \Eprint
  {https://arxiv.org/abs/1106.3880} {arXiv:1106.3880 [hep-ph]} \BibitemShut
  {NoStop}%
\bibitem [{\citenamefont {Jeong}\ \emph {et~al.}(2018)\citenamefont {Jeong},
  \citenamefont {Palomares-Ruiz}, \citenamefont {Reno},\ and\ \citenamefont
  {Sarcevic}}]{Jeong:2018yts}%
  \BibitemOpen
  \bibfield  {author} {\bibinfo {author} {\bibfnamefont {Y.~S.}\ \bibnamefont
  {Jeong}}, \bibinfo {author} {\bibfnamefont {S.}~\bibnamefont
  {Palomares-Ruiz}}, \bibinfo {author} {\bibfnamefont {M.~H.}\ \bibnamefont
  {Reno}},\ and\ \bibinfo {author} {\bibfnamefont {I.}~\bibnamefont
  {Sarcevic}},\ }\bibfield  {title} {\bibinfo {title} {{Probing secret
  interactions of eV-scale sterile neutrinos with the diffuse supernova
  neutrino background}},\ }\href
  {https://doi.org/10.1088/1475-7516/2018/06/019} {\bibfield  {journal}
  {\bibinfo  {journal} {JCAP}\ }\textbf {\bibinfo {volume} {06}},\ \bibinfo
  {pages} {019}},\ \Eprint {https://arxiv.org/abs/1803.04541} {arXiv:1803.04541
  [hep-ph]} \BibitemShut {NoStop}%
\bibitem [{\citenamefont {Chang}\ \emph {et~al.}(2023)\citenamefont {Chang},
  \citenamefont {Esteban}, \citenamefont {Beacom}, \citenamefont {Thompson},\
  and\ \citenamefont {Hirata}}]{Chang:2022aas}%
  \BibitemOpen
  \bibfield  {author} {\bibinfo {author} {\bibfnamefont {P.-W.}\ \bibnamefont
  {Chang}}, \bibinfo {author} {\bibfnamefont {I.}~\bibnamefont {Esteban}},
  \bibinfo {author} {\bibfnamefont {J.~F.}\ \bibnamefont {Beacom}}, \bibinfo
  {author} {\bibfnamefont {T.~A.}\ \bibnamefont {Thompson}},\ and\ \bibinfo
  {author} {\bibfnamefont {C.~M.}\ \bibnamefont {Hirata}},\ }\bibfield  {title}
  {\bibinfo {title} {{Toward Powerful Probes of Neutrino Self-Interactions in
  Supernovae}},\ }\href {https://doi.org/10.1103/PhysRevLett.131.071002}
  {\bibfield  {journal} {\bibinfo  {journal} {Phys. Rev. Lett.}\ }\textbf
  {\bibinfo {volume} {131}},\ \bibinfo {pages} {071002} (\bibinfo {year}
  {2023})},\ \Eprint {https://arxiv.org/abs/2206.12426} {arXiv:2206.12426
  [hep-ph]} \BibitemShut {NoStop}%
\bibitem [{\citenamefont {Fiorillo}\ \emph
  {et~al.}(2023{\natexlab{a}})\citenamefont {Fiorillo}, \citenamefont
  {Raffelt},\ and\ \citenamefont {Vitagliano}}]{Fiorillo:2023cas}%
  \BibitemOpen
  \bibfield  {author} {\bibinfo {author} {\bibfnamefont {D.~F.~G.}\
  \bibnamefont {Fiorillo}}, \bibinfo {author} {\bibfnamefont {G.}~\bibnamefont
  {Raffelt}},\ and\ \bibinfo {author} {\bibfnamefont {E.}~\bibnamefont
  {Vitagliano}},\ }\href@noop {} {\bibinfo {title} {{Supernova Emission of
  Secretly Interacting Neutrino Fluid: Theoretical Foundations}}} (\bibinfo
  {year} {2023}{\natexlab{a}}),\ \Eprint {https://arxiv.org/abs/2307.15122}
  {arXiv:2307.15122 [hep-ph]} \BibitemShut {NoStop}%
\bibitem [{\citenamefont {Fiorillo}\ \emph
  {et~al.}(2023{\natexlab{b}})\citenamefont {Fiorillo}, \citenamefont
  {Raffelt},\ and\ \citenamefont {Vitagliano}}]{Fiorillo:2023ytr}%
  \BibitemOpen
  \bibfield  {author} {\bibinfo {author} {\bibfnamefont {D.~F.~G.}\
  \bibnamefont {Fiorillo}}, \bibinfo {author} {\bibfnamefont {G.}~\bibnamefont
  {Raffelt}},\ and\ \bibinfo {author} {\bibfnamefont {E.}~\bibnamefont
  {Vitagliano}},\ }\href@noop {} {\bibinfo {title} {{Large Neutrino Secret
  Interactions, Small Impact on Supernovae}}} (\bibinfo {year}
  {2023}{\natexlab{b}}),\ \Eprint {https://arxiv.org/abs/2307.15115}
  {arXiv:2307.15115 [hep-ph]} \BibitemShut {NoStop}%
\bibitem [{\citenamefont {Ahlgren}\ \emph {et~al.}(2013)\citenamefont
  {Ahlgren}, \citenamefont {Ohlsson},\ and\ \citenamefont
  {Zhou}}]{Ahlgren:2013wba}%
  \BibitemOpen
  \bibfield  {author} {\bibinfo {author} {\bibfnamefont {B.}~\bibnamefont
  {Ahlgren}}, \bibinfo {author} {\bibfnamefont {T.}~\bibnamefont {Ohlsson}},\
  and\ \bibinfo {author} {\bibfnamefont {S.}~\bibnamefont {Zhou}},\ }\bibfield
  {title} {\bibinfo {title} {{Comment on \textquotedblleft{}Is Dark Matter with
  Long-Range Interactions a Solution to All Small-Scale Problems of
  \ensuremath{\Lambda} Cold Dark Matter Cosmology?\textquotedblright{}}},\
  }\href {https://doi.org/10.1103/PhysRevLett.111.199001} {\bibfield  {journal}
  {\bibinfo  {journal} {Phys. Rev. Lett.}\ }\textbf {\bibinfo {volume} {111}},\
  \bibinfo {pages} {199001} (\bibinfo {year} {2013})},\ \Eprint
  {https://arxiv.org/abs/1309.0991} {arXiv:1309.0991 [hep-ph]} \BibitemShut
  {NoStop}%
\bibitem [{\citenamefont {Huang}\ \emph {et~al.}(2018)\citenamefont {Huang},
  \citenamefont {Ohlsson},\ and\ \citenamefont {Zhou}}]{Huang:2017egl}%
  \BibitemOpen
  \bibfield  {author} {\bibinfo {author} {\bibfnamefont {G.-y.}\ \bibnamefont
  {Huang}}, \bibinfo {author} {\bibfnamefont {T.}~\bibnamefont {Ohlsson}},\
  and\ \bibinfo {author} {\bibfnamefont {S.}~\bibnamefont {Zhou}},\ }\bibfield
  {title} {\bibinfo {title} {{Observational Constraints on Secret Neutrino
  Interactions from Big Bang Nucleosynthesis}},\ }\href
  {https://doi.org/10.1103/PhysRevD.97.075009} {\bibfield  {journal} {\bibinfo
  {journal} {Phys. Rev. D}\ }\textbf {\bibinfo {volume} {97}},\ \bibinfo
  {pages} {075009} (\bibinfo {year} {2018})},\ \Eprint
  {https://arxiv.org/abs/1712.04792} {arXiv:1712.04792 [hep-ph]} \BibitemShut
  {NoStop}%
\bibitem [{\citenamefont {Venzor}\ \emph {et~al.}(2021)\citenamefont {Venzor},
  \citenamefont {P\'erez-Lorenzana},\ and\ \citenamefont
  {De-Santiago}}]{Venzor:2020ova}%
  \BibitemOpen
  \bibfield  {author} {\bibinfo {author} {\bibfnamefont {J.}~\bibnamefont
  {Venzor}}, \bibinfo {author} {\bibfnamefont {A.}~\bibnamefont
  {P\'erez-Lorenzana}},\ and\ \bibinfo {author} {\bibfnamefont
  {J.}~\bibnamefont {De-Santiago}},\ }\bibfield  {title} {\bibinfo {title}
  {{Bounds on neutrino-scalar nonstandard interactions from big bang
  nucleosynthesis}},\ }\href {https://doi.org/10.1103/PhysRevD.103.043534}
  {\bibfield  {journal} {\bibinfo  {journal} {Phys. Rev. D}\ }\textbf {\bibinfo
  {volume} {103}},\ \bibinfo {pages} {043534} (\bibinfo {year} {2021})},\
  \Eprint {https://arxiv.org/abs/2009.08104} {arXiv:2009.08104 [hep-ph]}
  \BibitemShut {NoStop}%
\bibitem [{\citenamefont {Ng}\ and\ \citenamefont {Beacom}(2014)}]{Ng:2014pca}%
  \BibitemOpen
  \bibfield  {author} {\bibinfo {author} {\bibfnamefont {K.~C.~Y.}\
  \bibnamefont {Ng}}\ and\ \bibinfo {author} {\bibfnamefont {J.~F.}\
  \bibnamefont {Beacom}},\ }\bibfield  {title} {\bibinfo {title} {{Cosmic
  neutrino cascades from secret neutrino interactions}},\ }\href
  {https://doi.org/10.1103/PhysRevD.90.065035} {\bibfield  {journal} {\bibinfo
  {journal} {Phys. Rev. D}\ }\textbf {\bibinfo {volume} {90}},\ \bibinfo
  {pages} {065035} (\bibinfo {year} {2014})},\ \bibinfo {note} {[Erratum:
  Phys.Rev.D 90, 089904 (2014)]},\ \Eprint {https://arxiv.org/abs/1404.2288}
  {arXiv:1404.2288 [astro-ph.HE]} \BibitemShut {NoStop}%
\bibitem [{\citenamefont {Ioka}\ and\ \citenamefont
  {Murase}(2014)}]{Ioka:2014kca}%
  \BibitemOpen
  \bibfield  {author} {\bibinfo {author} {\bibfnamefont {K.}~\bibnamefont
  {Ioka}}\ and\ \bibinfo {author} {\bibfnamefont {K.}~\bibnamefont {Murase}},\
  }\bibfield  {title} {\bibinfo {title} {{IceCube PeV\textendash{}EeV neutrinos
  and secret interactions of neutrinos}},\ }\href
  {https://doi.org/10.1093/ptep/ptu090} {\bibfield  {journal} {\bibinfo
  {journal} {PTEP}\ }\textbf {\bibinfo {volume} {2014}},\ \bibinfo {pages}
  {061E01} (\bibinfo {year} {2014})},\ \Eprint
  {https://arxiv.org/abs/1404.2279} {arXiv:1404.2279 [astro-ph.HE]}
  \BibitemShut {NoStop}%
\bibitem [{\citenamefont {Cherry}\ \emph {et~al.}(2016)\citenamefont {Cherry},
  \citenamefont {Friedland},\ and\ \citenamefont {Shoemaker}}]{Cherry:2016jol}%
  \BibitemOpen
  \bibfield  {author} {\bibinfo {author} {\bibfnamefont {J.~F.}\ \bibnamefont
  {Cherry}}, \bibinfo {author} {\bibfnamefont {A.}~\bibnamefont {Friedland}},\
  and\ \bibinfo {author} {\bibfnamefont {I.~M.}\ \bibnamefont {Shoemaker}},\
  }\href@noop {} {\bibinfo {title} {{Short-baseline neutrino oscillations,
  Planck, and IceCube}}} (\bibinfo {year} {2016}),\ \Eprint
  {https://arxiv.org/abs/1605.06506} {arXiv:1605.06506 [hep-ph]} \BibitemShut
  {NoStop}%
\bibitem [{\citenamefont {Bilenky}\ \emph {et~al.}(1993)\citenamefont
  {Bilenky}, \citenamefont {Bilenky},\ and\ \citenamefont
  {Santamaria}}]{Bilenky:1992xn}%
  \BibitemOpen
  \bibfield  {author} {\bibinfo {author} {\bibfnamefont {M.~S.}\ \bibnamefont
  {Bilenky}}, \bibinfo {author} {\bibfnamefont {S.~M.}\ \bibnamefont
  {Bilenky}},\ and\ \bibinfo {author} {\bibfnamefont {A.}~\bibnamefont
  {Santamaria}},\ }\bibfield  {title} {\bibinfo {title} {{Invisible width of
  the Z boson and 'secret' neutrino-neutrino interactions}},\ }\href
  {https://doi.org/10.1016/0370-2693(93)90703-K} {\bibfield  {journal}
  {\bibinfo  {journal} {Phys. Lett. B}\ }\textbf {\bibinfo {volume} {301}},\
  \bibinfo {pages} {287} (\bibinfo {year} {1993})}\BibitemShut {NoStop}%
\bibitem [{\citenamefont {Bardin}\ \emph {et~al.}(1970)\citenamefont {Bardin},
  \citenamefont {Bilenky},\ and\ \citenamefont {Pontecorvo}}]{Bardin:1970wq}%
  \BibitemOpen
  \bibfield  {author} {\bibinfo {author} {\bibfnamefont {D.~Y.}\ \bibnamefont
  {Bardin}}, \bibinfo {author} {\bibfnamefont {S.~M.}\ \bibnamefont
  {Bilenky}},\ and\ \bibinfo {author} {\bibfnamefont {B.}~\bibnamefont
  {Pontecorvo}},\ }\bibfield  {title} {\bibinfo {title} {{On the nu - nu
  interaction}},\ }\href {https://doi.org/10.1016/0370-2693(70)90602-7}
  {\bibfield  {journal} {\bibinfo  {journal} {Phys. Lett. B}\ }\textbf
  {\bibinfo {volume} {32}},\ \bibinfo {pages} {121} (\bibinfo {year}
  {1970})}\BibitemShut {NoStop}%
\bibitem [{\citenamefont {Bilenky}\ and\ \citenamefont
  {Santamaria}(1999)}]{Bilenky:1999dn}%
  \BibitemOpen
  \bibfield  {author} {\bibinfo {author} {\bibfnamefont {M.~S.}\ \bibnamefont
  {Bilenky}}\ and\ \bibinfo {author} {\bibfnamefont {A.}~\bibnamefont
  {Santamaria}},\ }\bibfield  {title} {\bibinfo {title} {{'Secret' neutrino
  interactions}},\ }in\ \href@noop {} {\emph {\bibinfo {booktitle} {{Neutrino
  Mixing: Meeting in Honor of Samoil Bilenky's 70th Birthday}}}}\ (\bibinfo
  {year} {1999})\ pp.\ \bibinfo {pages} {50--61},\ \Eprint
  {https://arxiv.org/abs/hep-ph/9908272} {arXiv:hep-ph/9908272} \BibitemShut
  {NoStop}%
\bibitem [{\citenamefont {Brdar}\ \emph {et~al.}(2020)\citenamefont {Brdar},
  \citenamefont {Lindner}, \citenamefont {Vogl},\ and\ \citenamefont
  {Xu}}]{Brdar:2020nbj}%
  \BibitemOpen
  \bibfield  {author} {\bibinfo {author} {\bibfnamefont {V.}~\bibnamefont
  {Brdar}}, \bibinfo {author} {\bibfnamefont {M.}~\bibnamefont {Lindner}},
  \bibinfo {author} {\bibfnamefont {S.}~\bibnamefont {Vogl}},\ and\ \bibinfo
  {author} {\bibfnamefont {X.-J.}\ \bibnamefont {Xu}},\ }\bibfield  {title}
  {\bibinfo {title} {{Revisiting neutrino self-interaction constraints from $Z$
  and $\tau$ decays}},\ }\href {https://doi.org/10.1103/PhysRevD.101.115001}
  {\bibfield  {journal} {\bibinfo  {journal} {Phys. Rev. D}\ }\textbf {\bibinfo
  {volume} {101}},\ \bibinfo {pages} {115001} (\bibinfo {year} {2020})},\
  \Eprint {https://arxiv.org/abs/2003.05339} {arXiv:2003.05339 [hep-ph]}
  \BibitemShut {NoStop}%
\bibitem [{\citenamefont {Lyu}\ \emph {et~al.}(2021)\citenamefont {Lyu},
  \citenamefont {Stamou},\ and\ \citenamefont {Wang}}]{Lyu:2020lps}%
  \BibitemOpen
  \bibfield  {author} {\bibinfo {author} {\bibfnamefont {K.-F.}\ \bibnamefont
  {Lyu}}, \bibinfo {author} {\bibfnamefont {E.}~\bibnamefont {Stamou}},\ and\
  \bibinfo {author} {\bibfnamefont {L.-T.}\ \bibnamefont {Wang}},\ }\bibfield
  {title} {\bibinfo {title} {{Self-interacting neutrinos: Solution to Hubble
  tension versus experimental constraints}},\ }\href
  {https://doi.org/10.1103/PhysRevD.103.015004} {\bibfield  {journal} {\bibinfo
   {journal} {Phys. Rev. D}\ }\textbf {\bibinfo {volume} {103}},\ \bibinfo
  {pages} {015004} (\bibinfo {year} {2021})},\ \Eprint
  {https://arxiv.org/abs/2004.10868} {arXiv:2004.10868 [hep-ph]} \BibitemShut
  {NoStop}%
\bibitem [{\citenamefont {Lessa}\ and\ \citenamefont
  {Peres}(2007)}]{Lessa:2007up}%
  \BibitemOpen
  \bibfield  {author} {\bibinfo {author} {\bibfnamefont {A.~P.}\ \bibnamefont
  {Lessa}}\ and\ \bibinfo {author} {\bibfnamefont {O.~L.~G.}\ \bibnamefont
  {Peres}},\ }\bibfield  {title} {\bibinfo {title} {{Revising limits on
  neutrino-Majoron couplings}},\ }\href
  {https://doi.org/10.1103/PhysRevD.75.094001} {\bibfield  {journal} {\bibinfo
  {journal} {Phys. Rev. D}\ }\textbf {\bibinfo {volume} {75}},\ \bibinfo
  {pages} {094001} (\bibinfo {year} {2007})},\ \Eprint
  {https://arxiv.org/abs/hep-ph/0701068} {arXiv:hep-ph/0701068} \BibitemShut
  {NoStop}%
\bibitem [{\citenamefont {Bakhti}\ and\ \citenamefont
  {Farzan}(2017)}]{Bakhti:2017jhm}%
  \BibitemOpen
  \bibfield  {author} {\bibinfo {author} {\bibfnamefont {P.}~\bibnamefont
  {Bakhti}}\ and\ \bibinfo {author} {\bibfnamefont {Y.}~\bibnamefont
  {Farzan}},\ }\bibfield  {title} {\bibinfo {title} {{Constraining secret gauge
  interactions of neutrinos by meson decays}},\ }\href
  {https://doi.org/10.1103/PhysRevD.95.095008} {\bibfield  {journal} {\bibinfo
  {journal} {Phys. Rev. D}\ }\textbf {\bibinfo {volume} {95}},\ \bibinfo
  {pages} {095008} (\bibinfo {year} {2017})},\ \Eprint
  {https://arxiv.org/abs/1702.04187} {arXiv:1702.04187 [hep-ph]} \BibitemShut
  {NoStop}%
\bibitem [{\citenamefont {Arcadi}\ \emph {et~al.}(2019)\citenamefont {Arcadi},
  \citenamefont {Heeck}, \citenamefont {Heizmann}, \citenamefont {Mertens},
  \citenamefont {Queiroz}, \citenamefont {Rodejohann}, \citenamefont
  {Slez\'ak},\ and\ \citenamefont {Valerius}}]{Arcadi:2018xdd}%
  \BibitemOpen
  \bibfield  {author} {\bibinfo {author} {\bibfnamefont {G.}~\bibnamefont
  {Arcadi}}, \bibinfo {author} {\bibfnamefont {J.}~\bibnamefont {Heeck}},
  \bibinfo {author} {\bibfnamefont {F.}~\bibnamefont {Heizmann}}, \bibinfo
  {author} {\bibfnamefont {S.}~\bibnamefont {Mertens}}, \bibinfo {author}
  {\bibfnamefont {F.~S.}\ \bibnamefont {Queiroz}}, \bibinfo {author}
  {\bibfnamefont {W.}~\bibnamefont {Rodejohann}}, \bibinfo {author}
  {\bibfnamefont {M.}~\bibnamefont {Slez\'ak}},\ and\ \bibinfo {author}
  {\bibfnamefont {K.}~\bibnamefont {Valerius}},\ }\bibfield  {title} {\bibinfo
  {title} {{Tritium beta decay with additional emission of new light bosons}},\
  }\href {https://doi.org/10.1007/JHEP01(2019)206} {\bibfield  {journal}
  {\bibinfo  {journal} {JHEP}\ }\textbf {\bibinfo {volume} {01}},\ \bibinfo
  {pages} {206}},\ \Eprint {https://arxiv.org/abs/1811.03530} {arXiv:1811.03530
  [hep-ph]} \BibitemShut {NoStop}%
\bibitem [{\citenamefont {Blinov}\ \emph {et~al.}(2019)\citenamefont {Blinov},
  \citenamefont {Kelly}, \citenamefont {Krnjaic},\ and\ \citenamefont
  {McDermott}}]{Blinov:2019gcj}%
  \BibitemOpen
  \bibfield  {author} {\bibinfo {author} {\bibfnamefont {N.}~\bibnamefont
  {Blinov}}, \bibinfo {author} {\bibfnamefont {K.~J.}\ \bibnamefont {Kelly}},
  \bibinfo {author} {\bibfnamefont {G.~Z.}\ \bibnamefont {Krnjaic}},\ and\
  \bibinfo {author} {\bibfnamefont {S.~D.}\ \bibnamefont {McDermott}},\
  }\bibfield  {title} {\bibinfo {title} {{Constraining the Self-Interacting
  Neutrino Interpretation of the Hubble Tension}},\ }\href
  {https://doi.org/10.1103/PhysRevLett.123.191102} {\bibfield  {journal}
  {\bibinfo  {journal} {Phys. Rev. Lett.}\ }\textbf {\bibinfo {volume} {123}},\
  \bibinfo {pages} {191102} (\bibinfo {year} {2019})},\ \Eprint
  {https://arxiv.org/abs/1905.02727} {arXiv:1905.02727 [astro-ph.CO]}
  \BibitemShut {NoStop}%
\bibitem [{\citenamefont {Blas}\ \emph {et~al.}(2011)\citenamefont {Blas},
  \citenamefont {Lesgourgues},\ and\ \citenamefont {Tram}}]{Blas:2011rf}%
  \BibitemOpen
  \bibfield  {author} {\bibinfo {author} {\bibfnamefont {D.}~\bibnamefont
  {Blas}}, \bibinfo {author} {\bibfnamefont {J.}~\bibnamefont {Lesgourgues}},\
  and\ \bibinfo {author} {\bibfnamefont {T.}~\bibnamefont {Tram}},\ }\bibfield
  {title} {\bibinfo {title} {{The Cosmic Linear Anisotropy Solving System
  (CLASS) II: Approximation schemes}},\ }\href
  {https://doi.org/10.1088/1475-7516/2011/07/034} {\bibfield  {journal}
  {\bibinfo  {journal} {JCAP}\ }\textbf {\bibinfo {volume} {07}},\ \bibinfo
  {pages} {034}},\ \Eprint {https://arxiv.org/abs/1104.2933} {arXiv:1104.2933
  [astro-ph.CO]} \BibitemShut {NoStop}%
\bibitem [{\citenamefont {Chudaykin}\ \emph {et~al.}(2020)\citenamefont
  {Chudaykin}, \citenamefont {Ivanov}, \citenamefont {Philcox},\ and\
  \citenamefont {Simonovi\'c}}]{Chudaykin:2020aoj}%
  \BibitemOpen
  \bibfield  {author} {\bibinfo {author} {\bibfnamefont {A.}~\bibnamefont
  {Chudaykin}}, \bibinfo {author} {\bibfnamefont {M.~M.}\ \bibnamefont
  {Ivanov}}, \bibinfo {author} {\bibfnamefont {O.~H.~E.}\ \bibnamefont
  {Philcox}},\ and\ \bibinfo {author} {\bibfnamefont {M.}~\bibnamefont
  {Simonovi\'c}},\ }\bibfield  {title} {\bibinfo {title} {{Nonlinear
  perturbation theory extension of the Boltzmann code CLASS}},\ }\href
  {https://doi.org/10.1103/PhysRevD.102.063533} {\bibfield  {journal} {\bibinfo
   {journal} {Phys. Rev. D}\ }\textbf {\bibinfo {volume} {102}},\ \bibinfo
  {pages} {063533} (\bibinfo {year} {2020})},\ \Eprint
  {https://arxiv.org/abs/2004.10607} {arXiv:2004.10607 [astro-ph.CO]}
  \BibitemShut {NoStop}%
\bibitem [{\citenamefont {Audren}\ \emph {et~al.}(2013)\citenamefont {Audren},
  \citenamefont {Lesgourgues}, \citenamefont {Benabed},\ and\ \citenamefont
  {Prunet}}]{Audren:2012wb}%
  \BibitemOpen
  \bibfield  {author} {\bibinfo {author} {\bibfnamefont {B.}~\bibnamefont
  {Audren}}, \bibinfo {author} {\bibfnamefont {J.}~\bibnamefont {Lesgourgues}},
  \bibinfo {author} {\bibfnamefont {K.}~\bibnamefont {Benabed}},\ and\ \bibinfo
  {author} {\bibfnamefont {S.}~\bibnamefont {Prunet}},\ }\bibfield  {title}
  {\bibinfo {title} {{Conservative Constraints on Early Cosmology: an
  illustration of the Monte Python cosmological parameter inference code}},\
  }\href {https://doi.org/10.1088/1475-7516/2013/02/001} {\bibfield  {journal}
  {\bibinfo  {journal} {JCAP}\ }\textbf {\bibinfo {volume} {1302}},\ \bibinfo
  {pages} {001}},\ \Eprint {https://arxiv.org/abs/1210.7183} {arXiv:1210.7183
  [astro-ph.CO]} \BibitemShut {NoStop}%
\bibitem [{\citenamefont {Brinckmann}\ and\ \citenamefont
  {Lesgourgues}(2019)}]{Brinckmann:2018cvx}%
  \BibitemOpen
  \bibfield  {author} {\bibinfo {author} {\bibfnamefont {T.}~\bibnamefont
  {Brinckmann}}\ and\ \bibinfo {author} {\bibfnamefont {J.}~\bibnamefont
  {Lesgourgues}},\ }\bibfield  {title} {\bibinfo {title} {{MontePython 3:
  boosted MCMC sampler and other features}},\ }\href
  {https://doi.org/10.1016/j.dark.2018.100260} {\bibfield  {journal} {\bibinfo
  {journal} {Phys. Dark Univ.}\ }\textbf {\bibinfo {volume} {24}},\ \bibinfo
  {pages} {100260} (\bibinfo {year} {2019})},\ \Eprint
  {https://arxiv.org/abs/1804.07261} {arXiv:1804.07261 [astro-ph.CO]}
  \BibitemShut {NoStop}%
\bibitem [{\citenamefont {Skilling}(2006)}]{Skilling:2006gxv}%
  \BibitemOpen
  \bibfield  {author} {\bibinfo {author} {\bibfnamefont {J.}~\bibnamefont
  {Skilling}},\ }\bibfield  {title} {\bibinfo {title} {{Nested sampling for
  general Bayesian computation}},\ }\href {https://doi.org/10.1214/06-BA127}
  {\bibfield  {journal} {\bibinfo  {journal} {Bayesian Analysis}\ }\textbf
  {\bibinfo {volume} {1}},\ \bibinfo {pages} {833} (\bibinfo {year}
  {2006})}\BibitemShut {NoStop}%
\bibitem [{\citenamefont {Sch\"oneberg}\ \emph {et~al.}(2022)\citenamefont
  {Sch\"oneberg}, \citenamefont {Franco~Abell\'an}, \citenamefont
  {P\'erez~S\'anchez}, \citenamefont {Witte}, \citenamefont {Poulin},\ and\
  \citenamefont {Lesgourgues}}]{Schoneberg:2021qvd}%
  \BibitemOpen
  \bibfield  {author} {\bibinfo {author} {\bibfnamefont {N.}~\bibnamefont
  {Sch\"oneberg}}, \bibinfo {author} {\bibfnamefont {G.}~\bibnamefont
  {Franco~Abell\'an}}, \bibinfo {author} {\bibfnamefont {A.}~\bibnamefont
  {P\'erez~S\'anchez}}, \bibinfo {author} {\bibfnamefont {S.~J.}\ \bibnamefont
  {Witte}}, \bibinfo {author} {\bibfnamefont {V.}~\bibnamefont {Poulin}},\ and\
  \bibinfo {author} {\bibfnamefont {J.}~\bibnamefont {Lesgourgues}},\
  }\bibfield  {title} {\bibinfo {title} {{The H0 Olympics: A fair ranking of
  proposed models}},\ }\href {https://doi.org/10.1016/j.physrep.2022.07.001}
  {\bibfield  {journal} {\bibinfo  {journal} {Phys. Rept.}\ }\textbf {\bibinfo
  {volume} {984}},\ \bibinfo {pages} {1} (\bibinfo {year} {2022})},\ \Eprint
  {https://arxiv.org/abs/2107.10291} {arXiv:2107.10291 [astro-ph.CO]}
  \BibitemShut {NoStop}%
\bibitem [{\citenamefont {Gelman}\ and\ \citenamefont
  {Rubin}(1992)}]{10.1214/ss/1177011136}%
  \BibitemOpen
  \bibfield  {author} {\bibinfo {author} {\bibfnamefont {A.}~\bibnamefont
  {Gelman}}\ and\ \bibinfo {author} {\bibfnamefont {D.~B.}\ \bibnamefont
  {Rubin}},\ }\bibfield  {title} {\bibinfo {title} {{Inference from Iterative
  Simulation Using Multiple Sequences}},\ }\href
  {https://doi.org/10.1214/ss/1177011136} {\bibfield  {journal} {\bibinfo
  {journal} {Statistical Science}\ }\textbf {\bibinfo {volume} {7}},\ \bibinfo
  {pages} {457 } (\bibinfo {year} {1992})}\BibitemShut {NoStop}%
\bibitem [{\citenamefont {Cartis}\ \emph {et~al.}(2019)\citenamefont {Cartis},
  \citenamefont {Fiala}, \citenamefont {Marteau},\ and\ \citenamefont
  {Roberts}}]{10.1145/3338517}%
  \BibitemOpen
  \bibfield  {author} {\bibinfo {author} {\bibfnamefont {C.}~\bibnamefont
  {Cartis}}, \bibinfo {author} {\bibfnamefont {J.}~\bibnamefont {Fiala}},
  \bibinfo {author} {\bibfnamefont {B.}~\bibnamefont {Marteau}},\ and\ \bibinfo
  {author} {\bibfnamefont {L.}~\bibnamefont {Roberts}},\ }\bibfield  {title}
  {\bibinfo {title} {Improving the flexibility and robustness of model-based
  derivative-free optimization solvers},\ }\bibfield  {journal} {\bibinfo
  {journal} {ACM Trans. Math. Softw.}\ }\textbf {\bibinfo {volume} {45}},\
  \href {https://doi.org/10.1145/3338517} {10.1145/3338517} (\bibinfo {year}
  {2019})\BibitemShut {NoStop}%
\bibitem [{\citenamefont {{Buchner}}\ \emph {et~al.}(2014)\citenamefont
  {{Buchner}}, \citenamefont {{Georgakakis}}, \citenamefont {{Nandra}},
  \citenamefont {{Hsu}}, \citenamefont {{Rangel}}, \citenamefont {{Brightman}},
  \citenamefont {{Merloni}}, \citenamefont {{Salvato}}, \citenamefont
  {{Donley}},\ and\ \citenamefont {{Kocevski}}}]{Buchner:pymul}%
  \BibitemOpen
  \bibfield  {author} {\bibinfo {author} {\bibfnamefont {J.}~\bibnamefont
  {{Buchner}}}, \bibinfo {author} {\bibfnamefont {A.}~\bibnamefont
  {{Georgakakis}}}, \bibinfo {author} {\bibfnamefont {K.}~\bibnamefont
  {{Nandra}}}, \bibinfo {author} {\bibfnamefont {L.}~\bibnamefont {{Hsu}}},
  \bibinfo {author} {\bibfnamefont {C.}~\bibnamefont {{Rangel}}}, \bibinfo
  {author} {\bibfnamefont {M.}~\bibnamefont {{Brightman}}}, \bibinfo {author}
  {\bibfnamefont {A.}~\bibnamefont {{Merloni}}}, \bibinfo {author}
  {\bibfnamefont {M.}~\bibnamefont {{Salvato}}}, \bibinfo {author}
  {\bibfnamefont {J.}~\bibnamefont {{Donley}}},\ and\ \bibinfo {author}
  {\bibfnamefont {D.}~\bibnamefont {{Kocevski}}},\ }\bibfield  {title}
  {\bibinfo {title} {{X-ray spectral modelling of the AGN obscuring region in
  the CDFS: Bayesian model selection and catalogue}},\ }\href
  {https://doi.org/10.1051/0004-6361/201322971} {\bibfield  {journal} {\bibinfo
   {journal} {\aap}\ }\textbf {\bibinfo {volume} {564}},\ \bibinfo {eid} {A125}
  (\bibinfo {year} {2014})},\ \Eprint {https://arxiv.org/abs/1402.0004}
  {arXiv:1402.0004 [astro-ph.HE]} \BibitemShut {NoStop}%
\bibitem [{\citenamefont {Feroz}\ and\ \citenamefont
  {Hobson}(2008)}]{Feroz:2007kg}%
  \BibitemOpen
  \bibfield  {author} {\bibinfo {author} {\bibfnamefont {F.}~\bibnamefont
  {Feroz}}\ and\ \bibinfo {author} {\bibfnamefont {M.~P.}\ \bibnamefont
  {Hobson}},\ }\bibfield  {title} {\bibinfo {title} {{Multimodal nested
  sampling: an efficient and robust alternative to MCMC methods for
  astronomical data analysis}},\ }\href
  {https://doi.org/10.1111/j.1365-2966.2007.12353.x} {\bibfield  {journal}
  {\bibinfo  {journal} {Mon. Not. Roy. Astron. Soc.}\ }\textbf {\bibinfo
  {volume} {384}},\ \bibinfo {pages} {449} (\bibinfo {year} {2008})},\ \Eprint
  {https://arxiv.org/abs/0704.3704} {arXiv:0704.3704 [astro-ph]} \BibitemShut
  {NoStop}%
\bibitem [{\citenamefont {Feroz}\ \emph {et~al.}(2009)\citenamefont {Feroz},
  \citenamefont {Hobson},\ and\ \citenamefont {Bridges}}]{Feroz:2008xx}%
  \BibitemOpen
  \bibfield  {author} {\bibinfo {author} {\bibfnamefont {F.}~\bibnamefont
  {Feroz}}, \bibinfo {author} {\bibfnamefont {M.~P.}\ \bibnamefont {Hobson}},\
  and\ \bibinfo {author} {\bibfnamefont {M.}~\bibnamefont {Bridges}},\
  }\bibfield  {title} {\bibinfo {title} {{MultiNest: an efficient and robust
  Bayesian inference tool for cosmology and particle physics}},\ }\href
  {https://doi.org/10.1111/j.1365-2966.2009.14548.x} {\bibfield  {journal}
  {\bibinfo  {journal} {Mon. Not. Roy. Astron. Soc.}\ }\textbf {\bibinfo
  {volume} {398}},\ \bibinfo {pages} {1601} (\bibinfo {year} {2009})},\ \Eprint
  {https://arxiv.org/abs/0809.3437} {arXiv:0809.3437 [astro-ph]} \BibitemShut
  {NoStop}%
\bibitem [{\citenamefont {{Higson}}\ \emph {et~al.}(2018)\citenamefont
  {{Higson}}, \citenamefont {{Handley}}, \citenamefont {{Hobson}},\ and\
  \citenamefont {{Lasenby}}}]{Higson:2018aaa}%
  \BibitemOpen
  \bibfield  {author} {\bibinfo {author} {\bibfnamefont {E.}~\bibnamefont
  {{Higson}}}, \bibinfo {author} {\bibfnamefont {W.}~\bibnamefont {{Handley}}},
  \bibinfo {author} {\bibfnamefont {M.}~\bibnamefont {{Hobson}}},\ and\
  \bibinfo {author} {\bibfnamefont {A.}~\bibnamefont {{Lasenby}}},\ }\bibfield
  {title} {\bibinfo {title} {{Sampling Errors in Nested Sampling Parameter
  Estimation}},\ }\href {https://doi.org/10.1214/17-BA1075} {\bibfield
  {journal} {\bibinfo  {journal} {Bayesian Analysis}\ }\textbf {\bibinfo
  {volume} {13}},\ \bibinfo {pages} {873} (\bibinfo {year} {2018})},\ \Eprint
  {https://arxiv.org/abs/1703.09701} {arXiv:1703.09701 [stat.ME]} \BibitemShut
  {NoStop}%
\bibitem [{\citenamefont {Higson}\ \emph {et~al.}(2019)\citenamefont {Higson},
  \citenamefont {Handley}, \citenamefont {Hobson},\ and\ \citenamefont
  {Lasenby}}]{Higson:2018cqj}%
  \BibitemOpen
  \bibfield  {author} {\bibinfo {author} {\bibfnamefont {E.}~\bibnamefont
  {Higson}}, \bibinfo {author} {\bibfnamefont {W.}~\bibnamefont {Handley}},
  \bibinfo {author} {\bibfnamefont {M.}~\bibnamefont {Hobson}},\ and\ \bibinfo
  {author} {\bibfnamefont {A.}~\bibnamefont {Lasenby}},\ }\bibfield  {title}
  {\bibinfo {title} {{Nestcheck: diagnostic tests for nested sampling
  calculations}},\ }\href {https://doi.org/10.1093/mnras/sty3090} {\bibfield
  {journal} {\bibinfo  {journal} {Mon. Not. Roy. Astron. Soc.}\ }\textbf
  {\bibinfo {volume} {483}},\ \bibinfo {pages} {2044} (\bibinfo {year}
  {2019})},\ \Eprint {https://arxiv.org/abs/1804.06406} {arXiv:1804.06406
  [stat.CO]} \BibitemShut {NoStop}%
\bibitem [{\citenamefont {Fowlie}\ \emph {et~al.}(2021)\citenamefont {Fowlie},
  \citenamefont {Handley},\ and\ \citenamefont {Su}}]{Fowlie:2020gfd}%
  \BibitemOpen
  \bibfield  {author} {\bibinfo {author} {\bibfnamefont {A.}~\bibnamefont
  {Fowlie}}, \bibinfo {author} {\bibfnamefont {W.}~\bibnamefont {Handley}},\
  and\ \bibinfo {author} {\bibfnamefont {L.}~\bibnamefont {Su}},\ }\bibfield
  {title} {\bibinfo {title} {{Nested sampling with plateaus}},\ }\href
  {https://doi.org/10.1093/mnras/stab590} {\bibfield  {journal} {\bibinfo
  {journal} {Mon. Not. Roy. Astron. Soc.}\ }\textbf {\bibinfo {volume} {503}},\
  \bibinfo {pages} {1199} (\bibinfo {year} {2021})},\ \Eprint
  {https://arxiv.org/abs/2010.13884} {arXiv:2010.13884 [stat.CO]} \BibitemShut
  {NoStop}%
\bibitem [{\citenamefont {{Buchner}}(2023)}]{Buchner:2021}%
  \BibitemOpen
  \bibfield  {author} {\bibinfo {author} {\bibfnamefont {J.}~\bibnamefont
  {{Buchner}}},\ }\bibfield  {title} {\bibinfo {title} {{Nested Sampling
  Methods}},\ }\href {https://doi.org/10.1214/23-SS144} {\bibfield  {journal}
  {\bibinfo  {journal} {Statistics Surveys}\ }\textbf {\bibinfo {volume}
  {17}},\ \bibinfo {pages} {169} (\bibinfo {year} {2023})},\ \Eprint
  {https://arxiv.org/abs/2101.09675} {arXiv:2101.09675 [stat.CO]} \BibitemShut
  {NoStop}%
\bibitem [{\citenamefont {Ashton}\ \emph {et~al.}(2022)\citenamefont {Ashton}
  \emph {et~al.}}]{Ashton:2022grj}%
  \BibitemOpen
  \bibfield  {author} {\bibinfo {author} {\bibfnamefont {G.}~\bibnamefont
  {Ashton}} \emph {et~al.},\ }\bibfield  {title} {\bibinfo {title} {{Nested
  sampling for physical scientists}},\ }\bibfield  {journal} {\bibinfo
  {journal} {Nature}\ }\textbf {\bibinfo {volume} {2}},\ \href
  {https://doi.org/10.1038/s43586-022-00121-x} {10.1038/s43586-022-00121-x}
  (\bibinfo {year} {2022}),\ \Eprint {https://arxiv.org/abs/2205.15570}
  {arXiv:2205.15570 [stat.CO]} \BibitemShut {NoStop}%
\bibitem [{\citenamefont {Philcox}\ and\ \citenamefont
  {Ivanov}(2022)}]{Philcox:2021kcw}%
  \BibitemOpen
  \bibfield  {author} {\bibinfo {author} {\bibfnamefont {O.~H.~E.}\
  \bibnamefont {Philcox}}\ and\ \bibinfo {author} {\bibfnamefont {M.~M.}\
  \bibnamefont {Ivanov}},\ }\bibfield  {title} {\bibinfo {title} {{BOSS DR12
  full-shape cosmology: \ensuremath{\Lambda}CDM constraints from the
  large-scale galaxy power spectrum and bispectrum monopole}},\ }\href
  {https://doi.org/10.1103/PhysRevD.105.043517} {\bibfield  {journal} {\bibinfo
   {journal} {Phys. Rev. D}\ }\textbf {\bibinfo {volume} {105}},\ \bibinfo
  {pages} {043517} (\bibinfo {year} {2022})},\ \Eprint
  {https://arxiv.org/abs/2112.04515} {arXiv:2112.04515 [astro-ph.CO]}
  \BibitemShut {NoStop}%
\bibitem [{\citenamefont {Chudaykin}\ \emph {et~al.}(2021)\citenamefont
  {Chudaykin}, \citenamefont {Dolgikh},\ and\ \citenamefont
  {Ivanov}}]{Chudaykin:2020ghx}%
  \BibitemOpen
  \bibfield  {author} {\bibinfo {author} {\bibfnamefont {A.}~\bibnamefont
  {Chudaykin}}, \bibinfo {author} {\bibfnamefont {K.}~\bibnamefont {Dolgikh}},\
  and\ \bibinfo {author} {\bibfnamefont {M.~M.}\ \bibnamefont {Ivanov}},\
  }\bibfield  {title} {\bibinfo {title} {{Constraints on the curvature of the
  Universe and dynamical dark energy from the Full-shape and BAO data}},\
  }\href {https://doi.org/10.1103/PhysRevD.103.023507} {\bibfield  {journal}
  {\bibinfo  {journal} {Phys. Rev. D}\ }\textbf {\bibinfo {volume} {103}},\
  \bibinfo {pages} {023507} (\bibinfo {year} {2021})},\ \Eprint
  {https://arxiv.org/abs/2009.10106} {arXiv:2009.10106 [astro-ph.CO]}
  \BibitemShut {NoStop}%
\bibitem [{\citenamefont {Ivanov}\ \emph {et~al.}(2022)\citenamefont {Ivanov},
  \citenamefont {Philcox}, \citenamefont {Simonovi\'c}, \citenamefont
  {Zaldarriaga}, \citenamefont {Nischimichi},\ and\ \citenamefont
  {Takada}}]{Ivanov:2021fbu}%
  \BibitemOpen
  \bibfield  {author} {\bibinfo {author} {\bibfnamefont {M.~M.}\ \bibnamefont
  {Ivanov}}, \bibinfo {author} {\bibfnamefont {O.~H.~E.}\ \bibnamefont
  {Philcox}}, \bibinfo {author} {\bibfnamefont {M.}~\bibnamefont
  {Simonovi\'c}}, \bibinfo {author} {\bibfnamefont {M.}~\bibnamefont
  {Zaldarriaga}}, \bibinfo {author} {\bibfnamefont {T.}~\bibnamefont
  {Nischimichi}},\ and\ \bibinfo {author} {\bibfnamefont {M.}~\bibnamefont
  {Takada}},\ }\bibfield  {title} {\bibinfo {title} {{Cosmological constraints
  without nonlinear redshift-space distortions}},\ }\href
  {https://doi.org/10.1103/PhysRevD.105.043531} {\bibfield  {journal} {\bibinfo
   {journal} {Phys. Rev. D}\ }\textbf {\bibinfo {volume} {105}},\ \bibinfo
  {pages} {043531} (\bibinfo {year} {2022})},\ \Eprint
  {https://arxiv.org/abs/2110.00006} {arXiv:2110.00006 [astro-ph.CO]}
  \BibitemShut {NoStop}%
\bibitem [{\citenamefont {Philcox}(2021)}]{Philcox:2020vbm}%
  \BibitemOpen
  \bibfield  {author} {\bibinfo {author} {\bibfnamefont {O.~H.~E.}\
  \bibnamefont {Philcox}},\ }\bibfield  {title} {\bibinfo {title} {{Cosmology
  without window functions: Quadratic estimators for the galaxy power
  spectrum}},\ }\href {https://doi.org/10.1103/PhysRevD.103.103504} {\bibfield
  {journal} {\bibinfo  {journal} {Phys. Rev. D}\ }\textbf {\bibinfo {volume}
  {103}},\ \bibinfo {pages} {103504} (\bibinfo {year} {2021})},\ \Eprint
  {https://arxiv.org/abs/2012.09389} {arXiv:2012.09389 [astro-ph.CO]}
  \BibitemShut {NoStop}%
\bibitem [{\citenamefont {Eisenstein}\ \emph {et~al.}(2011)\citenamefont
  {Eisenstein} \emph {et~al.}}]{SDSS:2011jap}%
  \BibitemOpen
  \bibfield  {author} {\bibinfo {author} {\bibfnamefont {D.~J.}\ \bibnamefont
  {Eisenstein}} \emph {et~al.} (\bibinfo {collaboration} {SDSS}),\ }\bibfield
  {title} {\bibinfo {title} {{SDSS-III: Massive Spectroscopic Surveys of the
  Distant Universe, the Milky Way Galaxy, and Extra-Solar Planetary Systems}},\
  }\href {https://doi.org/10.1088/0004-6256/142/3/72} {\bibfield  {journal}
  {\bibinfo  {journal} {Astron. J.}\ }\textbf {\bibinfo {volume} {142}},\
  \bibinfo {pages} {72} (\bibinfo {year} {2011})},\ \Eprint
  {https://arxiv.org/abs/1101.1529} {arXiv:1101.1529 [astro-ph.IM]}
  \BibitemShut {NoStop}%
\bibitem [{\citenamefont {Dawson}\ \emph {et~al.}(2013)\citenamefont {Dawson}
  \emph {et~al.}}]{BOSS:2012dmf}%
  \BibitemOpen
  \bibfield  {author} {\bibinfo {author} {\bibfnamefont {K.~S.}\ \bibnamefont
  {Dawson}} \emph {et~al.} (\bibinfo {collaboration} {BOSS}),\ }\bibfield
  {title} {\bibinfo {title} {{The Baryon Oscillation Spectroscopic Survey of
  SDSS-III}},\ }\href {https://doi.org/10.1088/0004-6256/145/1/10} {\bibfield
  {journal} {\bibinfo  {journal} {Astron. J.}\ }\textbf {\bibinfo {volume}
  {145}},\ \bibinfo {pages} {10} (\bibinfo {year} {2013})},\ \Eprint
  {https://arxiv.org/abs/1208.0022} {arXiv:1208.0022 [astro-ph.CO]}
  \BibitemShut {NoStop}%
\bibitem [{\citenamefont {Alam}\ \emph {et~al.}(2017)\citenamefont {Alam} \emph
  {et~al.}}]{BOSS:2016wmc}%
  \BibitemOpen
  \bibfield  {author} {\bibinfo {author} {\bibfnamefont {S.}~\bibnamefont
  {Alam}} \emph {et~al.} (\bibinfo {collaboration} {BOSS}),\ }\bibfield
  {title} {\bibinfo {title} {{The clustering of galaxies in the completed
  SDSS-III Baryon Oscillation Spectroscopic Survey: cosmological analysis of
  the DR12 galaxy sample}},\ }\href {https://doi.org/10.1093/mnras/stx721}
  {\bibfield  {journal} {\bibinfo  {journal} {Mon. Not. Roy. Astron. Soc.}\
  }\textbf {\bibinfo {volume} {470}},\ \bibinfo {pages} {2617} (\bibinfo {year}
  {2017})},\ \Eprint {https://arxiv.org/abs/1607.03155} {arXiv:1607.03155
  [astro-ph.CO]} \BibitemShut {NoStop}%
\bibitem [{\citenamefont {Philcox}\ \emph {et~al.}(2020)\citenamefont
  {Philcox}, \citenamefont {Ivanov}, \citenamefont {Simonovi\'c},\ and\
  \citenamefont {Zaldarriaga}}]{Philcox:2020vvt}%
  \BibitemOpen
  \bibfield  {author} {\bibinfo {author} {\bibfnamefont {O.~H.~E.}\
  \bibnamefont {Philcox}}, \bibinfo {author} {\bibfnamefont {M.~M.}\
  \bibnamefont {Ivanov}}, \bibinfo {author} {\bibfnamefont {M.}~\bibnamefont
  {Simonovi\'c}},\ and\ \bibinfo {author} {\bibfnamefont {M.}~\bibnamefont
  {Zaldarriaga}},\ }\bibfield  {title} {\bibinfo {title} {{Combining Full-Shape
  and BAO Analyses of Galaxy Power Spectra: A 1.6\textbackslash{}\%
  CMB-independent constraint on H$_0$}},\ }\href
  {https://doi.org/10.1088/1475-7516/2020/05/032} {\bibfield  {journal}
  {\bibinfo  {journal} {JCAP}\ }\textbf {\bibinfo {volume} {05}},\ \bibinfo
  {pages} {032}},\ \Eprint {https://arxiv.org/abs/2002.04035} {arXiv:2002.04035
  [astro-ph.CO]} \BibitemShut {NoStop}%
\bibitem [{\citenamefont {Kitaura}\ \emph {et~al.}(2016)\citenamefont {Kitaura}
  \emph {et~al.}}]{Kitaura:2015uqa}%
  \BibitemOpen
  \bibfield  {author} {\bibinfo {author} {\bibfnamefont {F.-S.}\ \bibnamefont
  {Kitaura}} \emph {et~al.},\ }\bibfield  {title} {\bibinfo {title} {{The
  clustering of galaxies in the SDSS-III Baryon Oscillation Spectroscopic
  Survey: mock galaxy catalogues for the BOSS Final Data Release}},\ }\href
  {https://doi.org/10.1093/mnras/stv2826} {\bibfield  {journal} {\bibinfo
  {journal} {Mon. Not. Roy. Astron. Soc.}\ }\textbf {\bibinfo {volume} {456}},\
  \bibinfo {pages} {4156} (\bibinfo {year} {2016})},\ \Eprint
  {https://arxiv.org/abs/1509.06400} {arXiv:1509.06400 [astro-ph.CO]}
  \BibitemShut {NoStop}%
\bibitem [{\citenamefont {Rodr\'\i{}guez-Torres}\ \emph
  {et~al.}(2016)\citenamefont {Rodr\'\i{}guez-Torres} \emph
  {et~al.}}]{Rodriguez-Torres:2015vqa}%
  \BibitemOpen
  \bibfield  {author} {\bibinfo {author} {\bibfnamefont {S.~A.}\ \bibnamefont
  {Rodr\'\i{}guez-Torres}} \emph {et~al.},\ }\bibfield  {title} {\bibinfo
  {title} {{The clustering of galaxies in the SDSS-III Baryon Oscillation
  Spectroscopic Survey: modelling the clustering and halo occupation
  distribution of BOSS CMASS galaxies in the Final Data Release}},\ }\href
  {https://doi.org/10.1093/mnras/stw1014} {\bibfield  {journal} {\bibinfo
  {journal} {Mon. Not. Roy. Astron. Soc.}\ }\textbf {\bibinfo {volume} {460}},\
  \bibinfo {pages} {1173} (\bibinfo {year} {2016})},\ \Eprint
  {https://arxiv.org/abs/1509.06404} {arXiv:1509.06404 [astro-ph.CO]}
  \BibitemShut {NoStop}%
\bibitem [{\citenamefont {Beutler}\ \emph {et~al.}(2011)\citenamefont
  {Beutler}, \citenamefont {Blake}, \citenamefont {Colless}, \citenamefont
  {Jones}, \citenamefont {Staveley-Smith}, \citenamefont {Campbell},
  \citenamefont {Parker}, \citenamefont {Saunders},\ and\ \citenamefont
  {Watson}}]{Beutler:2011hx}%
  \BibitemOpen
  \bibfield  {author} {\bibinfo {author} {\bibfnamefont {F.}~\bibnamefont
  {Beutler}}, \bibinfo {author} {\bibfnamefont {C.}~\bibnamefont {Blake}},
  \bibinfo {author} {\bibfnamefont {M.}~\bibnamefont {Colless}}, \bibinfo
  {author} {\bibfnamefont {D.~H.}\ \bibnamefont {Jones}}, \bibinfo {author}
  {\bibfnamefont {L.}~\bibnamefont {Staveley-Smith}}, \bibinfo {author}
  {\bibfnamefont {L.}~\bibnamefont {Campbell}}, \bibinfo {author}
  {\bibfnamefont {Q.}~\bibnamefont {Parker}}, \bibinfo {author} {\bibfnamefont
  {W.}~\bibnamefont {Saunders}},\ and\ \bibinfo {author} {\bibfnamefont
  {F.}~\bibnamefont {Watson}},\ }\bibfield  {title} {\bibinfo {title} {{The 6dF
  Galaxy Survey: Baryon Acoustic Oscillations and the Local Hubble Constant}},\
  }\href {https://doi.org/10.1111/j.1365-2966.2011.19250.x} {\bibfield
  {journal} {\bibinfo  {journal} {Mon. Not. Roy. Astron. Soc.}\ }\textbf
  {\bibinfo {volume} {416}},\ \bibinfo {pages} {3017} (\bibinfo {year}
  {2011})},\ \Eprint {https://arxiv.org/abs/1106.3366} {arXiv:1106.3366
  [astro-ph.CO]} \BibitemShut {NoStop}%
\bibitem [{\citenamefont {Ross}\ \emph {et~al.}(2015)\citenamefont {Ross},
  \citenamefont {Samushia}, \citenamefont {Howlett}, \citenamefont {Percival},
  \citenamefont {Burden},\ and\ \citenamefont {Manera}}]{Ross:2014qpa}%
  \BibitemOpen
  \bibfield  {author} {\bibinfo {author} {\bibfnamefont {A.~J.}\ \bibnamefont
  {Ross}}, \bibinfo {author} {\bibfnamefont {L.}~\bibnamefont {Samushia}},
  \bibinfo {author} {\bibfnamefont {C.}~\bibnamefont {Howlett}}, \bibinfo
  {author} {\bibfnamefont {W.~J.}\ \bibnamefont {Percival}}, \bibinfo {author}
  {\bibfnamefont {A.}~\bibnamefont {Burden}},\ and\ \bibinfo {author}
  {\bibfnamefont {M.}~\bibnamefont {Manera}},\ }\bibfield  {title} {\bibinfo
  {title} {{The clustering of the SDSS DR7 main Galaxy sample \textendash{} I.
  A 4 per cent distance measure at $z = 0.15$}},\ }\href
  {https://doi.org/10.1093/mnras/stv154} {\bibfield  {journal} {\bibinfo
  {journal} {Mon. Not. Roy. Astron. Soc.}\ }\textbf {\bibinfo {volume} {449}},\
  \bibinfo {pages} {835} (\bibinfo {year} {2015})},\ \Eprint
  {https://arxiv.org/abs/1409.3242} {arXiv:1409.3242 [astro-ph.CO]}
  \BibitemShut {NoStop}%
\bibitem [{\citenamefont {Lewis}(2019)}]{Lewis:2019xzd}%
  \BibitemOpen
  \bibfield  {author} {\bibinfo {author} {\bibfnamefont {A.}~\bibnamefont
  {Lewis}},\ }\bibfield  {title} {\bibinfo {title} {{GetDist: a Python package
  for analysing Monte Carlo samples}},\ }\href {https://getdist.readthedocs.io}
  {\  (\bibinfo {year} {2019})},\ \Eprint {https://arxiv.org/abs/1910.13970}
  {arXiv:1910.13970 [astro-ph.IM]} \BibitemShut {NoStop}%
\bibitem [{\citenamefont {Aghanim}\ \emph
  {et~al.}(2020{\natexlab{b}})\citenamefont {Aghanim} \emph
  {et~al.}}]{Planck:2019nip}%
  \BibitemOpen
  \bibfield  {author} {\bibinfo {author} {\bibfnamefont {N.}~\bibnamefont
  {Aghanim}} \emph {et~al.} (\bibinfo {collaboration} {Planck}),\ }\bibfield
  {title} {\bibinfo {title} {{Planck 2018 results. V. CMB power spectra and
  likelihoods}},\ }\href {https://doi.org/10.1051/0004-6361/201936386}
  {\bibfield  {journal} {\bibinfo  {journal} {Astron. Astrophys.}\ }\textbf
  {\bibinfo {volume} {641}},\ \bibinfo {pages} {A5} (\bibinfo {year}
  {2020}{\natexlab{b}})},\ \Eprint {https://arxiv.org/abs/1907.12875}
  {arXiv:1907.12875 [astro-ph.CO]} \BibitemShut {NoStop}%
\bibitem [{\citenamefont {Abdalla}\ \emph {et~al.}(2022)\citenamefont {Abdalla}
  \emph {et~al.}}]{Abdalla:2022yfr}%
  \BibitemOpen
  \bibfield  {author} {\bibinfo {author} {\bibfnamefont {E.}~\bibnamefont
  {Abdalla}} \emph {et~al.},\ }\bibfield  {title} {\bibinfo {title} {{Cosmology
  intertwined: A review of the particle physics, astrophysics, and cosmology
  associated with the cosmological tensions and anomalies}},\ }\href
  {https://doi.org/10.1016/j.jheap.2022.04.002} {\bibfield  {journal} {\bibinfo
   {journal} {JHEAp}\ }\textbf {\bibinfo {volume} {34}},\ \bibinfo {pages} {49}
  (\bibinfo {year} {2022})},\ \Eprint {https://arxiv.org/abs/2203.06142}
  {arXiv:2203.06142 [astro-ph.CO]} \BibitemShut {NoStop}%
\bibitem [{\citenamefont {Cooke}\ \emph {et~al.}(2018)\citenamefont {Cooke},
  \citenamefont {Pettini},\ and\ \citenamefont {Steidel}}]{Cooke:2017cwo}%
  \BibitemOpen
  \bibfield  {author} {\bibinfo {author} {\bibfnamefont {R.~J.}\ \bibnamefont
  {Cooke}}, \bibinfo {author} {\bibfnamefont {M.}~\bibnamefont {Pettini}},\
  and\ \bibinfo {author} {\bibfnamefont {C.~C.}\ \bibnamefont {Steidel}},\
  }\bibfield  {title} {\bibinfo {title} {{One Percent Determination of the
  Primordial Deuterium Abundance}},\ }\href
  {https://doi.org/10.3847/1538-4357/aaab53} {\bibfield  {journal} {\bibinfo
  {journal} {Astrophys. J.}\ }\textbf {\bibinfo {volume} {855}},\ \bibinfo
  {pages} {102} (\bibinfo {year} {2018})},\ \Eprint
  {https://arxiv.org/abs/1710.11129} {arXiv:1710.11129 [astro-ph.CO]}
  \BibitemShut {NoStop}%
\bibitem [{\citenamefont {Aver}\ \emph {et~al.}(2015)\citenamefont {Aver},
  \citenamefont {Olive},\ and\ \citenamefont {Skillman}}]{Aver:2015iza}%
  \BibitemOpen
  \bibfield  {author} {\bibinfo {author} {\bibfnamefont {E.}~\bibnamefont
  {Aver}}, \bibinfo {author} {\bibfnamefont {K.~A.}\ \bibnamefont {Olive}},\
  and\ \bibinfo {author} {\bibfnamefont {E.~D.}\ \bibnamefont {Skillman}},\
  }\bibfield  {title} {\bibinfo {title} {{The effects of He I
  \ensuremath{\lambda}10830 on helium abundance determinations}},\ }\href
  {https://doi.org/10.1088/1475-7516/2015/07/011} {\bibfield  {journal}
  {\bibinfo  {journal} {JCAP}\ }\textbf {\bibinfo {volume} {07}},\ \bibinfo
  {pages} {011}},\ \Eprint {https://arxiv.org/abs/1503.08146} {arXiv:1503.08146
  [astro-ph.CO]} \BibitemShut {NoStop}%
\bibitem [{\citenamefont {Trotta}\ and\ \citenamefont
  {Melchiorri}(2005)}]{Trotta:2004ty}%
  \BibitemOpen
  \bibfield  {author} {\bibinfo {author} {\bibfnamefont {R.}~\bibnamefont
  {Trotta}}\ and\ \bibinfo {author} {\bibfnamefont {A.}~\bibnamefont
  {Melchiorri}},\ }\bibfield  {title} {\bibinfo {title} {{Indication for
  primordial anisotropies in the neutrino background from WMAP and SDSS}},\
  }\href {https://doi.org/10.1103/PhysRevLett.95.011305} {\bibfield  {journal}
  {\bibinfo  {journal} {Phys. Rev. Lett.}\ }\textbf {\bibinfo {volume} {95}},\
  \bibinfo {pages} {011305} (\bibinfo {year} {2005})},\ \Eprint
  {https://arxiv.org/abs/astro-ph/0412066} {arXiv:astro-ph/0412066}
  \BibitemShut {NoStop}%
\bibitem [{\citenamefont {Akaike}(1974)}]{Akaike:1974ddd}%
  \BibitemOpen
  \bibfield  {author} {\bibinfo {author} {\bibfnamefont {H.}~\bibnamefont
  {Akaike}},\ }\bibfield  {title} {\bibinfo {title} {A new look at the
  statistical model identification},\ }\href
  {https://doi.org/10.1109/TAC.1974.1100705} {\bibfield  {journal} {\bibinfo
  {journal} {IEEE Transactions on Automatic Control}\ }\textbf {\bibinfo
  {volume} {19}},\ \bibinfo {pages} {716} (\bibinfo {year} {1974})}\BibitemShut
  {NoStop}%
\bibitem [{\citenamefont {Hu}\ \emph {et~al.}(1998)\citenamefont {Hu},
  \citenamefont {Eisenstein},\ and\ \citenamefont {Tegmark}}]{Hu:1997mj}%
  \BibitemOpen
  \bibfield  {author} {\bibinfo {author} {\bibfnamefont {W.}~\bibnamefont
  {Hu}}, \bibinfo {author} {\bibfnamefont {D.~J.}\ \bibnamefont {Eisenstein}},\
  and\ \bibinfo {author} {\bibfnamefont {M.}~\bibnamefont {Tegmark}},\
  }\bibfield  {title} {\bibinfo {title} {{Weighing neutrinos with galaxy
  surveys}},\ }\href {https://doi.org/10.1103/PhysRevLett.80.5255} {\bibfield
  {journal} {\bibinfo  {journal} {Phys. Rev. Lett.}\ }\textbf {\bibinfo
  {volume} {80}},\ \bibinfo {pages} {5255} (\bibinfo {year} {1998})},\ \Eprint
  {https://arxiv.org/abs/astro-ph/9712057} {arXiv:astro-ph/9712057}
  \BibitemShut {NoStop}%
\bibitem [{\citenamefont {Lesgourgues}\ and\ \citenamefont
  {Pastor}(2006)}]{Lesgourgues:2006nd}%
  \BibitemOpen
  \bibfield  {author} {\bibinfo {author} {\bibfnamefont {J.}~\bibnamefont
  {Lesgourgues}}\ and\ \bibinfo {author} {\bibfnamefont {S.}~\bibnamefont
  {Pastor}},\ }\bibfield  {title} {\bibinfo {title} {{Massive neutrinos and
  cosmology}},\ }\href {https://doi.org/10.1016/j.physrep.2006.04.001}
  {\bibfield  {journal} {\bibinfo  {journal} {Phys. Rept.}\ }\textbf {\bibinfo
  {volume} {429}},\ \bibinfo {pages} {307} (\bibinfo {year} {2006})},\ \Eprint
  {https://arxiv.org/abs/astro-ph/0603494} {arXiv:astro-ph/0603494}
  \BibitemShut {NoStop}%
\bibitem [{\citenamefont {Brinckmann}\ \emph {et~al.}(2022)\citenamefont
  {Brinckmann}, \citenamefont {Chang}, \citenamefont {Du},\ and\ \citenamefont
  {LoVerde}}]{Brinckmann:2022ajr}%
  \BibitemOpen
  \bibfield  {author} {\bibinfo {author} {\bibfnamefont {T.}~\bibnamefont
  {Brinckmann}}, \bibinfo {author} {\bibfnamefont {J.~H.}\ \bibnamefont
  {Chang}}, \bibinfo {author} {\bibfnamefont {P.}~\bibnamefont {Du}},\ and\
  \bibinfo {author} {\bibfnamefont {M.}~\bibnamefont {LoVerde}},\ }\bibfield
  {title} {\bibinfo {title} {{Confronting interacting dark radiation scenarios
  with cosmological data}},\ }\href@noop {} {\  (\bibinfo {year} {2022})},\
  \Eprint {https://arxiv.org/abs/2212.13264} {arXiv:2212.13264 [astro-ph.CO]}
  \BibitemShut {NoStop}%
\bibitem [{\citenamefont {Aloni}\ \emph {et~al.}(2023)\citenamefont {Aloni},
  \citenamefont {Joseph}, \citenamefont {Schmaltz},\ and\ \citenamefont
  {Weiner}}]{Aloni:2023tff}%
  \BibitemOpen
  \bibfield  {author} {\bibinfo {author} {\bibfnamefont {D.}~\bibnamefont
  {Aloni}}, \bibinfo {author} {\bibfnamefont {M.}~\bibnamefont {Joseph}},
  \bibinfo {author} {\bibfnamefont {M.}~\bibnamefont {Schmaltz}},\ and\
  \bibinfo {author} {\bibfnamefont {N.}~\bibnamefont {Weiner}},\ }\bibfield
  {title} {\bibinfo {title} {{Dark Radiation from Neutrino Mixing after Big
  Bang Nucleosynthesis}},\ }\href@noop {} {\  (\bibinfo {year} {2023})},\
  \Eprint {https://arxiv.org/abs/2301.10792} {arXiv:2301.10792 [astro-ph.CO]}
  \BibitemShut {NoStop}%
\bibitem [{\citenamefont {Handley}(2019)}]{Handley:2019mfs}%
  \BibitemOpen
  \bibfield  {author} {\bibinfo {author} {\bibfnamefont {W.}~\bibnamefont
  {Handley}},\ }\bibfield  {title} {\bibinfo {title} {{anesthetic: nested
  sampling visualisation}},\ }\href {https://doi.org/10.21105/joss.01414}
  {\bibfield  {journal} {\bibinfo  {journal} {J. Open Source Softw.}\ }\textbf
  {\bibinfo {volume} {4}},\ \bibinfo {pages} {1414} (\bibinfo {year} {2019})},\
  \Eprint {https://arxiv.org/abs/1905.04768} {arXiv:1905.04768 [astro-ph.IM]}
  \BibitemShut {NoStop}%
\bibitem [{\citenamefont {Fowlie}\ \emph {et~al.}(2020)\citenamefont {Fowlie},
  \citenamefont {Handley},\ and\ \citenamefont {Su}}]{Fowlie:2020mzs}%
  \BibitemOpen
  \bibfield  {author} {\bibinfo {author} {\bibfnamefont {A.}~\bibnamefont
  {Fowlie}}, \bibinfo {author} {\bibfnamefont {W.}~\bibnamefont {Handley}},\
  and\ \bibinfo {author} {\bibfnamefont {L.}~\bibnamefont {Su}},\ }\bibfield
  {title} {\bibinfo {title} {{Nested sampling cross-checks using order
  statistics}},\ }\href {https://doi.org/10.1093/mnras/staa2345} {\bibfield
  {journal} {\bibinfo  {journal} {Mon. Not. Roy. Astron. Soc.}\ }\textbf
  {\bibinfo {volume} {497}},\ \bibinfo {pages} {5256} (\bibinfo {year}
  {2020})},\ \Eprint {https://arxiv.org/abs/2006.03371} {arXiv:2006.03371
  [stat.CO]} \BibitemShut {NoStop}%
\bibitem [{\citenamefont {Campello}\ \emph {et~al.}(2013)\citenamefont
  {Campello}, \citenamefont {Moulavi},\ and\ \citenamefont
  {Sander}}]{Campello:2013aaa}%
  \BibitemOpen
  \bibfield  {author} {\bibinfo {author} {\bibfnamefont {R.~J. G.~B.}\
  \bibnamefont {Campello}}, \bibinfo {author} {\bibfnamefont {D.}~\bibnamefont
  {Moulavi}},\ and\ \bibinfo {author} {\bibfnamefont {J.}~\bibnamefont
  {Sander}},\ }\bibfield  {title} {\bibinfo {title} {Density-based clustering
  based on hierarchical density estimates},\ }in\ \href@noop {} {\emph
  {\bibinfo {booktitle} {Advances in Knowledge Discovery and Data Mining}}},\
  \bibinfo {editor} {edited by\ \bibinfo {editor} {\bibfnamefont
  {J.}~\bibnamefont {Pei}}, \bibinfo {editor} {\bibfnamefont {V.~S.}\
  \bibnamefont {Tseng}}, \bibinfo {editor} {\bibfnamefont {L.}~\bibnamefont
  {Cao}}, \bibinfo {editor} {\bibfnamefont {H.}~\bibnamefont {Motoda}},\ and\
  \bibinfo {editor} {\bibfnamefont {G.}~\bibnamefont {Xu}}}\ (\bibinfo
  {publisher} {Springer Berlin Heidelberg},\ \bibinfo {address} {Berlin,
  Heidelberg},\ \bibinfo {year} {2013})\ pp.\ \bibinfo {pages}
  {160--172}\BibitemShut {NoStop}%
\bibitem [{\citenamefont {Campello}\ \emph {et~al.}(2015)\citenamefont
  {Campello}, \citenamefont {Moulavi}, \citenamefont {Zimek},\ and\
  \citenamefont {Sander}}]{Campello:2015aaa}%
  \BibitemOpen
  \bibfield  {author} {\bibinfo {author} {\bibfnamefont {R.~J. G.~B.}\
  \bibnamefont {Campello}}, \bibinfo {author} {\bibfnamefont {D.}~\bibnamefont
  {Moulavi}}, \bibinfo {author} {\bibfnamefont {A.}~\bibnamefont {Zimek}},\
  and\ \bibinfo {author} {\bibfnamefont {J.}~\bibnamefont {Sander}},\
  }\bibfield  {title} {\bibinfo {title} {Hierarchical density estimates for
  data clustering, visualization, and outlier detection},\ }\bibfield
  {journal} {\bibinfo  {journal} {ACM Trans. Knowl. Discov. Data}\ }\textbf
  {\bibinfo {volume} {10}},\ \href {https://doi.org/10.1145/2733381}
  {10.1145/2733381} (\bibinfo {year} {2015})\BibitemShut {NoStop}%
\bibitem [{\citenamefont {McInnes}\ and\ \citenamefont
  {Healy}(2017)}]{McInnes:2017aaa}%
  \BibitemOpen
  \bibfield  {author} {\bibinfo {author} {\bibfnamefont {L.}~\bibnamefont
  {McInnes}}\ and\ \bibinfo {author} {\bibfnamefont {J.}~\bibnamefont
  {Healy}},\ }\bibfield  {title} {\bibinfo {title} {Accelerated hierarchical
  density based clustering},\ }in\ \href
  {https://doi.org/10.1109/icdmw.2017.12} {\emph {\bibinfo {booktitle} {2017
  IEEE International Conference on Data Mining Workshops (ICDMW)}}}\ (\bibinfo
  {publisher} {IEEE},\ \bibinfo {year} {2017})\BibitemShut {NoStop}%
\bibitem [{\citenamefont {Swendsen}\ and\ \citenamefont
  {Wang}(1986)}]{Swendsen:1986}%
  \BibitemOpen
  \bibfield  {author} {\bibinfo {author} {\bibfnamefont {R.}~\bibnamefont
  {Swendsen}}\ and\ \bibinfo {author} {\bibfnamefont {J.-S.}\ \bibnamefont
  {Wang}},\ }\bibfield  {title} {\bibinfo {title} {Replica monte carlo
  simulation of spin-glasses},\ }\href
  {https://doi.org/10.1103/PhysRevLett.57.2607} {\bibfield  {journal} {\bibinfo
   {journal} {Physical review letters}\ }\textbf {\bibinfo {volume} {57}},\
  \bibinfo {pages} {2607} (\bibinfo {year} {1986})}\BibitemShut {NoStop}%
\bibitem [{\citenamefont {Freedman}\ and\ \citenamefont
  {Diaconis}(1981)}]{Freedman1981OnTH}%
  \BibitemOpen
  \bibfield  {author} {\bibinfo {author} {\bibfnamefont {D.~A.}\ \bibnamefont
  {Freedman}}\ and\ \bibinfo {author} {\bibfnamefont {P.}~\bibnamefont
  {Diaconis}},\ }\bibfield  {title} {\bibinfo {title} {On the histogram as a
  density estimator:l2 theory},\ }\href
  {https://api.semanticscholar.org/CorpusID:14437088} {\bibfield  {journal}
  {\bibinfo  {journal} {Zeitschrift f{\"u}r Wahrscheinlichkeitstheorie und
  Verwandte Gebiete}\ }\textbf {\bibinfo {volume} {57}},\ \bibinfo {pages}
  {453} (\bibinfo {year} {1981})}\BibitemShut {NoStop}%
\end{thebibliography}%

\end{document}